\documentclass[
 aps,prb,
 amsmath,amssymb,
 reprint,
]{revtex4-2}

\usepackage[dvipdfmx]{graphicx}
\usepackage{color}
\usepackage{dcolumn}
\usepackage{bm}

\usepackage{midpage}
\usepackage{here}

\begin{document}


\title
{
Two liquid states of distinguishable helium-4: the
existence of another non-superfluid frozen  by heating
}

\author{Momoko Tsujimoto}
 \altaffiliation[]{Present address: KYOCERA Communication Systems Co., Ltd., Takeda-tobadono, Fushimi-ku, Kyoto 612-8450, Japan.}
\author{Kenichi Kinugawa}%
\email{Author to whom correspondence should be addressed.  
kinugawa@cc.nara-wu.ac.jp}
\affiliation{ 
Department of Chemistry, Graduate School of Humanities and Sciences, Nara Women's University, Nara 630-8506, Japan
}

\begin{abstract}

We show   
 that there can  exist two liquid states  in distinguishable helium-4 ($^4$He)
obeying Boltzmann statistics
 by  path integral centroid molecular dynamics (CMD) simulations. 
This is an indication of quantum liquid polyamorphism induced by  nuclear quantum effect.
For 0.08-3.3 K and 1-500 bar, we extensively conducted the isothermal-isobaric CMD simulations  to explore  not only  possible states and state diagram but the state 
  characteristics.  
  The distinguishable $^4$He  below 25 bar does not freeze down to 0.1 K even though it includes no Bosonic exchange effect and therefore no Bose condensation.
One liquid state, low quantum-dispersion liquid (LQDL), is nearly identical  to normal
  liquid He-I of real $^4$He.
The other is high quantum-dispersion liquid (HQDL) consisting of
atoms with longer quantum wavelength.
This is another  non-superfluid existing 
 below 0.5 K or the 
  temperatures of LQDL.
  The HQDL is also a low-entropy and fragile liquid to  exhibit, unlike conventional liquids, rather gas-like relaxation of 
  velocity autocorrelation function, while there the atoms diffuse without noticeable contribution 
  from quantum tunneling.
  The LQDL-HQDL transition is not
  a thermodynamic phase transition but a continuous  crossover accompanied by 
  the change of the expansion factor of quantum wavelength.
  Freezing of HQDL into the low quantum-dispersion amorphous solid
  occurs by heating from  0.2 to 0.3 K at 40-50 bar, while this $P$-$T$ condition coincides with the Kim-Chan normal-supersolid phase boundary of real $^4$He.
The obtained state diagram was  compared with that of the  confined subnano-scale $^4$He systems  where
Bosonic correlation is considerably suppressed.
\end{abstract}

\maketitle

\section{\label{sec:introduction}INTRODUCTION}

Is it a true postulation that every substance should freeze into its solid phase as it is cooled down toward 
absolute zero temperature?
While classical systems inevitably freeze by losing kinetic energy proportional to temperature,
 anti-freezing of quantum systems is benefited  by promotion of atomic diffusivity due to
  increasing quantum  atomic fluctuation down to zero temperature.
Indeed, we know an exceptionally  non-freezing quantum substance, helium-4 ($^4$He), which  retains
 the superfluid (He-II) phase  which never freezes down to zero temperature under atmospheric 
pressure.
The cause of  superfluidity includes: (i) the nuclear quantum effect (NQE), i.e., 
the atomic wave nature parametrized as the de Broglie thermal wavelength
$\lambda_{\rm{dB}}:=\hbar\sqrt{2\pi/mk_{\rm{B}}T}$ ($m$ is the atomic mass and $k_{\rm{B}}$ is the Boltzmann constant), 
(ii) the weakness of the van der Waals interatomic forces, 
and (iii) the atomic indistinguishableness due to the   Bosonic  permutation \cite{feynman1972}.
Primitively, the $\rm{\lambda}$-type isobaric heat capacity  $C_P$ at the  transition between normal liquid (He-I) and 
He-II  of $^4$He is explained in terms of the Bose-Einstein condensation (BEC) 
of ideal Bose gas \cite{london1938,london1938pr,pitaevskii2016}.  
This leads to a conventional understanding  that the  Bose statistics is the key to the 
$\lambda$ transition and causes the BEC through which non-freezing of this quantum substance is achieved.
Then,  if there were solely the NQE and no Bosonic exchange were imposed on atoms, could such a 
distinguishable $^4$He model avoid freezing to retain the  normal liquid phase 
when cooled?
According to Feynman \cite{feynman1972}, it was argued that there was no difference between the ground state wave function of an ensemble of Bosons and 
of distinguishable particles \cite{boninsegni2012}.
This indicates that 
 distinguishable $^4$He at atmospheric pressure  is  liquid at $T=0$
 as is the Bosonic  system.
As  discussed \cite{boninsegni2012}, 
in the textbooks \cite{feynman1972,huang1987}, 
the non-freezing property of $^4$He
was explained by considering only the two factors
(i) the NQE of individual atoms and (ii)
the weak interatomic forces \cite{huang1987}.
However, Boninsegni et al. reported  that  
the third factor, (iii) the Bosonic exchange among $N$ atoms, was indispensable to retaining the non-freezing property by being in superfluid state;
the crystal of distinguishable $^4$He melted 
 by introducing the Bosonic correlation \cite{boninsegni2012}.
Therefore, the state of distinguishable $^4$He would not always be the same as
that of the Bosonic counterpart in general.
However, for a wider range of pressure and non-zero low temperature, 
it has  not yet been completely explored
what  kind of states emerges by the sole NQE in distinguishable $^4$He.

The non-freezing tendency of a quantum substance 
might be reflected to the ease of
residual diffusion \cite{bernu1987} of atoms in its quenched glassy solid.
In this connection, Markland et al. conducted the discretized path integral simulations based on the quantum-classical isomorphism, where quantum atoms are expressed
as  classical ring polymers (necklaces), to show the reentrant behavior of   
self-diffusion coefficient
$D$ in the glassy solid of a binary Lennard-Jones model.  In their study, 
increasing the 
quantumness parameter $\Lambda^{*}:=\hbar/\sqrt{mk_{\rm{B}}T\sigma^2}$
 ($\sigma$ is the closest interatomic distance) caused the  transition from  a glass state
involving trapped configurations of atomic necklaces 
(called {\it{trapped regime}}) 
 to another state with stretched configurations of atoms
({\it{tunneling regime}}) \cite{markland2011,markland2012}.
As  $\Lambda^{*}$ increased, 
  $D$  once decreased in the former regime  but it turned to increase in the latter \cite{markland2011,markland2012}.  This also suggested  
a  tendency toward  inverse melting (or {\it{melting by cooling}}) of this quantum 
system \cite{markland2012}.
In fact,  such an  atomic tunneling between local potential minima explains  the linear temperature
dependence of heat capacity in low-temperature 
glasses \cite{anderson1972,phillips1972}.

The existence of more than one  states in liquids
or glasses is called polyamorphism \cite{stanley2013,tanaka2020}, which  is, for example, observed for 
classical {\it{fragile}} liquids \cite{ediger1996} such as  water \cite{mishima3921998,mishima3961998} and phosphorous 
 \cite{katayama2004}. 
The existence of two liquid states, He-I and He-II,
 in $^4$He is certainly an example of liquid polyamorphism.
 However, the polyamorphism of the other  quantum systems has rarely been discovered so far 
 \cite{morales2010,nguyen2018,eltareb2022}. 
Recently, our path integral centroid molcular dynamics (CMD) simulations showed  a quantum polyamorphism of compressed distinguishable $^4$He glass above 3 K \cite{kinugawa2021}.
In good accordance with the two regimes reported by Markland et al. \cite{markland2011,markland2012},
compressed distinguishable  $^4$He also
 involved two glass states \cite{kinugawa2021}.
 The two  states were distinguished by the difference in atomic quantum wavelength
$\lambda_{\rm{quantum}}$ which is equivalent to two times {\it{radius of gyration}}
of atomic necklaces \cite{kinugawa2021}. 
Then we call the two states, depending on $\lambda_{\rm{quantum}}$,  the low quantum dispersion amorphous solid (LQDA) and  the high quantum dispersion solid (HQDA) \cite{kinugawa2021} 
\footnote{Here we should mention the abbreviation of these names. 
In Ref.~\onlinecite{kinugawa2021}, we abbreviated them as ``LDA'' and ``HDA'', respectively. 
However, since these  may be confused with the low and high density amorphous state of classical 
systems (e.g., water \cite{mishima3921998,mishima3961998}), we presently  call  them LQDA and HQDA instead.}
The LQDA and HQDA are  equivalent to  the
trapped  and the tunneling regimes
in  the quantum Lennard-Jones model \cite{markland2011,markland2012}, respectively.
The LQDA-HQDA transition was reversibly caused    by varying temperature or pressure, while this was 
  akin to 
the coil-globule transition of classical polymers 
\cite{lifshitz1978,maffi2012,wu1998,podewitz2019,ma1995,simmons2008,simmons2010,sherman2006,matsuyama1991}.
In accordance with the reports by Markland et al. \cite{markland2011,markland2012}, 
 increasing the quantumness parameter by  compression or  cooling 
promoted the residual diffusion of atoms in  the HQDA state.
Thus,  the NQE is certainly a cause of the polyamorphism and the enhancement of 
 atomic diffusivity  in these quantum glasses.

For distinguishable $^4$He, we  anticipate that  
a liquid  polyamorphism may occur as the counterpart of  observed 
glass polyamorphism \cite{kinugawa2021},
since the glass structure  is produced by quenching the liquid configurations.  
Another  reason for this anticipation  
is that  an extrapolation 
of the LQDA-HQDA  boundary line  on  the 
 state diagram \cite{kinugawa2021} (which will be shown later in Fig.~\ref{fig:diagramlarge}
  in this paper) 
 would reach
 a particular region  ($T\simeq0.2$ K and $P\simeq26$ bars) at which 
 Kim and Chan experimentally observed the torsional oscillation anomaly suggesting 
 the existence of  {\it{supersolid}} \cite{kim2004}. 
Although   subsequent researches provided  more
 freedom in the interpretation of their measured results \cite{hallock2015}, 
 some unusual state  must be emerging in  real  $^4$He at this $P$-$T$ condition.
The third reason  
is that   the experimental melting curves of both $^4$He  and $^3$He are slightly downward 
convex possessing  shallow minima 
on the $P$-vs-$T$ plane ($^4$He: 0.8 K and 26 bars; $^3$He: 0.3 K and 29 bar) \cite{wilks1970}.  
 In fact, the existence of an minimum  of 
melting curve is  a thermodynamical implication of   liquid polyamorphism \cite{rapoport1967,stanley2013,tanaka2020}.
In general, when a state point on the $P$-$T$ plane moves across a convex melting curve,
even an  inverse freezing (freezing by heating) or an inverse melting (melting by cooling) occurs \cite{stillinger2001,stillinger2003,feeney2003,schupper2005,prestipino2007,greer2000}.
For these reasons,  it is expected  
that  two liquid states  emerge in 
distinguishable $^4$He including the occurrence of inverse melting.
Nevertheless, this issue  remains unresolved, while 
the phase diagram of distinguishable $^4$He has
never been revealed  except for a presumed  outline \cite{boninsegni2012}.

Motivated by these  questions, in the present study we  started to undertake
 the isothermal-isobaric CMD simulation of  the  distinguishable $^4$He model.  
 The surveyed range 
 spanned 0.08-3.3 K and 1-500 bar, involving   357   mesh points which covered the $P$-$T$ conditions of 
  He-I, He-II, and solid phase of real $^4$He.
 This is an ideal model  to investigate what kind of  states appears  in  quantum liquid 
 as a result of the  sole NQE without inclusion of Bosonic exchange correlation.

 This paper is organized as follows. Section ~\ref{sec:method}  describes the  CMD simulation
  method.   
 The results of the isothermal compression simulation are described in Sec.~\ref{sec:resultsofiosthermalcompression}. 
 At the beginning of this section,  we  provide the state diagram on the $P$-$T$ plane as a summary of
the  isothermal compression simulation.
 And then we show  the results of static and dynamic properties supporting the conclusion of this
 state diagram. 
  In Sec.~\ref{sec:onstatediagram}, we present the detailed analysis on the basis of
  the state diagram.
    The discussions are given in Sec.~\ref{sec:discussions}.  Finally, we describe the  
  conclusions in Sec.~\ref{sec:conclusions}.
  Appendix provides the results of another  CMD simulation conducted to examine  
 the occurrence of inverse freezing,
  and also gives comparisons of  the state diagram with that of the real confined $^4$He systems  and  reported  {\it{supersolid}}.

In the Supplementary Material, we provide  a substantial amount of additional figures and data.

\section{\label{sec:method}METHOD}

\subsection{\label{sec:CMDsimulation}Isothermal-isobaric CMD}
To explore the phase equilibrium of substances, the investigation should be undertaken for an 
 isothermal-isobaric ensemble because  thermodynamic phase is uniquely determined under a given condition of $P$ and $T$ \cite{landau1980}. 
For a quantum system, the discretized path integral representation   provides the quantum-classical isomorphism  that 
the partition function 
of  a quantum system consisting of $N$ quantum atoms is equivalent to that of a classical system composed of $N$ necklaces (ring polymers);  
each of necklaces consists of $N_{\rm{b}}$ beads connected with the springs \cite{chandler1981}.
Then the partition function of an isothermal-isobaric ensemble of $N$ distinguishable $^4$He atoms at inverse temperature 
$\beta=(k_{\rm{B}}T)^{-1}$ and external pressure $P_{\rm{ex}}$ is then 
\begin{widetext}
\begin{eqnarray}
\label{eq:partition}
\Delta_{NPT}=
\frac{1}{v}
{\int}
dV
\exp(-{\beta}P_{\rm{ex}}V)
\int\cdots\int
\prod^{N}_{i=1}
d{\it{\bf{r}}}_{{\rm{c}}i}\rho_{\rm{c}}
(\{{\it{\bf{r}}}_{{\rm{c}}i}\}),
\end{eqnarray}   
where $v$ is a certain quantity of volume \cite{frenkel2002}
and $\rho_{\rm{c}}$ is the centroid density,
\begin{eqnarray}
\label{eq:centroiddensity}
\rho_{\rm{c}}
\left(\{{\it{\bf{r}}}_{{\rm{c}}i}\}\right)&=&\frac{1}{N!}\lim_{N_{\rm{b}}\rightarrow\infty}\left(\frac{m}{2\pi\beta\hbar^2}\right)^{\frac{3NN_{\rm{b}}}{2}}\int\cdots\int\prod^{N}_{i=1}\prod^{N_{\rm{b}}}_{j=1}
d{\it{\bf{r}}}_i^{(j)} \delta({\it{\bf{r}}}_{{\rm{c}}i}-{\it{\bf{\bar{r}}}}_{i})\exp\left[-{\beta}H_{\rm{system}}(\{\it{\bf{r}}_{i}^{(j)}\})\right].
\end{eqnarray}  
Here, the system Hamiltonian $H_{\rm{system}}$ is \cite{chandler1981}
\begin{eqnarray}
\label{eq:hamiltonian}
H_{\rm{system}}
\left(
\{{\it{\bf{r}}}_{i}^{(j)}\}
\right)
=\sum_{i=1}^{N} \sum_{j=1}^{N_{\rm{b}}}
\frac{1}{2}k_{\rm{s}}\left({\it{\bf{r}}}_i^{(j)}-{\it{\bf{r}}}_i^{(j+1)}\right)^2
+\frac{1}{N_{\rm{b}}}
\Phi\left(\{\it{\bf{r}}_i^{(j)}\}\right), 
\end{eqnarray}   
\end{widetext}
$N_{\rm{b}}$ is the discretization number (the Trotter number),  $m$ is the atomic mass, 
${\it{\bf{r}}}_i^{(j)}$ is the position of the $j$-th bead (i.e., the atomic 
position at imaginary time $\tau_j=j\beta\hbar/N_{\rm{b}}$ $(0<{\tau_j}\leq\beta\hbar$)) of the $i$-th atom,
 ${\it{\bf{\bar{r}}}}_{i}$ 
is the imaginary-time averaged position, 
${\it{\bf{\bar{r}}}}_{i}=N_{\rm{b}}^{-1}\sum_{j=1}^{N_{\rm{b}}}{\it{\bf{r}}}_i^{(j)}$,
$k_{\rm{s}}$ is the Hooke constant, $k_{\rm{s}}=mN_{\rm{b}}/\beta^2\hbar^2$,
 and 
$\Phi$ is the system potential arising from pairwise interatomic interactions.
Since the cyclic boundary condition 
 ${\it{\bf{r}}}_i^{(N_{\rm{b}}+1)}= {\it{\bf{r}}}_i^{(1)}$
is imposed, the 
isomorphic classical system  consists of $N$ necklaces
where  $N_{\rm{b}}$ beads are sequentially connected through the springs with the Hooke constant 
$k_{\rm{s}}$.  Therefore, the springs become looser as the temperature falls.

On the basis of this representation, we conducted the isothermal-isobaric 
Nos\'{e}-Hoover-chain-Andersen-type
normal-mode CMD (NMCMD) simulations for a bulk $^4$He system of 256 atoms
contained in a cubic box under the periodic boundary condition.
The simulation technique was the same as our recent work \cite{kinugawa2021}.
The pair He-He interatomic potential was Aziz's
HFD-B3-FCI1 type \cite{aziz1995}, which was also adopted in our recent 
study \cite{kinugawa2021,takemoto2018}.  
The centroid equation of motion for an isothermal-isobaric ensemble is
\begin{eqnarray}
\label{eq:eom}
m{\it{\bf{\ddot{r}}}}_{{\rm{c}}i}
&=&
{\it{\bf{F}}}_{{\rm{c}}i}-m\dot{\xi}_1\dot{{\it{\bf{r}}}}_{{\rm{c}}i}-\frac{m}{N}\dot{\epsilon}\dot{\it{\bf{r}}}_{{\rm{c}}i},\;\;
i=1,2,\cdots,N,
\end{eqnarray}   
where $\dot{\xi}_1$ is the velocity of the first layer of the Nos\'e-Hoover chain thermostat,  
$\dot{\epsilon}$ is the velocity of the Andersen barostat, and ${\it{\bf{F}}}_{{\rm{c}}i}$ is the 
interatomic force arising from  $\Phi$,
\begin{eqnarray}
\label{eq:force}
{\it{\bf{F}}}_{{\rm{c}}i}
&=&
-\frac{{\partial}{H}_{\rm{system}}\left(
\{{\it{\bf{r}}}_{i}^{(j)}\}
\right)}
{{\partial}{\it{\bf{r}}}_{{\rm{c}}i}}\nonumber\\
&=&
-\frac{1}{N_{\rm{b}}}\sum_{j=1}^{N_{\rm{b}}}
\frac{{\partial}{\Phi}\left(
\{{\it{\bf{r}}}_{i}^{(l)}\}
\right)}
{{\partial}{\it{\bf{r}}}_{i}^{(j)}} 
\equiv
\frac{1}{N_{\rm{b}}}\sum_{j=1}^{N_{\rm{b}}}{\it{\bf{f}}}_i^{(j)}.
\end{eqnarray}   
Equation (\ref{eq:eom}) was numerically solved together with the equations of motion of 
the normal-mode-transformed bead coordinates by use of  the NMCMD algorithm \cite{kinugawa2021}.
The centroid time increment $\Delta{t}_{\rm{MD}}$  was taken as 0.1 fs. 
The time increment for the normal-mode propagation was taken equal to $\Delta{t}_{\rm{MD}}$, while the effective period 
 \cite{kinugawa2021} of the non-centroid normal-modes was set at 100 fs.  

     The system pressure was evaluated via the virial theorem \cite{martyna1999}.
   The system kinetic energy $K$  was evaluated from the centroid-reference virial estimator \cite{yamamoto2005,glaesemann2002,tuckerman1993},   
\begin{eqnarray}
\label{eq:kinetic}
K&=&\sum_{i=1}^{N}K_i\\\nonumber
&=&\sum_{i=1}^{N}\left(\frac{3}{2}k_{\rm{B}}T-\frac{1}{2N_{\rm{b}}}
\left\langle\sum_{j=1}^{N_{\rm{b}}}({\it{\bf{r}}}_i^{(j)}-{\it{\bf{r}}}_{{\rm{c}}i})\cdot
{\it{\bf{f}}}_i^{(j)}\right\rangle\right).
\end{eqnarray}   
 In Eq. ~(\ref{eq:kinetic}), the contributing term  $K_i$  is not dependent solely on atom $i$ since it consists of the averaged virial terms including interatomic forces ${\it{\bf{f}}}_i^{(j)}$.

 In our previous study \cite{kinugawa2021}, we confirmed that the CMD 
 simulation method accurately  reproduced   the measured isotherms of   $V$ vs $P$ and of  
the internal energy $U$ vs $P$ for real $^4$He
 in the temperature range where the Bose statistics is negligible.

\subsection{\label{sec:trotter}Check of the Trotter  number dependence}

First, we checked the Trotter number $N_{\rm{b}}$ dependence of energetic properties.
The results of the energetic properties at 0.1 K are shown in Fig. SA-1
in the Supplementary Material. In this figure, we can see a fair convergence of the energetic properties for $N_{\rm{b}}\geq{500}$, so that we adopted 
$N_{\rm{b}}$  as 500 in all the simulations in 
 the present study.   
 Figure SA-2 in the Supplementary Material shows the Trotter number dependence of
 energy-temperature relations obtained from the CMD simulation at 1 bar.
 In this figure, the energetic properties for $T\lesssim0.1$ K
depend fairly on $N_{\rm{b}}$ and the convergence of estimated energy values  is not good. 
For $T\lesssim0.1$ K, remarkable dropping of $H$, $U$, and $\Phi$
 is observed for $500\leq{N_{\rm{b}}}\leq{2000}$, though $K$ converges fairly well.
We see that $\Phi$, $U$, and $H$ were evaluated low too much at $T\lesssim{0.1}$ K when  
we adopted the setting of $N_{\rm{b}}\leq{2000}$.
However, there is a symptom that the temperature dependence plot 
 of each energy in the case of  $N_{\rm{b}}=3000$ is likely to converge asymptotically to a horizontal value as $T\rightarrow{0}$ K.
 The plot looks like a slight  S-curve in the range of 
 $0.01\leq{T}\leq{0.1}$ K.
 If we adopted the Trotter number as $N_{\rm{b}}>3000$, 
  more accurate results would be obtained even for $T\lesssim0.1$ K.
Although the low temperature properties at $T<0.1$ K are
intriguing, 
 we would rather save this subject for a future work.
 As for the results shown in this paper, we need to be careful that 
 the thermodynamic properties at $T\lesssim0.1$ K in this work,
 where we set  $N_{\rm{b}}=500$,  contains this degree of inaccuracy.
The $N_{\rm{b}}$-dependence of other properties is also shown in Figs. SA-3 and SA-4
in the Supplementary Material; this issue is discussed in Secs.~\ref{sec:RDF} 
and \ref{sec:wavelength}.

\subsection{\label{sec:simulationprocedure}Survey of pressure-temperature points by  isothermal compression}
We surveyed  possible states at 357 $P$-$T$ mesh 
points   in the range of $0.08\leq{T}\leq{3.3}$ K and
$1\leq{P}\leq{500}$ bar; Figure SA-5 in the Supplementary Material shows these points plotted on the $P$-$T$ 
 plane.
The procedure of the CMD simulation, shown schematically
 in Fig. SA-6 in the Supplementary Material, is as follows:

(1)  First,  the system was equilibrated  at 3.3 K and 1 bar, and then it was stepwise cooled toward 0.08 K, keeping the pressure at 1 bar.
 In this isobaric procedure, CMD was carried out spanning 100,000 steps (10 ps) at each temperature, and then the set temperature 
 was switched to the next lower temperature.
Then the system  was further equilibrated  spanning 200,000 steps (20 ps), followed by 
another 200,000 steps' production run, from which the physical properties at 1 bar given in the 
following sections were calculated.   

(2) Next, the isothermal compression  was  started from the end configuration of the 1 bar 
production run at each temperature.   At each temperature in the range of
0.08-3.3 K, the set hydrostatic pressure was increased to the next higher value
stepwise toward 500 bar. The pressure
was kept constant during each 
time interval of 50,000 steps, and increased at the end of each interval.
At each $P$-$T$ point we further conducted 150,000 steps' (15 ps) production run. 
The physical  properties for $P>1$ bar shown in the following
sections  were evaluated from each production run.

\section{\label{sec:resultsofiosthermalcompression} RESULTS}

   In this section, we show the  results of the CMD simulations.
   First, in Sec.~\ref{sec:statediagram}, we present the $P$-$T$ state diagram which was   derived from the computational results.  
   In Secs.~\ref{sec:VandE}-\ref{sec:diffusion}, we provide the results evidencing
   the conclusion of  state diagram, while the criterion for the identification of  states
   is  described in Sec.~\ref{sec:Identification}.
    Finally, the examination of atomic tunneling and 
   the velocity autocorrelation function are presented  in 
   Secs.~\ref{sec:tunneling} and \ref{sec:VAF}, respectively.

\subsection{\label{sec:statediagram}State diagram}

\begin{figure*}
\includegraphics[width=10cm]{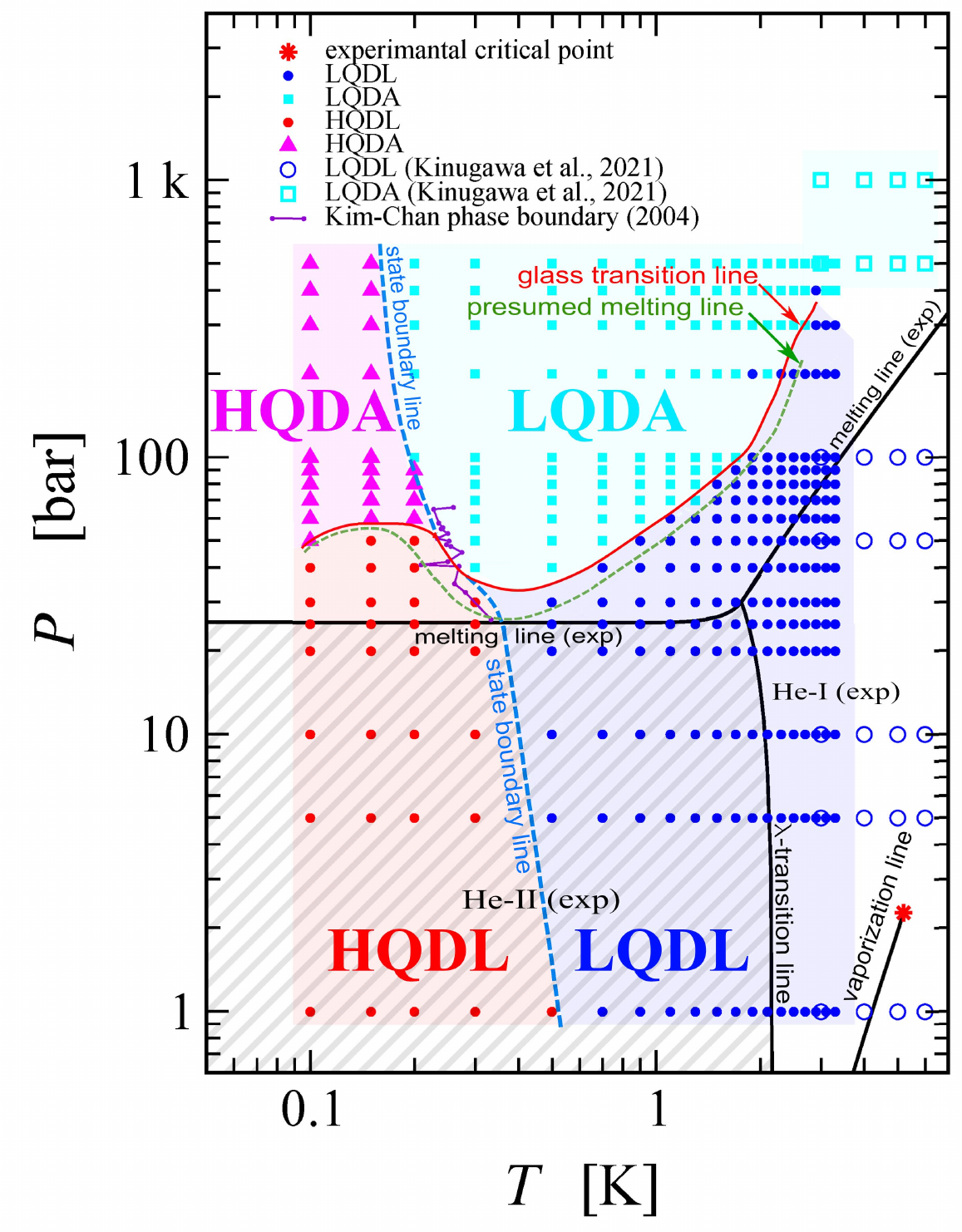}
\caption{\label{fig:statediagram} 
The state diagram of distinguishable $^4$He obtained from the isothermal compression by the CMD simulation.
The points at 1 bar were obtained from the isobaric cooling from the state at 3.3 K.
 LQDL (blue): low quantum dispersion liquid;
 HQDL (red): high quantum dispersion liquid;
 LQDA (cyan): low quantum dispersion amorphous solid (metastable); 
 HQDA (magenta): high quantum dispersion amorphous solid (metastable). 
Black solid lines are the experimental phase boundary lines of real $^4$He.
Experimentally, He-I and He-II (gray slant stripe) exist at  higher and lower temperatures than $\rm{\lambda}$ transition 
line (black), respectively, 
while the crystalline solid phase exists at pressures higher than the melting curve (black).
The open symbols denote
the states obtained from the isothermal compression in our previous study \cite{kinugawa2021}.  
The points of ``LQDL (Kinugawa et al., 2021)'' \cite{kinugawa2021} includes those
of supercritical fluid (SCF) state with low quantum dispersion, existing beyond the critical point (red star).
The  melting line (dashed green line) is presumed to be 
located at lower pressures than the glass transition line (red line)
because of overpressurization.
The Kim-Chan phase boundary denotes the locus of the emergence of non-classical rotational inertia 
which then suggested the existence of {\it{supersolid}} at lower temperatures
 below this boundary \cite{kim2004};  
see 
 Appendix \ref{sec:comexp}.
 }
\end{figure*}

\begin{figure*}
\includegraphics[width=12cm]{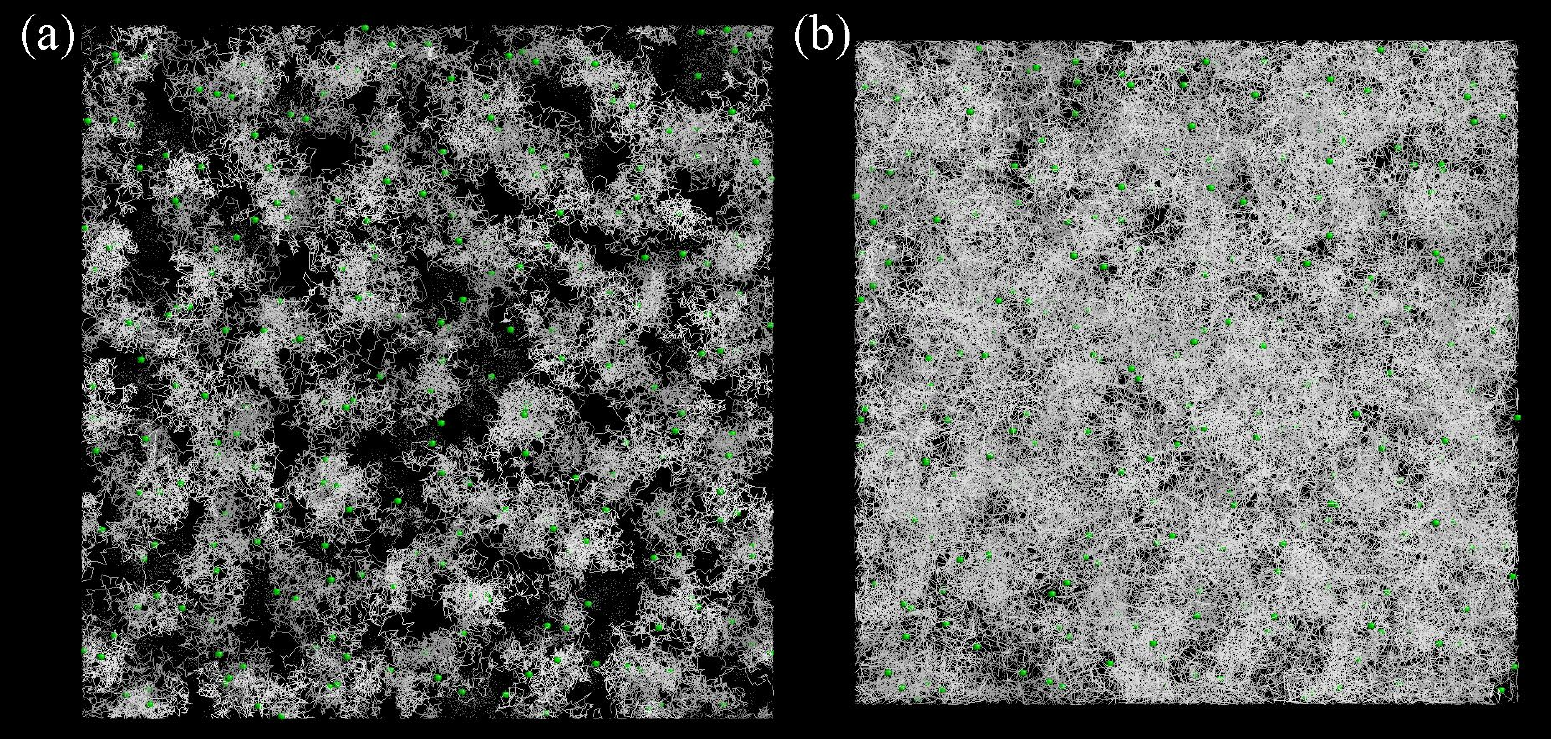}
\caption{\label{fig:snapshot} 
The {\it{xy}}-projected snapshots of the configuration of the $^4$He atomic necklaces and centroids. 
(a) low quantum dispersion liquid (LQDL) state at 1 bar and 1.5 K; (b) high quantum dispersion liquid (HQDL) state at 1 bar and 0.3 K.
Green spheres and white ones denote the centroids and the beads, respectively.  The drawn scales of the centroids and the beads
are arbitrary.}
\end{figure*}   

Figure~\ref{fig:statediagram} shows the $P$-$T$ state diagram derived  from the 
computational results in the present study, together with our
previous results  for higher $P$-$T$ conditions \cite{kinugawa2021}.  
Similar to the  compressed conditions \cite{kinugawa2021}, 
we can  observe the existence of two glassy states, LQDA and HQDA, also in the  $P$-$T$ range of the present study.
Moreover,  we can see the existence of two liquid states:
 low quantum dispersion liquid (LQDL)  and high quantum dispersion liquid (HQDL).
They are distinguished from each other 
mainly by the difference in  quantum dispersion of atoms; the 
characteristics of these two liquid states are described in 
later subsections.
The system at each $P$-$T$ point retained its state throughout the simulation and
no transition to another state was observed.
Evidently, LQDA and HQDA should be  metastable states which must eventually reach the 
 crystalline state at an equilibrium  even though  we  observed no state changes
 during the  simulation. 
As  LQDL exists also  at $P$-$T$ region of He-I at  temperatures higher than  
${\rm{\lambda}}$-transition line, it is considered as being nearly identical  to  He-I of real $^4$He.
Further analysis  regarding the state diagram will be described in Sec.~\ref{sec:onstatediagram}.

For visual understanding, in Fig.~\ref{fig:snapshot}
we show the snapshots of typical instantaneous configurations
of  LQDL and HQDL at 1 bar.  
Those for 40 bar are also 
shown in Fig. SA-7 in the Supplementary Material.
In these figures, we find that the  LQDA consists of localized necklaces representing low quantum dispersion (LQD) of 
$^4$He atoms, 
while  HQDA involves stretched necklaces representing  high quantum dispersion (HQD) of atoms.
However, the border of the two states is visually ambiguous.

\subsection{\label{sec:VandE}Thermodynamic properties}

\begin{figure}
\includegraphics[width=8cm]{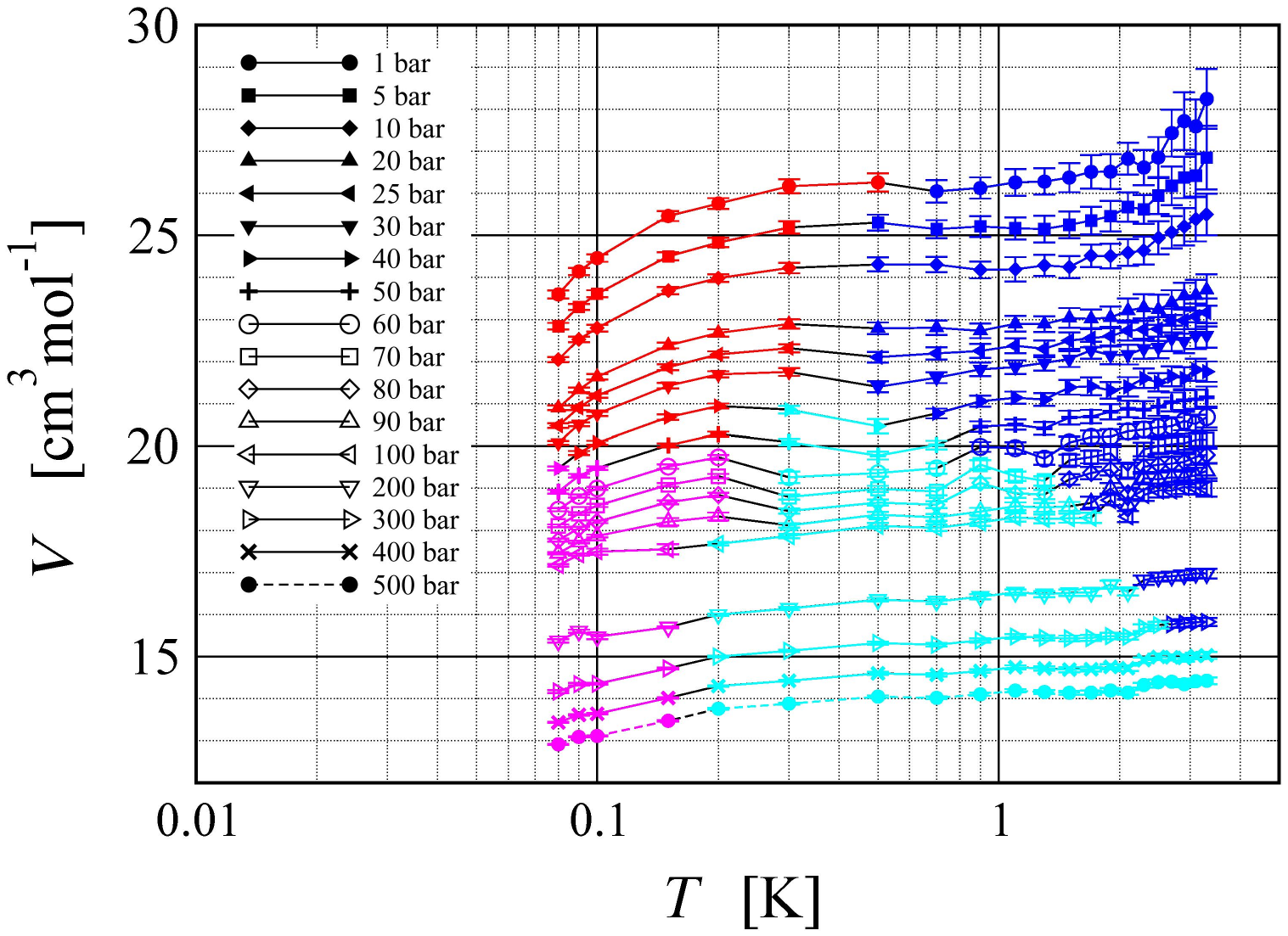}
\caption{\label{fig:V-T} 
The  molar volume vs temperature.
The lines denote 
isobars.  The colors of the symbols
are the same as those displayed in Fig.~\ref{fig:statediagram}: LQDL (blue), LQDA (cyan), HQDL (red), and HQDA (magenta).  
}
\end{figure}

\begin{figure}
\includegraphics[width=8cm]{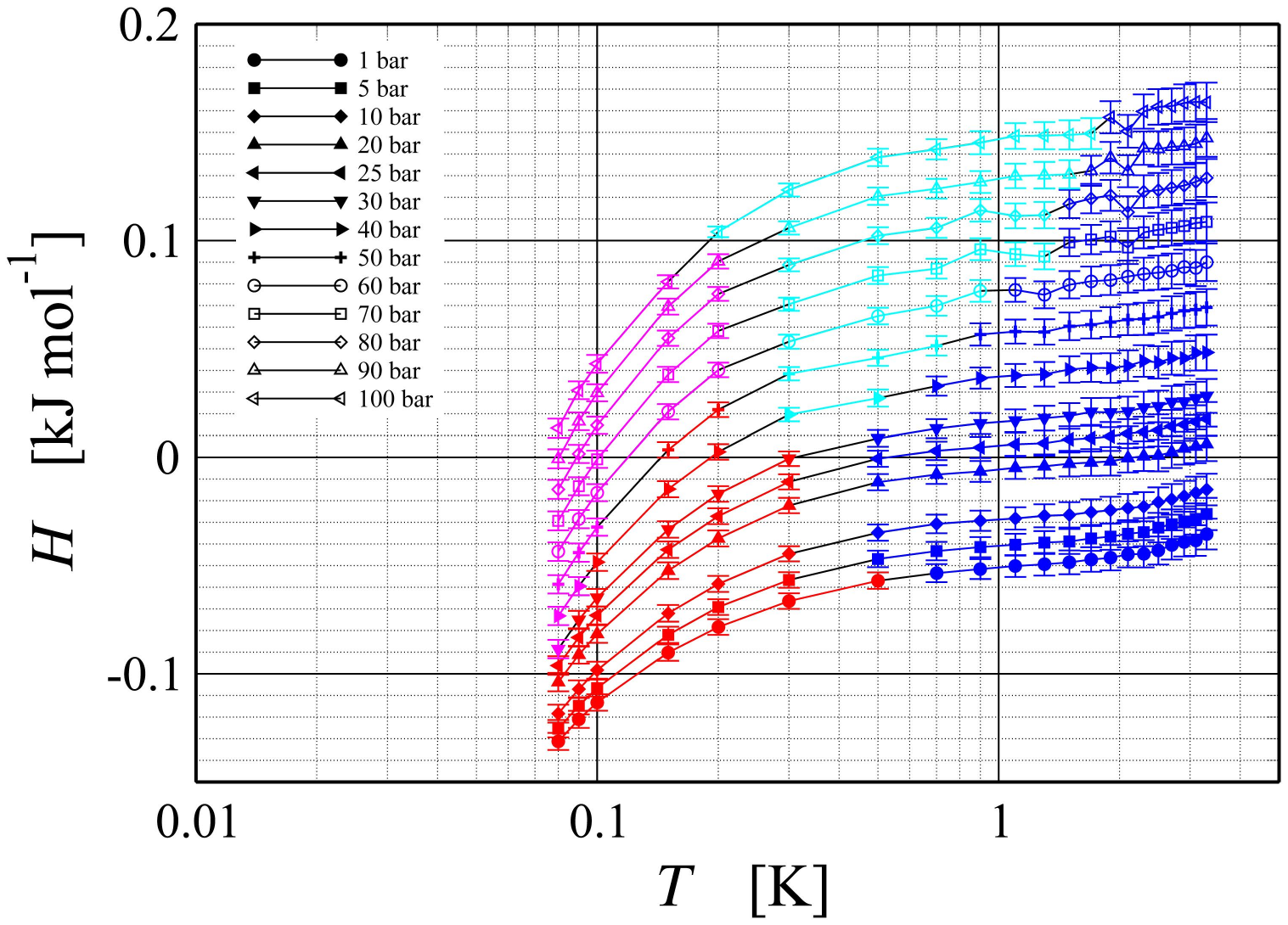}
\caption{\label{fig:H-T} 
The enthalpy vs temperature.
The lines denote isobars.
The colors of the symbols
are the same as those displayed in Fig.~\ref{fig:statediagram}: LQDL (blue), LQDA (cyan), HQDL (red), and HQDA (magenta).  }
\end{figure}

Figures \ref{fig:V-T} and \ref{fig:H-T} show the temperature dependences of molar volume $V$ and
enthalpy $H$, respectively. 
We provide the complete set of data of the  $H$-$T$ plot over the entire range of 1-500 bar
in Fig.
SA-8 in the Supplementary Material, while the temperature dependences of 
 $K$,  $U$, and $\Phi$ 
are shown in Figs. SA-9, SA-10, and SA-11, respectively.
In Figs.~\ref{fig:V-T} and \ref{fig:H-T}, 
we can see no discontinuity in each isobar of $V$-$T$ and $H$-$T$ plots, so that there is 
no  evidence to support the existence of the first-order phase transition.
In Fig.~\ref{fig:V-T}, we can see  that 
some curves have negative slopes on the occasion of the HQDL$\rightarrow$LQDL or the HQDA$\rightarrow$LQDA transition.
In contrast, in Fig.~\ref{fig:H-T},   each $H$-$T$ curve is an increasing  monotonous function of 
$T$, i.e.,  $(\partial{H}/\partial{T})_P=C_P>0$ (or $\Delta{H}>0$ for heating) over 
the whole temperature range.
Therefore, at a given pressure, HQDL existing at lower temperatures 
has lower entropy than LQDL or LQDA at higher temperatures, according to 
the relation $\Delta{S}={\int}C_PdT/T$.

In Fig.~\ref{fig:H-T}, the slope of the $H$-$T$ plot, $C_P$, is  larger at lower temperatures.  
 For example, the $C_P$ at 1 bar obtained from the curve fitting of 
 the $H$-$T$ relation is shown in Fig. SA-12 in the Supplementary 
 Material.  In this figure, we cannot see any indication of such  $C_P$ maximum  as 
 the ${\rm{\lambda}}$-type peak in real $^4$He
 and the cusp in the ideal Bose gas \cite{pitaevskii2016}. 
 Therefore, there are no evidences for supporting the second-order 
 phase transition between  LQDL and HQDL.
Consequently, each isobaric transition of  HQDL-LQDL, HQDL-LQDA, and 
HQDA-LQDA is considered as a continuous crossover, not the thermodynamic
phase transition.

There is a remarkable drop of $H$ at $T\lesssim{0.1}$ K  in Fig.~\ref{fig:H-T}, 
and equivalently, one sees the divergent increase of $C_P$ at the 
lowest temperature edge ($T\lesssim0.1$ K) in Fig. SA-12 in the 
Supplementary Material.  This is  an artifact due to  the insufficient 
 $N_{\rm{b}}$ (Sec.~\ref{sec:trotter}).
Though the computed values of, e.g., energetic properties, near 0.1 K are not completely accurate, the bulk of this work is hardly affected by the inaccuracy 
emerging at $T\lesssim0.1$ K.
 As described in Sec.~\ref{sec:trotter}, when we adopted the Trotter number $N_{\rm{b}}=3000$,
 which should yield more accurate results,  the $H$-$T$ plot within $0.01<T<0.1$ K exhibited a symptom of
 slight S-curve with a subtle inflection point at $\sim0.02$ K (see Fig. SA-2(e) in the Supplementary Material). 
 Then, the $C_P$  may have a 
 maximum at this temperature, and this would possibly imply another state transition.
However, we do not draw a conclusion at present; this subject will be resolved in a future study which must demand very heavy computation
with enormous $N_{\rm{b}}$.

\begin{figure*}
\includegraphics[width=12cm]{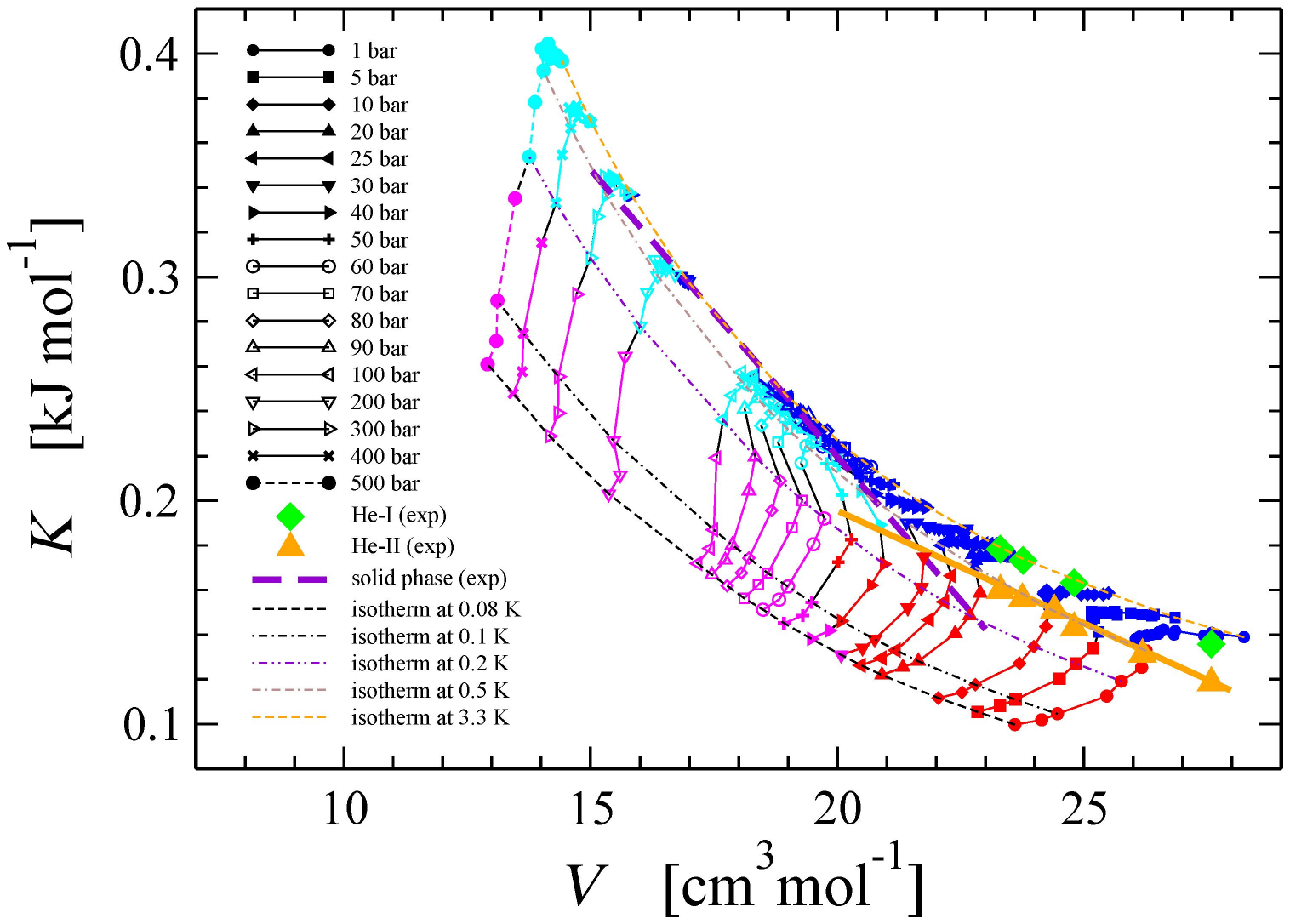}
\caption{\label{fig:kinvol} 
The kinetic energy vs molar volume.
 LQDL (blue), LQDA (cyan), HQDL (red), and HQDA (magenta). 
 All the experimental  data  are cited from Fig. 16 of Ref.~\onlinecite{glyde2011}; the measured temperature of normal fluid (He-I) and superfluid (He-II) 
 spans $2.1\leq{T}\leq{2.3}$ K and $0.045\leq{T}\leq{0.075}$ K, respectively~\cite{glyde2011}.  }
\end{figure*}

Figure ~\ref{fig:kinvol} shows the volume dependence  of  kinetic energy $K$ together with
the experimental relation of real bulk $^4$He.
The experimental data points \cite{glyde2011} 
of He-I, He-II, and solid phase likely lie on  a single universal line, irrespective of 
temperature.   
The  data points  of the
LQD states in the present work tends to gather to the vicinity of this universal curve, whereas those of 
the HQD states are clearly deviated from it and exhibit remarkable temperature dependence.
  This issue will be discussed  in Sec.~\ref{sec:general}

\subsection{\label{sec:RDF}Radial distribution functions}

\begin{figure}
\includegraphics[width=8cm]{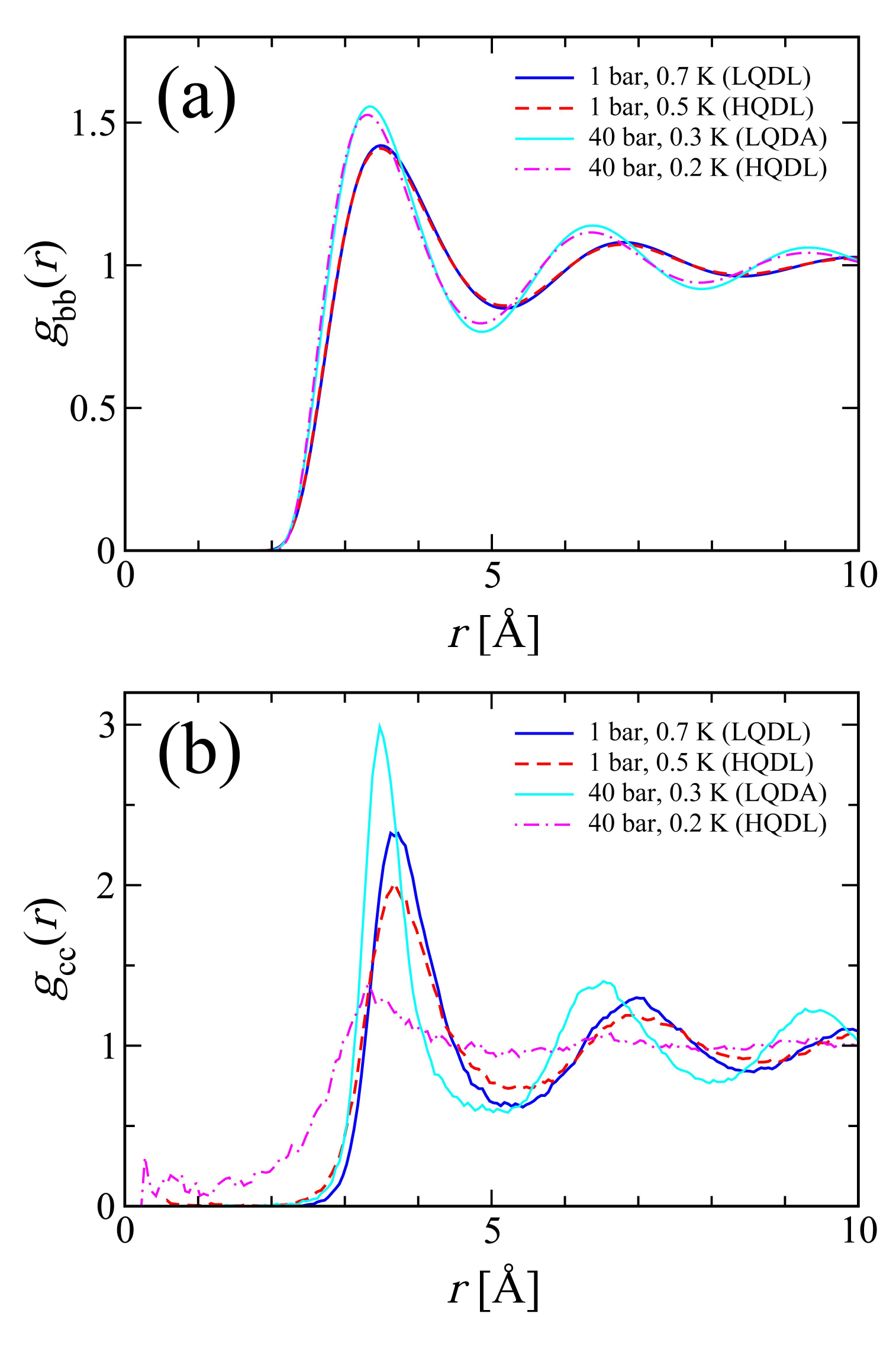}
\caption{\label{fig:rdf} 
The radial distribution functions at selected $P$-$T$ conditions.
(a) bead-bead radial distribution function $g_{\rm{bb}}$;
(b) centroid-centroid radial distribution function $g_{\rm{cc}}$.
The $g_{\rm{bb}}$ is observable by diffraction experiments, whereas the
$g_{\rm{cc}}$ is unobservable because atoms do not exist at centroid positions.
}
\end{figure}

  The radial distribution functions (RDFs) at  selected $P$-$T$ points are shown in Fig.~\ref{fig:rdf},  
 while the complete set of RDFs is shown in Figs. SB-1 and SB-2 in the Supplementary Material.  
In both the  bead-bead RDF $g_{\rm{bb}}$ and the  centroid-centroid  RDF $g_{\rm{cc}}$, we 
can recognize broad distributions, indicating that the system is in a liquid or glass state and denying
the existence of crystalline order at all $P$-$T$ conditions.
  However, in Fig.~\ref{fig:rdf}(b)
  and Fig. SB-1 in the Supplementary Material, the 
  $g_{\rm{cc}}$  exhibits remarkable structural change due to
  the LQD-HQD state transition  occurring in the range of 0.15-0.7 K.
 The peaks in $g_{\rm{cc}}$ of the HQD states
  are  collapsed and there is non-zero distribution  penetrated
 into 
 ${r}\sim{0}$ {\AA}.
 However, this is  not unnatural because the interatomic forces are 
not directly exerted on the centroid positions on which atoms do not exist.
Since the profile of  $g_{\rm{bb}}$ does not change significantly  even on the occasion of 
the LQD-HQD state transition, 
  structural change can be hardly detected by  diffraction experiments, in spite of
  drastic change in the $g_{\rm{cc}}$. 
 In fact, the non-zero distribution at ${r}\sim{0}$ {\AA} in  $g_{\rm{cc}}$  is 
  also characteristic of  the HQDA of compressed $^4$He glass \cite{kinugawa2021} 
   and the classical ring polymer systems \cite{goto2023,cai2022}.

We show the $g_{\rm{cc}}$ at 1 bar with  $N_{\rm{b}}=500$, 1500, and 
2000 in Fig. SA-4 in the
Supplementary Material.
For each of $N_{\rm{b}}$, we can see that
the LQD-HQD  structural transition  occurred at 0.5-0.7 K;
the structure of HQD state is characterized by 
 non-zero distribution of $g_{\rm{cc}}$ at the distances shorter than the first peak.
Consequently, the occurrence of the LQD-HQD transition is not artificially caused by the
insufficiency of $N_{\rm{b}}$ but attributed to an essential property of the system.

\subsection{\label{sec:wavelength}Quantum wavelength and expansion factor} 

\begin{figure}
\includegraphics[width=8cm]{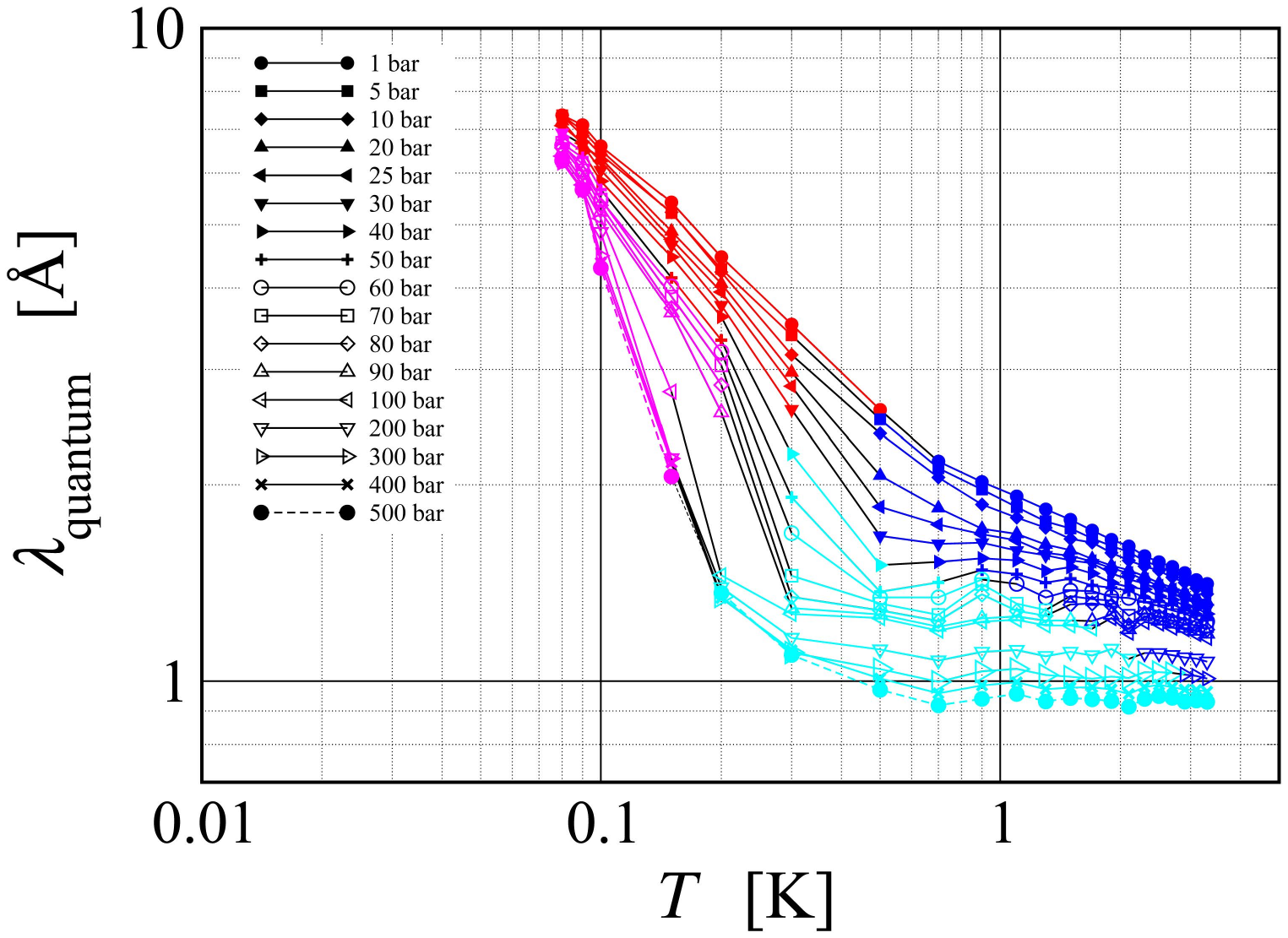}
\caption{\label{fig:lamT} 
A double logarithm plot of the quantum wavelength vs temperature.  The colors of the symbols
are the same as those displayed in Fig.~\ref{fig:statediagram}: LQDL (blue), LQDA (cyan), HQDL (red), and HQDA (magenta). 
}
\end{figure}

To quantify the spatial extension of atomic necklaces, i.e., the degree of quantum dispersion, we evaluated 
quantum wavelength \cite{takemoto2018,kinugawa2021} $\lambda_{\rm{quantum}}$ 
from
the square root of the imaginary-time mean square displacements (IMSD) \cite{nichols1984}. 
Namely, the IMSD  at imaginary time $\tau_n=n\beta\hbar/N_{\rm{b}}$ 
is 
\begin{eqnarray}
\label{eq:IMSD}
R^2_{\rm{IMSD}}(\tau_n)=
{\frac{1}{N}}\sum_{i=1}^{N}\left\langle|{\it{\bf{r}}}_i^{(j+n)}
-{\it{\bf{r}}}_i^{(j)}|^2\right\rangle_j,\;\;\\
0<{\tau_n}\leq\beta\hbar,\nonumber
\end{eqnarray}
from which we estimated 
the quantum wavelength 
$\lambda_{\rm{quantum}}$
defined as
\begin{eqnarray}
\label{eq:quantumwavelength}
\lambda_{\rm{quantum}}=\sqrt{R^2_{\rm{IMSD}}(\beta\hbar/2)}.
\end{eqnarray}
The IMSD exhibits a semiarc-like curve as a function of $\tau_n$ \cite{takemoto2018,kinugawa2021,nichols1984}.
The $\lambda_{\rm{quantum}}$ denotes an effective diameter of atomic necklaces and  
corresponds to   twice the   {\it{radius of gyration}} 
$R_{\rm{g}}$ 
in  polymer physics terminology \cite{rubinstein2003}.
 Figure \ref{fig:lamT} shows the temperature dependence of 
 $\lambda_{\rm{quantum}}$. 
   At each pressure,  $\lambda_{\rm{quantum}}$ tends to increase with lowering temperature.
  The HQDL exhibits significantly longer $\lambda_{\rm{quantum}}$ than the closest distances observed 
   as the first peaks  in $g_{\rm{bb}}$ in Fig.~\ref{fig:rdf}(a).
For  1 bar, to analyze the temperature dependence of 
$\lambda_{\rm{quantum}}$ \cite{takemoto2018},
\begin{eqnarray}
\label{eq:lambdatemp}
\lambda_{\rm{quantum}}\sim{T}^{\chi},
\end{eqnarray} 
 we  estimated the power $\chi$ by the least squares fitting.  
It yielded
\begin{eqnarray}
\label{eq:fit}
\chi=-0.28, \text{for LQDL at 1 bar},\nonumber\\
\chi=-0.58, \text{for HQDL at 1 bar}.
\end{eqnarray}
  Surprisingly, these values are nearly equal to  those of  the two detailed states
of supercritical fluid (SCF) $^4$He 
($-0.29\leq\chi\leq-0.26$ for liquid or liquid-like state, 
$-0.49\leq\chi\leq-0.46$ for the gas or gas-like state).\cite{takemoto2018}
In this sense, the LQD-HQD  boundary  in the $\lambda$-$T$ plot is similar to  the
Widom line \cite{takemoto2018} that separates the liquid-like and gas-like SCF states by the
$C_P$ maxima.
However, as seen in Sec.~\ref{sec:VandE}, there are no traces of such $C_P$  maxima in the present case.

\begin{figure}
\includegraphics[width=8cm]{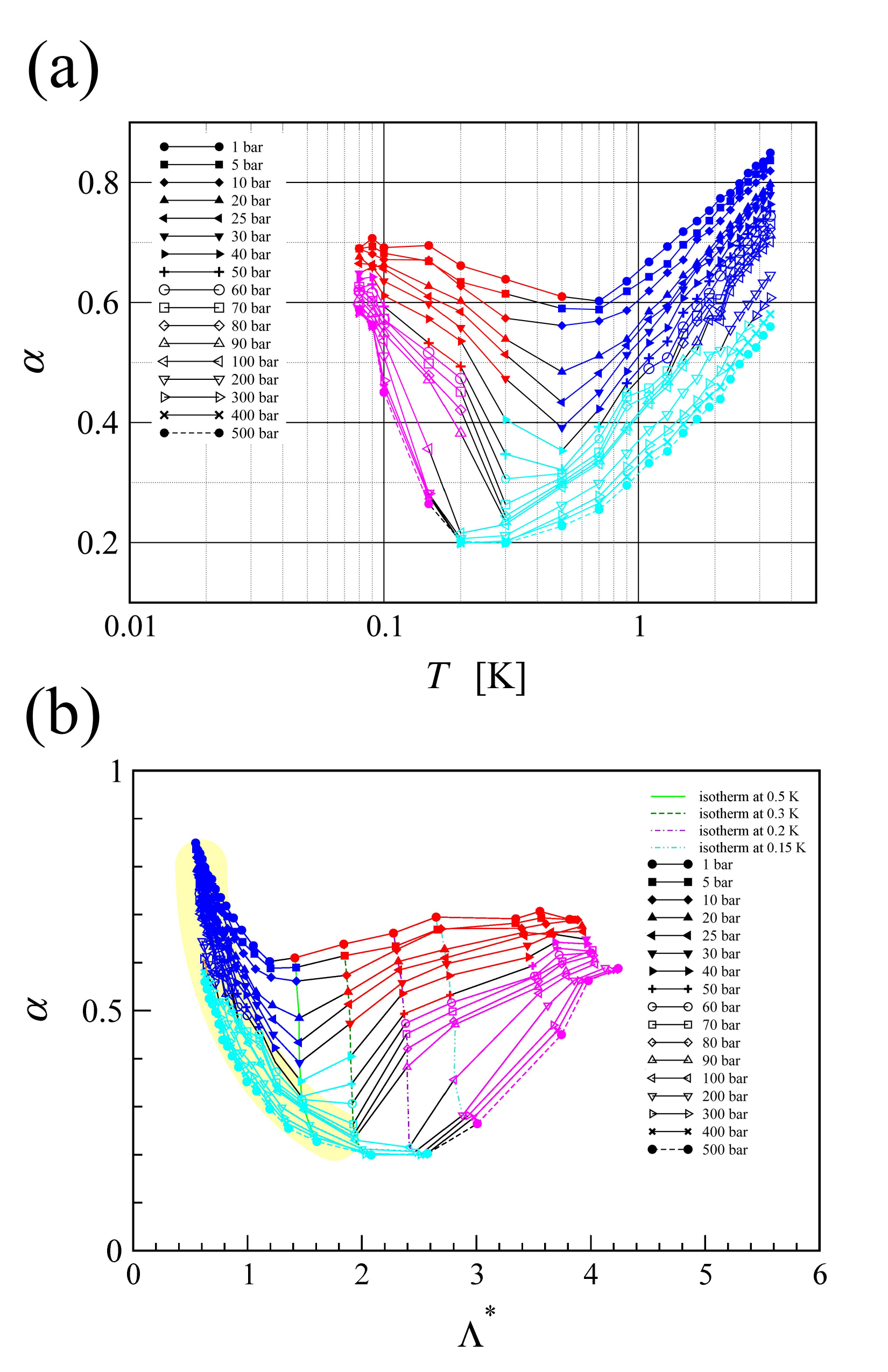}
\caption{\label{fig:expab} 
The  expansion factor vs (a) temperature and (b) quantumness parameter.
The colors of the symbols
are the same as those displayed in Fig.~\ref{fig:statediagram}: LQDL (blue), LQDA (cyan), HQDL (red), and HQDA (magenta).
In (b), when there is a relation between the de Broglie thermal
wavelength and the closest distance, $\lambda_{\rm{dB}}\geq\sigma$, we get $\Lambda^{*}\geq1/\sqrt{2\pi}=0.40$.
The yellow background shown in (b) denotes a possible universal curve for the LQD states.
}
\end{figure}

The state transition can be detected more clearly by use of  expansion factor  
$\alpha:=\lambda_{\rm{quantum}}/\lambda_{\rm{quantum}}^{(0)}$
where the denominator is the quantum wavelength of
a three-dimensional free particle, $\lambda_{\rm{quantum}}^{(0)}=(\hbar/2)\sqrt{3\beta/m}=\sqrt{3/8\pi}\lambda_{\rm{dB}}$ \cite{kinugawa2021}.
We can regard  $\alpha$ as being equivalent to 
  expansion factor in polymer physics, i.e., $\alpha\approx
R_{\rm{g}}/{{R_{\rm{g}}}^{\rm{(free)}}}$, 
where $R_{\rm{g}}^{\rm{(free)}}$
is the radius 
of gyration of  classical free ring polymer.
  Figure ~\ref{fig:expab}(a) shows the temperature dependence of  $\alpha$.
   In every isobar there exists a minimum or the tipping point locating   at 0.2-0.6 K.
  As the temperature lowers, $\alpha$ decreases in the LQDL and LQDA, whereas it reversely increases in the
  HQDL and HQDA.  Namely, each minimum point in this $\alpha$-$T$ plot is the transition point between 
  an LQD state and an HQD state.
 A  positive (negative)  slope of the $\alpha$-$T$ curve corresponds to an LQD state 
 (an HQD state), respectively.

Figure~\ref{fig:expab}(b) shows the replot of $\alpha$ shown in Fig.~\ref{fig:expab}(a) 
as a function 
of quantumness parameter $\Lambda^{*}$. Here the interatomic closest distance $\sigma$ was taken as
the distance of the first peak of $g_{\rm{bb}}$.  
In this figure,   $\Lambda^{*}$ of the HQD states are higher than those at tipping points.
Interestingly, the $\Lambda^{*}$  at the tipping points ($1\lesssim{\Lambda^{*}}\lesssim3$) 
is in good agreement with $\Lambda^{*}$ in the quantum binary Lennard-Jones system 
($1\lesssim\Lambda^{*}\lesssim2$) \cite{markland2011, markland2012}.   
Seemingly, in Fig.~\ref{fig:expab}(b), plotted points of LQD states tend to gather  on a
 single curve which is possibly located at yellow background region.  This suggests the universality of
the relation between $\alpha$ and $\Lambda^{*}$ of the LQD states, whereas the data points
of the HQD states are remarkably deviated from it.

We comment on the bead number dependence of $\lambda_{\rm{quantum}}$ and
$\alpha$.  The examination of this dependence 
was examined  by short CMD runs at 1 bar, and the graphs are shown in Fig. SA-3 in the
Supplementary Material.
In Fig. SA-3(a), clearly, remarkable increase of  $\lambda_{\rm{quantum}}$ with
lowering temperature toward 0.1 K is observed for every $N_{\rm{b}}$.  Further, in 
Fig. SA-3(b) the reentrant 
temperature-dependence of $\alpha$ passing the tipping points is  recognized 
for all $N_{\rm{b}}$ conditions.
Therefore, the occurrence of the LQD-HQD transition is essential and is not an artifact.

\subsection{\label{sec:diffusion}Apparent self-diffusion coefficient}

 In Figs. SB-3 and SB-4  in the Supplementary Material, we provide the  mean square displacement 
 (MSD) of atomic centroids,
\begin{eqnarray}
\label{eq:MSD}
R^2(t)=
{\frac{1}{N}}\sum_{i=1}^{N}\left\langle|{\it{\bf{r}}}_{{\rm{c}}i}(t+t_0)
-{\it{\bf{r}}}_{{\rm{c}}i}(t_0)|^2\right\rangle_{t_0}.
\end{eqnarray}
The  Einstein relation, $R^2(t)=6Dt$ $(t\rightarrow\infty)$, holds when atoms diffuse according to  Brownian motion
characteristic of liquid phase.
However, in the present study,  apparent self-diffusion coefficient $D$ was conveniently 
estimated by fitting the  $R^2(t)$  to this linear relation over  $0\leq{t}\leq{4}$ ps.
 Therefore, such an estimated $D$ is not a true self-diffusion coefficient but an apparent value merely reflecting the speed of atomic displacement.  
 When we get $R^{2}(t=4\,{\rm{ps}}){\leq}{0.5}$ {\AA}$^{2}$,
i.e., 
$D\leq{2}{\times}10^{-6}$ cm$^2$s$^{-1}$, 
 the system was regarded as being solidified, otherwise being in a liquid state;
the existence of gas phase was denied from the molar volume
 (Fig.~\ref{fig:V-T}).
Since there is no indication of crystalline structure as described in Sec.~\ref{sec:RDF},
the solidification in the present work means the glassification.
The temperature dependence of $D$ is shown in Fig.~\ref{fig:difpu} ($P\leq100$ bar) 
and Fig. SA-13 ($P\geq200$ bar) in the Supplementary Material. 
In Fig.~\ref{fig:difpu}, there is a  tendency that $D$ decreases with lowering temperature in LQDL and LQDA.
However, as for the isobars at $P{\geq}20$ bar, further drop of $T$ causes   $D$ to turn to increasing
on the occasion of the LQD$\rightarrow$HQD transition, and the increase  is  more drastic  
at higher pressures. 
The increase of $D$  occurring at the LQDA$\rightarrow$HQDA transition for $P\geq{60}$ bar indicates the reentrant behavior that the 
 residual diffusion is enhanced  by the glass-glass transition; this tendency agrees with the previous observation  of the  pressure or temperaure-induced reentrant tendency of $D$ \cite{markland2011,markland2012,kinugawa2021}. 
However, we note that
the  liquid-liquid transition from LQDL to HQDL  also increases $D$  at $20\leq{P}\leq{30}$ bar, indicating the reentrant behavior, 
though this is not recognized at lower pressures, $1\leq{P}\leq{10}$ bar.
We also find that the temperature lowering of LQDA  at 40 and 50 bars leads to melting into the HQDL state
({\it{melting by cooling}}).

Finally, we provide the  $D$-$\Lambda^{*}$ relation in Fig. SA-14 in the Supplementary Material.
Compared with Fig. ~\ref{fig:difpu}, the plotted points of LQD states seemingly
tend to be located on a single line.  This 
suggests the possible universality of the   $D$-$\Lambda^{*}$ 
relation underlying the LQD states.

\begin{figure}
\includegraphics[width=8cm]{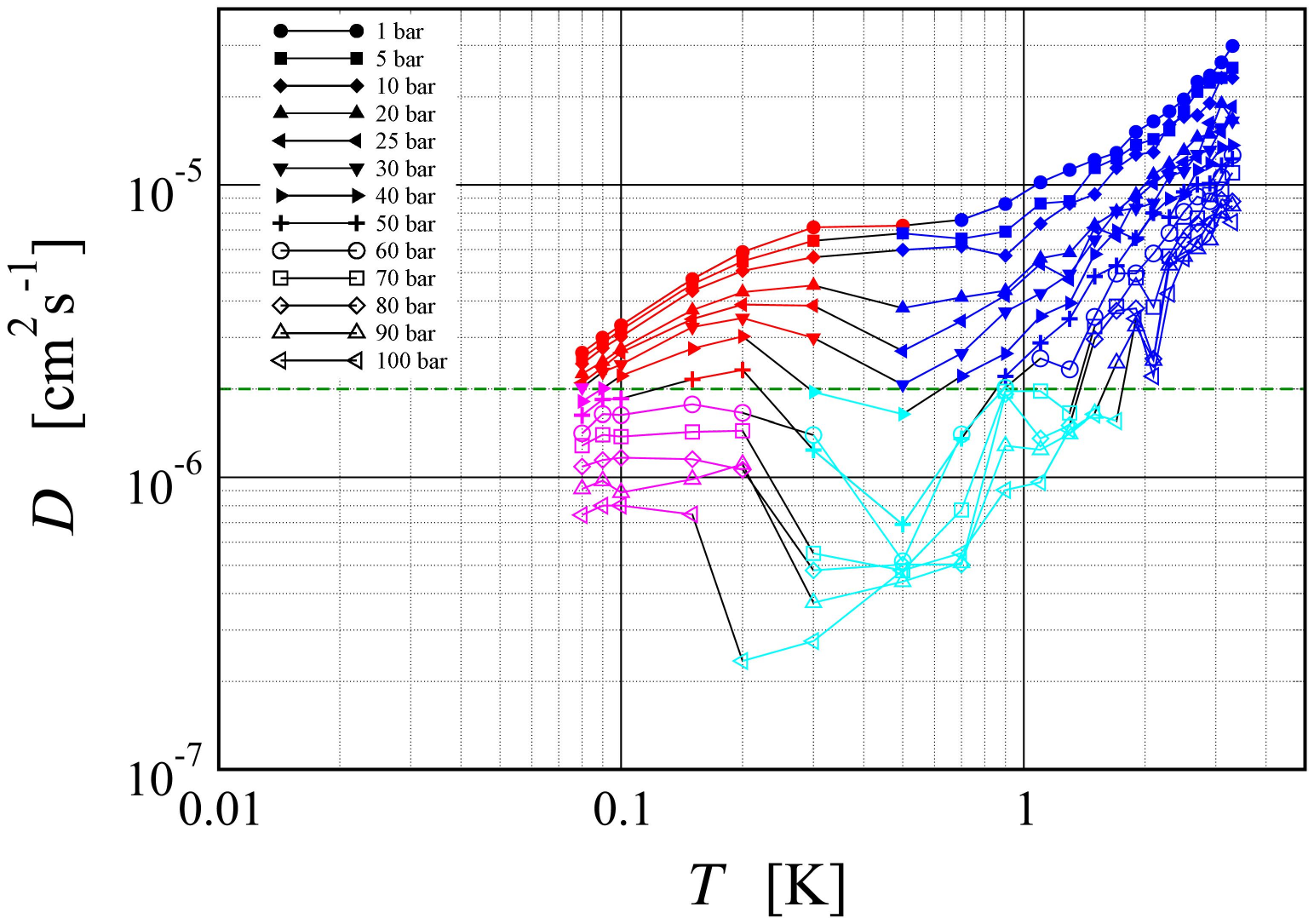}
\caption{\label{fig:difpu} 
The temperature dependence of apparent self-diffusion coefficient below 100 bar.
The colors of the symbols
are the same as those displayed in Fig.~\ref{fig:statediagram}: LQDL (blue), LQDA (cyan), HQDL (red), and HQDA (magenta).
Green horizontal dashed line denotes the threshold between liquids and glasses ($D={2}{\times}10^{-6}$ cm$^2$s$^{-1}$).   
}
\end{figure}

\subsection{\label{sec:Identification}Identification of states} 
On the basis of the results presented in Secs.~\ref{sec:VandE}-~\ref{sec:diffusion},
we identified the state at each $P$-$T$ point by following the criteria
(1)-(3):

(1) When there is non-zero distribution of $g_{\rm{cc}}$  at $r$ shorter
than the closest distance observed in  $g_{\rm{bb}}$, 
this evidences the HQD state, otherwise the state is classified into the LQD state (Sec.~\ref{sec:RDF}) 

(2)  The increasing and decreasing $\alpha$ with isobaric lowering of temperature corresponds to
HQD and LQD state, respectively (Sec.~\ref{sec:wavelength}).

(3)  The apparent self-diffusion coefficient  $D\geq{2}{\times}10^{-6}$ cm$^2$s$^{-1}$ corresponds to
a liquid state, otherwise the state is regarded as a glassy solid (Sec.~\ref{sec:diffusion}).  There 
 are no  crystal states (Sec.~\ref{sec:RDF}).

The conclusion of such identification of the system state was summarized in Fig.~\ref{fig:statediagram}. 
Here, after plotting the identified points on the $P$-$T$ plane, 
the boundary lines were drawn passing between two adjacent points representing the two different states. 
In this figure, we discarded the data points at  $T<0.1$ K from the plot because we cannot deny possible 
artifact due to the insufficiency of the number of beads ($N_{\rm{b}}=500$) at these lowermost temperatures.

\subsection{\label{sec:tunneling}Quantum tunneling of atoms}

\begin{figure}
\includegraphics[width=8cm]{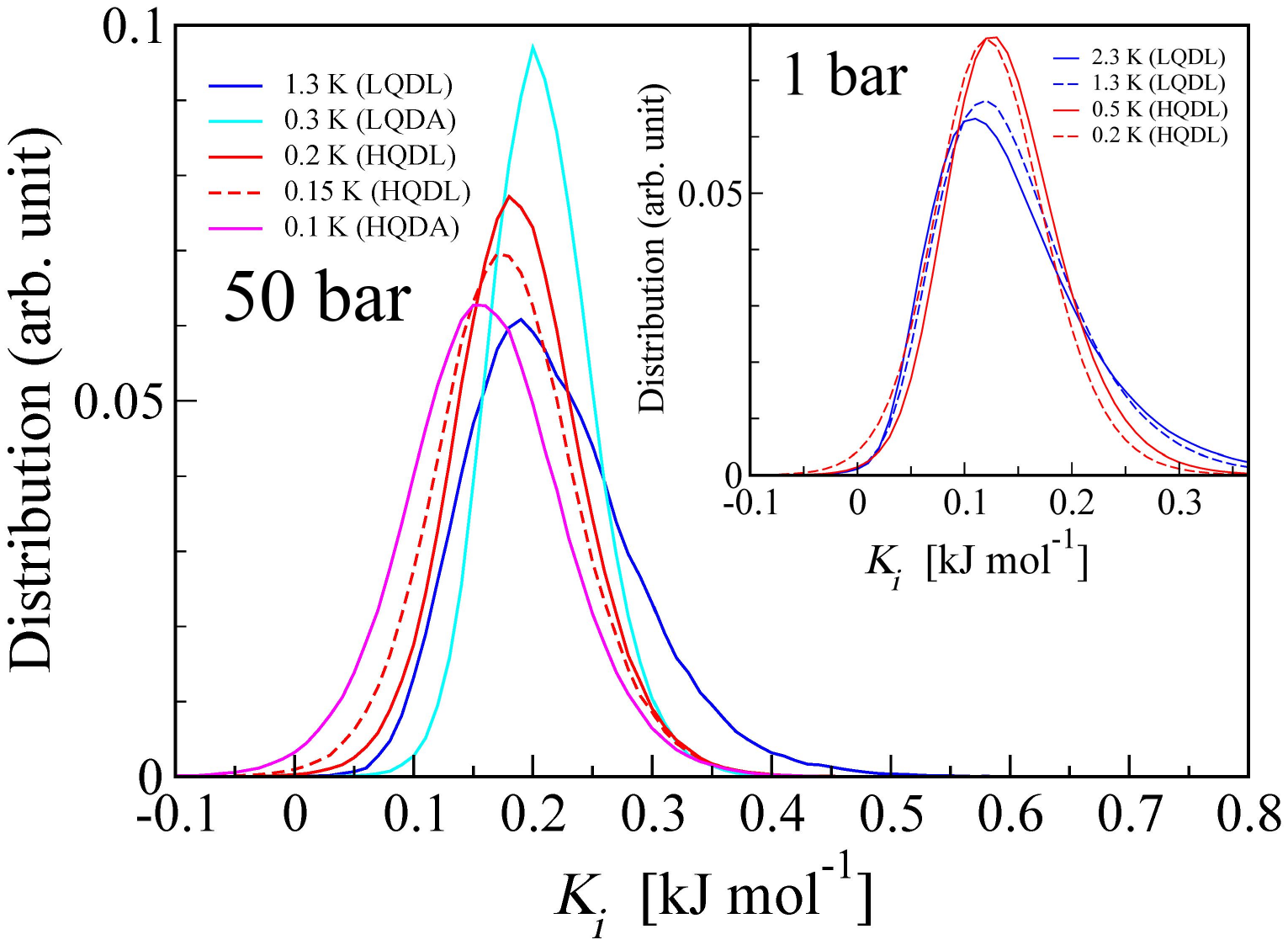}
\caption{\label{fig:puki50} 
The probability distributions of contributing term $K_i$ at 50 bar and 1 bar.  The main graph and the inset show the distributions at 50 bar and 1 bar, respectively.
The colors of the curves
are the same as those displayed in Fig.~\ref{fig:statediagram}: LQDL (blue), LQDA (cyan), HQDL (red), and HQDA (magenta).}
\end{figure}

Figure~\ref{fig:puki50} shows the probability distributions of contributing term $K_i$ 
(Eq. (\ref{eq:kinetic})).  
As we showed in our previous paper \cite{kinugawa2021},
negative $K_i$ denotes the existence of tunneling atoms whose necklaces
bestride on saddle points of the potential surface.
We can  see no negative $K_i$ for all the presented 
conditions in the
LQD states.  There is only a small portion of non-zero distribution of negative $K_i$ in the
HQD states, in particular, at lowermost temperatures.  This indicates that atomic tunneling is rare even in HQD states. 
The  distribution of negative $K_i$ we see here  is 
much less than  that in compressed HQDA \cite{kinugawa2021}  
where the atomic tunneling occurs as a consequence of the
explosive expansion of necklaces induced  by far higher  pressure. 
Therefore, we can conclude  that $^4$He atoms  diffuse in each of HQDL and LQDL with {\it{almost}} no quantum tunneling, while 
this is in contrast to the highly compressed HQDA where the residual diffusion is enhanced   by the tunneling effect 
 \cite{kinugawa2021}.

Among the graphs at 50 bar, the HQDA 
at 0.1 K and 50 bar has the largest population of negative $K_i$.
This seems to be in harmony with the  double-well tunneling model to explain the linear
 temperature-dependence of heat capacity, $C_V{\sim}{T}$,  of quantum glasses 
near $T\sim0$ \cite{anderson1972,phillips1972}. 
The existence of a low concentration of tunneling atoms in HQDL at lowermost temperatures
 likely suggests  a possible deviation from the heat capacity in the phonon regime, $C_V{\sim}T^3$,
  in this liquid state near $T{\sim}0$  if it exists.

\subsection{\label{sec:VAF}Centroid velocity autocorrelation function}

\begin{figure}
\includegraphics[width=8cm]{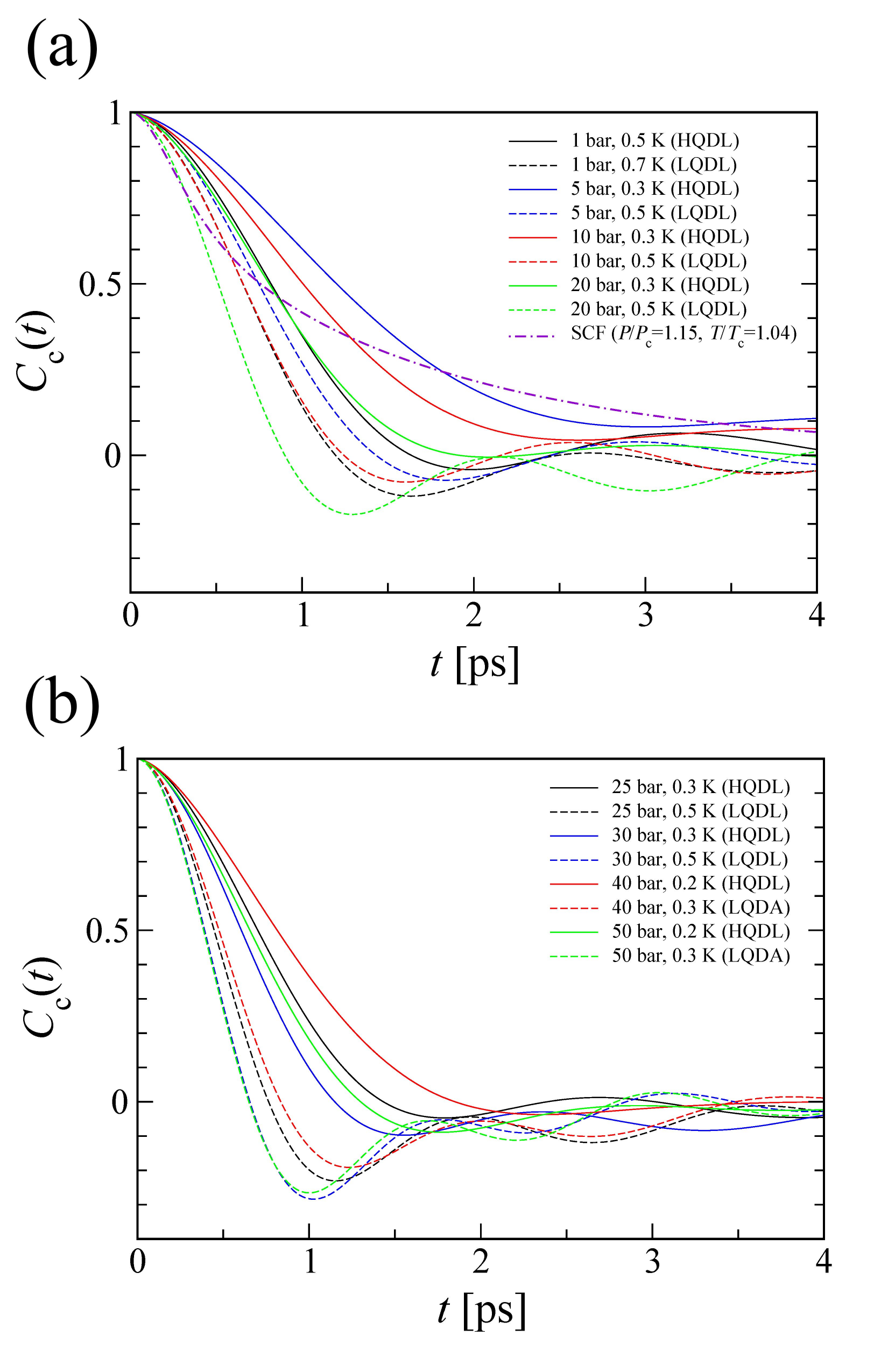}
\caption{\label{fig:vaf} 
The centroid velocity autocorrelation functions at the neighboring $P$-$T$ points  
on the both sides of the HQD-LQD boundary.
(a) $P=1-20$ bar; (b) $P=25-50$ bar.
Solid lines: HQDL; dashed lines: LQDL or LQDA.
The two points of both sides at the same pressure are drawn in the same color.
For comparison, in (a) we also plot the curve for the supercritical fluid state (SCF) at $P/P_{\rm{c}}=1.15$ and $T/T_{\rm{c}}=1.04$ 
($P_{\rm{c}}$ and $T_{\rm{c}}$ is critical pressure and temperature, respectively) \cite{takemoto2018}.
}
\end{figure}

 Figure~\ref{fig:vaf} shows the centroid velocity autocorrelation function (VAF) at the 
 $P$-$T$  points on the both sides of each  HQDL-LQDL and HQDL-LQDA boundary.
 Clearly, the VAFs of the LQDL exhibit an oscillatory decay 
characteristic of  liquid state,
whereas  the decay in  HQDL is rather monotonous  or damped oscillatory.
This distinction  is similar to the clearer dynamic crossover at  the Frenkel line in the SCFs \cite{brazhkin2013}.
For SCFs above the Frenkel line the VAF exhibits a monotonous decay and the state is
identified as a gas-like SCF, while below the line it exhibits an oscillatory decay by which  
the state is regarded as a liquid-like SCF \cite{brazhkin2013,takemoto2018}.

For example, in Fig.~\ref{fig:vaf}(b),  in the occasion of the HQDL-LQDA isobaric transition (0.2$\leftrightarrow$0.3 K) at 40-50 bar, the VAF changes its profile as if the Frenkel line is crossed over.
Therefore, from the VAF point of view,  we can say that 
the HQD-LQD boundary line is akin to the Frenkel line.
However, HQDL
is neither gas nor gas-like SCF because molar volume is not so much expanded
(Fig.~\ref{fig:V-T}).
The gentle  relaxation in HQDL is a consequence of atomic dynamics 
driven by mild centroid  force ${\textbf{F}}_{{\rm{c}}i}$
(Eq. (~\ref{eq:force})), i.e., the  remarkably weak  interatomic force smoothed out by 
stretched necklace configurations.
Thus, the atoms in HQDL diffuse  without remarkable oscillation and noticeable tunneling
(Sec.~\ref{sec:tunneling}).
 Further analysis of dynamical properties of this novel liquid
 is to be reported in our next paper.

\section{\label{sec:onstatediagram}ANALYSIS BASED ON THE STATE DIAGRAM}
 
In this section, we analyze the state diagram shown in  Fig.~\ref{fig:statediagram}, referring to
the physical properties provided in the last section.

First, we find that  at $P\leq{25}$ bar the present system  
does not freeze down to 0.08 K. 
This indicates 
the ease of melting of the present Boltzmann system at low $T$ and low $P$
even though there is not Bosonic permutation effect leading to the emergence of BEC.
The NQE solely caused the liquid state to 
 persist down to 0.08 K in this pressure range.

We presumed that the melting line was located at somewhat  lower pressures than the glass transition line.
This is because 
the overpressurization should be caused  by 
unavoidable 
hysteresis due to  unrealistically  rapid compression  in computer  simulation.
In addition, for real $^4$He, the ease of overpressurization inherent
in this substance was experimentally reported \cite{caupin2003,werner2004}.
 The obtained glass transition line in Fig.~\ref{fig:statediagram} 
 is not the ideal glass transition line 
 (the Kauzmann curve) \cite{stillinger2001, stillinger2003}.

In Fig.~\ref{fig:statediagram}, the  presumed melting curve  drawn in parallel with the glass transition line is downward convex  and has a  minimum at  $T_{\rm{min}}\cong0.4$ K and  $P_{\rm{min}}\cong25$ bar. 
In general, a melting line with an  extremum can be related to  the occurrence of liquid polyamorphism 
 \cite{rapoport1967};  two liquid states may 
exist at higher and lower temperatures than $T_{\rm{min}}$.  
In accordance with this, the LQDL-HQDL boundary line
emanates from the vicinity of the minimum of 
the presumed melting curve. 
This boundary line 
is located in parallel with the experimental $\lambda$ transition line, and is shifted low by about 1.8 K. 
Boninsegni et al. presumed the phase diagram of distinguishable $^4$He,
where the  solid-liquid boundary was also downward 
convex \cite{boninsegni2012}.
In fact, the experimental melting curve of $^4$He  ($^3$He) has a shallow minimum  at 
$T_{\rm{min}}=0.8$ K and $P_{\rm{min}}=26$ bar for $^4$He  (0.3 K and 29 bar for $^3$He).
The melting line in Fig.~\ref{fig:statediagram} is much more emphasized convex  
than the real system.

When a state point on the $P$-$T$ plane moves across the HQDL-LQDA boundary line  within 
the range of $0.2\leq{T}\leq0.3$ K and  $40\leq{P}\leq50$ bar,
  {\it{melting by cooling}} 
({\it{freezing by heating}}) or
exothermic melting (endothermic freezing) occurs; these transitions are  alternatively  called 
inverse melting (freezing) \cite{stillinger2001,stillinger2003,feeney2003,greer2000,prestipino2007,tombari2005,schupper2005}.
Thus,  the state diagram evidences  
the occurrence of isobaric  inverse melting (freezing).
In order to confirm it, however, such inverse freezing of HQDL by heating was examined in this study, and the
results are shown in  Appendix \ref{sec:freezing}; we succeeded in  observing
the inverse freezing in this test simulation.

In  Fig.~\ref{fig:V-T}, we can see that there is a temperature range with negative slopes
of the {\it{V-T}} line at $20\leq{P}\leq{90}$ bar.
This indicates negative isobaric thermal expansion,  
\begin{eqnarray}
\label{eq:nte}
\alpha_P=\frac{1}{V}\biggl(\frac{\partial{V}}{\partial{T}}\biggr)_P
=-\frac{1}{V}\biggl(\frac{\partial{S}}{\partial{P}}\biggr)_T<0.
\end{eqnarray}  
Therefore, entropy increases  by the isothermal compression from LQDL
to LQDA.  
The negative thermal expansion is also true for real 
solid $^4$He \cite{hoffer1976,goldstein1961}.

The Clausius-Clapeyron equation denoting the slope of melting line is expressed as 
\begin{eqnarray}
\label{eq:clapeyron}
\frac{dP_{\rm{m}}(T)}{dT}=\frac{{\Delta}S_{1\rightarrow{2}}}{{\Delta}V_{1\rightarrow{2}}}=\frac{{\Delta}H_{1\rightarrow{2}}}{T_{\rm{m}}{\Delta}V_{1\rightarrow{2}}},
\end{eqnarray}  
where ${\Delta}X_{1\rightarrow{2}}:=X_2-X_1$ is the change of $X$ in the process from state 1 to 2 at the melting 
temperature $T_{\rm{m}}$.   
At the minimum of melting curve  ($P_{\rm{min}}$, $T_{\rm{min}}$), we get  ${\Delta}S=0$, indicating that 
the entropy of solid and liquid is equal.  Therefore,  we regard this point as
the  Kauzmann point 
at which the melting curve and the ideal
glass transition line, yet unrevealed,
 cross with each other \cite{stillinger2001,stillinger2003}.

We further proceed
analysis of the results
by rereading Eq. (\ref{eq:clapeyron})
as a pseudo-Clapeyron equation
which approximately holds for
the slope of glass transition line \cite{lima2018},
\begin{eqnarray}
\label{eq:pclapeyron}
\frac{dP_{\rm{m}}(T)}{dT}\cong\frac{dP_{\rm{g}}(T)}{dT}\cong\frac{{\Delta}H_{1\rightarrow{2}}}
{T_{\rm{g}}{\Delta}V_{1\rightarrow{2}}},
\end{eqnarray}  
where $P_{\rm{g}}$ and $T_{\rm{g}}$ are the glass transition pressure and temperature, respectively,
while  ${\Delta}V_{1\rightarrow{2}}$ and 
${\Delta}H_{1\rightarrow{2}}$ are the volume and enthalpy change
over a finite temperature width $\Delta{T}_{1\rightarrow{2}}$
   riding on the glass transition line,  $T_1<T_{\rm{g}}<T_2$, respectively.
 As shown in Ref.~\onlinecite{lima2018}, 
this approximated 
relation is never far from reality,
because the difference between  thermodynamic property $X$ of a glass and the crystal is, in general, 
 small enough 
and the slopes of the  glass transition line and of the melting curve are nearly equal.
A similar  treatment was carried out in  our previous study where 
 the pseudo-Clapeyron equation was applied to the Widom line, which  we regarded 
 as the pseudo-boiling line separating the gas-like and liquid-like SCF \cite{takemoto2018}.
In Fig.~\ref{fig:V-T}, we can see 
${\Delta}V=V_{\rm{LQDA, 0.3 K}}-V_{\rm{HQDL, 0.2 K}}<0$  at $0.2\leq{T_{g}}\leq0.3$ K for the isobars 
of 40 and 50 bars, while in Fig.~\ref{fig:H-T}   enthalpy change with increasing $T$ is always 
positive, 
${\Delta}H>0$,  
 for any isobars.  
Accordingly,  Eq. (\ref{eq:pclapeyron}) yields  ${dP_{\rm{m}}}/{dT}\cong{dP_{\rm{g}}}/{dT}<0$, 
coinciding with negative  slope of the left half  ($T<T_{\rm{min}}$) 
of the glass transition line and presumed melting 
line in Fig.~\ref{fig:statediagram}.
Therefore,   the presumed  equilibrium crystal phase overlapping on the convex 
region of LQDA at 0.3 K and 40-50 bars 
should have larger entropy than the HQDL at 0.2 K and 40-50 bars.
On the other hand,  the slope of the right half  ($T>T_{\rm{min}}$) of 
the glass transition line and the presumed melting line   are  positive,
$dP_{\rm{m}}/dT\cong{dP_{\rm{g}}}/{dT}>0$, indicating the normal melting (melting by heating).
This is because we can see, in the range of 
 $0.4{\leq}T{\leq}3.3$ K
and $40{\leq}P{\leq}500$ bar,
${\Delta}H_{{\rm{solid (LQDA)}}\rightarrow{\rm{liquid (LQDL)}}}>0$ (Fig.~\ref{fig:H-T})
and
${\Delta}V_{{\rm{solid (LQDA)}}\rightarrow{\rm{liquid (LQDL)}}}>0$ (Fig.~\ref{fig:V-T}).
Thus, the analysis based on  the pseudo-Clapeyron equation 
evidences the downward convex curves of 
the glass transition line  and the presumed melting line shown in Fig.~\ref{fig:statediagram}.

\begin{figure*}
\includegraphics[width=10cm]{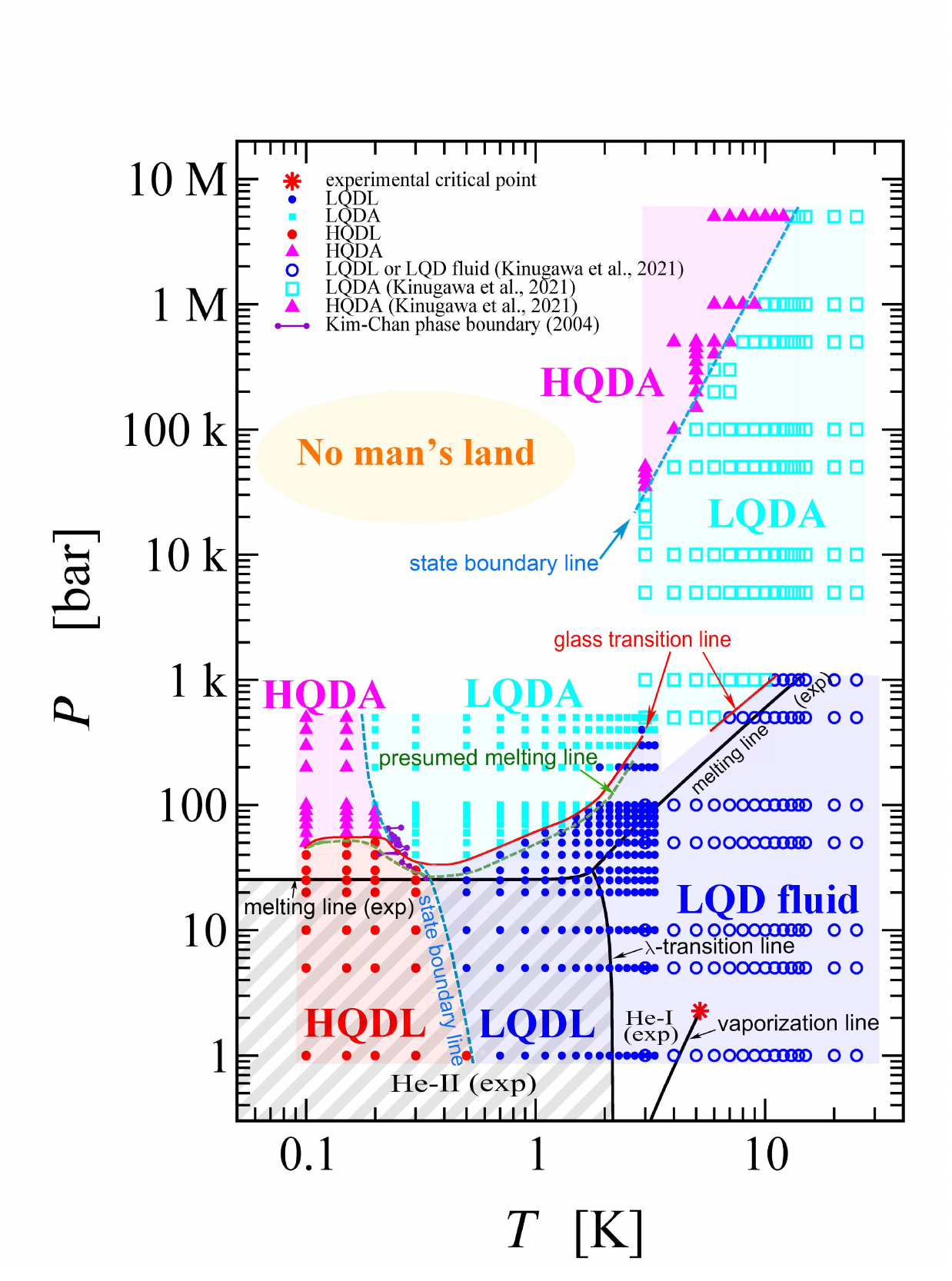}
\caption{\label{fig:diagramlarge} 
The state diagram of distinguishable $^4$He obtained from isothermal compression
over wider $P$-$T$ range.
 LQDL (blue): low quantum dispersion liquid;
 HQDL (red): high quantum dispersion liquid;
 LQDA (cyan): low quantum dispersion amorphous solid (metastable);
 HQDA (magenta): high quantum dispersion amorphous solid (metastable).
 ``LQD fluid" includes LQDL, 
 gas (in LQD state), and supercritical fluid 
 (in LQD state) existing beyond the critical point (red star). 
Black solid lines are the experimental phase boundary lines of real $^4$He.
Experimentally, He-I and He-II (gray slant stripe) exist at  higher and lower temperatures than $\rm{\lambda}$ transition 
line (black), respectively, 
while the crystalline solid phase exists at pressures higher than the melting curve (black).
The open symbols are the points explored by the isothermal compression simulation 
in our previous study \cite{kinugawa2021}.  
 The Kim-Chan boundary line is cited from Ref.~\onlinecite{kim2004}.
}
\end{figure*}

As described in Sec.~\ref{sec:introduction}, the present system should be liquid at $T=0$.
Since this should not violate the third law of thermodynamics,
the system must be in a single state without degeneration \cite{huang1987,johari2001}.
At $T=0$, therefore, the liquid  phase must have zero entropy equal to the crystal phase
existing at higher pressures,
so that the presumed melting line should be horizontal owing to the Clausius-Clapeyron 
equation, Eq. (\ref{eq:clapeyron}).
Although this line was drawn down to 0.1 K in Fig.~\ref{fig:statediagram},
at present,  it is unknown  whether this becomes horizontal even at non-zero 
temperatures such as 0.01 or 0.001 K before reaching $T=0$.
This issue would be revealed in a future work,
together with exploring the possibility of another liquid polymorph at
such lower temperatures.

Figure ~\ref{fig:diagramlarge} shows the extended state diagram  which includes
the data points at $3\leq{T}\leq25$ K provided in our previous paper \cite{kinugawa2021}.
Since  the LQDL covers the  $P$-$T$  conditions of He-I of real $^4$He, this state is considered as
being nearly identical  to He-I. 
Indeed, in our previous works, we showed that the properties of He-I were well reproduced by 
the CMD simulations obeying  Boltzmann statistics \cite{imaoka2017,takemoto2018,kinugawa2021}.
However, LQDL  is not completely
 identical to He-I because it does not involve  Bosonic permutation effect 
 though it must be negligibly weak in  He-I \cite{takemoto2018}. 
In this figure, we can see that the HQD states, in general,  exist at higher $P$ and lower $T$, i.e.,
at higher quantumness condition
 than LQD states.

Finally, it should be noted that any  state diagram including metastable states
is apt to depend considerably on the production process in
the simulations, as we saw in our previous work \cite{kinugawa2021}.
The production history, time increment \cite{kinugawa2021}, and 
adopted algorithm would affect the emergence of states.
Therefore, we should consider that the state diagrams, Figs.~\ref{fig:statediagram} and 
~\ref{fig:diagramlarge}, are not absolutely definitive
and another production procedure would possibly yield
a modified  state diagram.

\section{\label{sec:discussions}DISCUSSIONS}

\subsection{\label{sec:onstatediaram}On state transition}

As shown in Sec.~\ref{sec:VandE}, there is no discontinuity in the
$H$-$T$ plot and no maximum  in the $C_P$-$T$ relation  even when 
 the LQD-HQD state transitions occur.  
 This denies that the LQD-HQD transitions are  the first- or second-order phase transitions
 and supports that they are  continuous crossovers.
This  is in contrast to the fact that the ideal Bose gas exhibits a cusp at the 
BEC temperature $T_{\rm{C}}$ in the  $C_V$-$T$ plot \cite{huang1987} and that 
even the ideal quantum Boltzmann gas possesses a peak as well \cite{markham2020}.
For ideal Bose gas, as the number of atoms decreases from the thermodynamic limit 
($N\rightarrow\infty$), the cusp of the 
 $C_V$-$T$ plot becomes smeared-out \cite{glaum2007}. 
 This indicates a finite-size effect;
 in the $C_P$-$T$ relation of the confined $^4$He, not the $\lambda$-like peak but
   the blunted shape  emerges \cite{yamamoto2008,tani2022,gasparini2008}.
However, the lack of maximum in the  $C_P$-$T$ plot in this study is not because of 
the size effect.  Although the number of atoms in the present simulations is finite ($N=256$),
the periodic boundary condition ensures to reproduce  the bulk properties \cite{takemoto2018,kinugawa2021,imaoka2017}.
Therefore, the smooth change of $H$ and $V$ with varying temperature in the  LQDL-HQDL 
transition is 
 considered as being a bulk property.  
 This indicates that this transition is continuous crossover, not the
 thermodynamic transition.

In Sec.~\ref{sec:wavelength}, the analysis of temperature dependence of quantum wavelength $\lambda_{\rm{quantum}}$ showed 
the change of  the power $\chi$ in Eq. (\ref{eq:lambdatemp}) on the occasion of LQDL-HQDL transition at 1 bar.
The change of $\chi$  also happens when the state point crosses over
the Widom line in the SCF state of $^4$He.  
However, there is no appearance of the $C_P$ maximum on the LQDL-HQDL transition
(Sec.~\ref{sec:VandE}).
On the other hand, also at the LQDL-HQDL transition, the VAF exhibits the profile change  which is similar to 
that on the Frenkel line of the $^4$He SCF (Sec.~\ref{sec:VAF}).
Originally, this line is not  a thermodynamic  boundary but a dynamic boundary.
The present LQDL-HQDL boundary is not the thermodynamic boundary because of the lack of $C_P$ maximum.  In this sense,  it is closer to the Frenkel line.
However, we again note that both LQDL and HQDL are not SCF but liquid states without expanded volume.

\subsection{\label{sec:general}State metastability and statistical mechanical effect}

In Fig.~\ref{fig:kinvol},  we can see that 
the data points of  He-I, He-II, and solid phase of real $^4$He lie on a universal line.
This line, in other words, denotes the $K$-$V$ relation which corresponds to
the global minimum of the free energy,
since He-I, He-II, and solids phase are all at thermodynamic equilibrium.
However, the points of  HQDL are extensively 
scattered on the plane and are downward deviated from the universal line.
Therefore, if we introduced the Bosonic permutation to the present system,
these deviated   points of HQDL would be  metastable points.
This  means that the HQDL can exist only as a metastable state (if it can exist), not as an equilibrium state in reality (i.e., in the regime of Bose statistics).

However, we would mention that, plausibly,  HQDL is metastable even in  Boltzmann statistics.
One main reason is that the HQDL-LQDL transition 
 by varying temperature 
is  continuous without any indication of
thermodynamic transition. 
The LQDL-to-HQDL transition  by 
isobaric cooling should be 
similar to a glass transition in that a metastable state is formed from an equilibrium state
 without discontinuous change of enthalpy.
 Both transitions should depend on the production process and hysteresis.
The second reason is, being intuitive, that 
the wide scatter of  $K$-$V$ points of HQDL  seemingly 
reflects the ease of emergence of many configurations of stretched flexible necklaces.
This can be plausible because the springs of atomic necklaces are remarkably 
loosed by low temperatures.
There should be many local free-energy minima corresponding to necklace configurations in 
such stretched flexible {\it{polymer}} system.

In fact, we examined another CMD simulation of isothermal decompression of the
crystal of distinguishable $^4$He for 0.1-0.9 K, starting from 100 bar (the pressure at experimental solid phase region) \footnote{M. Tsujimoto and K. Kinugawa, to be
submitted.}.
Even down to 1 bar,  the crystal state persisted at the  $P$-$T$ conditions where the LQDL or HQDL were 
produced in the main simulation of this study.
We obtained, for example, the enthalpy at 0.5 K and 1 bar, $H=-0.057\pm0.004$ kJmol$^{\rm{-1}}$ 
for HQDL
and $-0.058\pm0.004$ kJmol$^{\rm{-1}}$ for the crystal.
Namely, the enthalpy difference is negligible within the standard deviation.
Therefore, we cannot say that the crystal is energetically stabler than HQDL.
Rather, this crystal state should be regarded as being as metastable as HQDL.
Accordingly, both   HQDL and the crystal obtained from the decompression 
must be metastable states under Boltzmann statistics.
Every  state  produced
in the experimental He-II region by the Boltzmann-type simulations 
 should plausibly be metastable and correspond
to  each local free-energy minimum.
As a result, during the simulations 
each state was allowed to exist quasi-stably as it was.
If we introduced Bose statistics to this system, 
these local minima would be converged into a 
global free-energy minimum, which corresponds to the true thermodynamic 
equilibrium state, He-II.
This should be akin to the gelation of sol, 
which is equivalent to the emergence of BEC by polymerizing the necklaces \cite{ftanaka2006}.
It seems to be difficult to judge which metastable state is the
stablest and at the thermodynamic equilibrium 
in the Boltzmann regime, because energetic difference is quite negligible
and estimation of entropy is burdensome. 
Further examination of the metastability would be a future subject.
Regardless of 
the discussion of 
metastability, it is important 
that  one metastable state called HQDL below 25 bar succeeded in non-freezing down to
 0.1 K in the present investigation.

As   in our  isothermal decompression test, the retention of crystal state  at 0.5 K and 24.3 cm$^3$mol$^{-1}$  was  
 reported for distinguishable $^4$He \cite{boninsegni2012}. 
This  can   also be attributed to the above-mentioned  metastability of
each  state in  distinguishable $^4$He.
The retention of the crystalline states, observed in  our 
and  their simulations, seems to be
nontrivial,
considering  that distinguishable $^4$He should be liquid at
$T=0$ as described in Sec.~\ref{sec:introduction}.  However, simple
extrapolation from the zero-temperature limit
does not predict  what state or liquid polymorph   emerges 
 at a given  $P$-$T$ point for $T>0$.

The flexibility of necklaces and metastability of each state points  endorse high  {\it{fragility}} 
 \cite{stillinger2003,ediger1996} of this liquid state.
Because of metastability of HQDL, it is probable that another production procedure
 makes the state boundary lines
shown in Fig.~\ref{fig:statediagram} be shifted on the $P$-$T$ plane.
It is also probable that 
 the  properties shown in Sec.~\ref{sec:resultsofiosthermalcompression} 
are somewhat deviated from the present.

One also sees  in Fig.~\ref{fig:kinvol} that the  $K$ of  the HQD states is lower than 
the experimental universal line.  
This suggests that atomic necklaces
of HQD are more stretched than    even He-II.
This can be attributed to the lack of the Bosonic correlation.
In fact, introducing the Bosonic permutation effectively corresponds to the addition of attractive 
correlation to the system \cite{huang1987}. It is known that this 
 makes a VAF be more oscillatory than that in the Boltzmann regime \cite{nakayama2005}.
 Therefore, the VAF of HQDL   shown in Fig.~\ref{fig:vaf}
 should be less oscillatory than the real He-II obeying Bose statistics.
On the other hand, seemingly,  the lower kinetic energy of HQDL than LQDL  
is in accordance with more  gas-like diffusion.
 
It is anticipated that  the off-diagonal 
density matrix $\rho(r)(=\rho(|{\bf{r}}-{\bf{r}}\prime|))$, i.e.,  the end-to-end distribution of  
isomorphic open chain polymer, may possess a non-zero distribution 
at  $r\lesssim\lambda_{\rm{quantum}}$.   The momentum distribution function 
$\tilde{\rho}(k)$, i.e.,  the Fourier transform of  $\rho(r)$, should then possess a sharp distribution 
at $k\sim{0}$.  This is the situation close to the BEC where $\tilde{\rho}(k)$ has an infinite peak at $k=0$ (momentum degeneration).
Although we did not evaluate  $\rho(r)$ in this study, long $\lambda_{\rm{quantum}}$ in HQD states 
is considered as a precursor of the BEC, prior to beginning 
the polymerization by introducing the atomic permutation.
Such polymerized open chains would exhibit even  longer
end-to-end distribution in $\rho(r)$ \cite{boninsegni2006}.

In Appendix \ref{sec:comexp},
we compared the present state diagram  with the experimental results of confined sub-nano
scale systems and {\it{supersolid}} of real $^4$He.

\section{\label{sec:conclusions}CONCLUSIONS}

In the present study, the path integral CMD simulation for distinguishable $^4$He obeying Boltzmann statistics  was conducted  in the range of  $0.08\leq{T}\leq{3.3}$ K and 
$1\leq{P}\leq{500}$ bar.  The $P$-$T$ thermodynamic conditions were
strictly  controlled by means of the isothermal-isobaric CMD method to provide 
computational results with high  quantitativeness.
On the basis of the isothermal compression simulations, we
revealed the state diagram of this ideal model system on a $P$-$T$ plane
($0.1\leq{T}\leq{3.3}$ K and $1\leq{P}\leq{500}$ bar).
The conclusions are summarized as follows:

(1) Even though this system does not include the Bosonic permutation effect, 
it does not freeze  down to 0.08 K when the pressure is  below 25 bar.
The non-freezing property of $^4$He is  fulfilled by sole NQE,
 not needing to introduce Bose statistics.

(2) There can exist two liquid states  in distinguishable $^4$He.
One liquid state is  LQDL which is nearly identical   to normal liquid He-I.
The other is  HQDL, which consists of  atoms with wider  spatial extension
 of necklaces  and  exists at lower temperatures ($T\leq0.5$ K) than LQDL. 
There can be two glass states, LQDA and HQDA, corresponding to the two liquid states,
and they were the same as those discovered at higher temperatures and pressures in our previous work.
These glass states are considered as metastable states which should eventually become 
the crystal phase as the thermodynamic equilibrium state.

(3) The HQDL is a non-superfluid state with lower entropy than the solid (LQDA) existing at higher temperatures.
It can exist  only as a metastable state 
under Bose statistics  (if it  exists).
Plausibly,  it is also metastable even in the Boltzmann regime.
The HQDL  is  a {\it{fragile}} liquid involving many configurations of  stretched flexible necklaces.  This fragility benefits the occurrence of liquid polyamorphism
in distinguishable $^4$He.
The  velocity autocorrelation function of  HQDL  
exhibits such a relaxation as observed in  the gas-like supercritical fluid above the Frenkel line.  
In HQDL, the atoms diffuse  without  noticeable contribution from quantum 
tunneling.

(4)  The presumed melting line  on the $P$-$T$ plane
 is  downward convex and has a minimum point at $T\simeq{0.4}$ K and $P\simeq{25}$ bar, from which
the LQDL-HQDL boundary line emanates, as an indication of  liquid polyamorphism. 
Since there is no $C_P$ maximum at this transition, the LQDL-HQDL transition  is not a 
thermodynamic phase transition but a continuous crossover.
Observable structural change  by the transition is smooth and continuous.
Thus, the existence of these two liquids is not conventional classical polyamorphism 
but the  {\it{quantum polyamorphism}} characterized by the change of atomic
quantum dispersion.
 The temperature dependence of  quantum wavelength  and of its expansion factor remarkably changes
 on the occasion of  the LQDL-HQDL transition.

(5) At 40-50 bar, the HQDL undergoes the 
inverse freezing (endothermic {\it{freezing by heating}}) into LQDA  when it is heated from 0.2 K to 0.3 K.  This inverse freezing is accompanied with the change of spatial extension 
of atomic necklaces.  This resembles the coil-globule transition in classical polymer systems.

(6) There are only negligible portions of tunneling atoms in all the states: HQDL, LQDL, HQDA, and LQDA.  

\;

 Finally,  in  Appendix \ref{sec:comexp}, the state diagram of this study was compared with 
the experimental reports of confined systems and 
{\it{supersolid}} of real $^4$He.
 A part of   $P$-$T$ region of HQDL 
overlaps with that of the real non-superfluid states (the LBEC and the BEC-like low-entropy state) 
of  $^4$He confined in nanopores in meso-porous materials,
which 
should be more Boltzmann-like (or suppressed Bosonic) than the bulk.
Further, the $P$-$T$ region of the HQDL causing inverse freezing almost overlaps with 
the reported {\it{supersolid}} region neighboring to 
the Kim-Chan state boundary.
These comparisons provide a future subject  as to a question 
whether HQDL has some essential relevance to these real systems.

This is the first investigation that  revealed the state diagram, the existence of liquid polyamorphism,
and inverse freezing in  distinguishable $^4$He without inclusion of Bosonic statistical
effect.
As a future subject, it
 is  intriguing 
to  investigate whether the sole NQE would generate another 
liquid 
polymorph
  when the system is cooled down to
 far lower temperatures such as 1 mK 
before reaching $T=0$,
at which distinguishable $^4$He should be  a liquid. 
It is also unrevealed whether even more ``classical'' systems such as Ne and Ar can melt at 
extremely low temperatures solely by the NQE  when the temperature is extremely lowered.
The universality of 
 quantum liquid polyamorphism
  induced by NQE is still 
an open question.

\appendix

\section{\label{sec:freezing}TEST OF FREEZING BY HEATING}

In Appendix \ref{sec:freezing}, we provide the results of another CMD simulation to test 
the  inverse freezing.
In Fig.~\ref{fig:statediagram}, we can see that an isobaric heating of the  HQDL at 40 bar, 
starting from 0.2 K toward 0.3 K,  may induce  solidification or inverse freezing ({\it{freezing by heating}}).  
To check this, starting from 
the end configuration of the  production  run at 0.2 K and 40 bar, we  conducted three isobaric CMD runs.
One of them was carried out keeping the temperature at 0.2 K, while the other runs were the process of 
raising the set  temperature to 0.3 or
to  0.4 K.

\begin{figure}
\includegraphics[width=8cm]{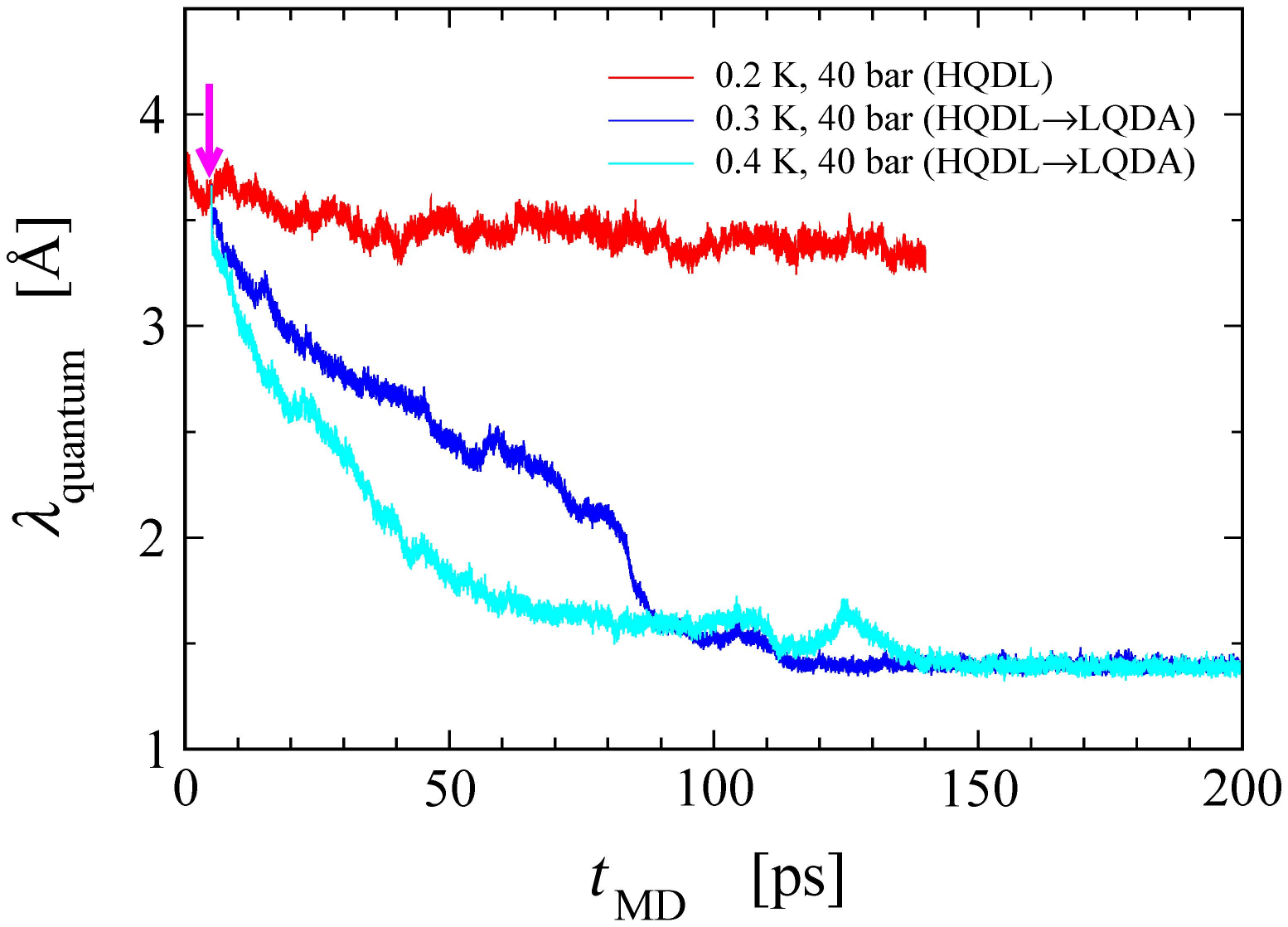}
\caption{\label{fig:lamfreeze}
The MD time evolution of quantum wavelength  of $^4$He atoms by isobaric heating test at 40 bar.
The graphs show the results of three CMD runs with different control of temperature: (1) keeping the temperature at 0.2 K throughout the run; 
(2) raising temperature  to 0.3 K at $t_{\rm{MD}}=5$ ps (pointed by a magenta arrow);
(3) raising temperature  to 0.4 K at $t_{\rm{MD}}=5$ ps (pointed by a magenta arrow).}
\end{figure}

The results of the MD time evolution of $\lambda_{\rm{quantum}}$ are shown in 
Fig. \ref{fig:lamfreeze}.
 In this figure,  $\lambda_{\rm{quantum}}$  retains  about 
 3.5 {\AA} when the temperature is kept at 0.2 K.  In contrast, 
 it  starts to be shortened right after the temperature switching to 0.3 or 0.4 K
 at $t_{\rm{MD}}=5$ ps.  
The quantum wavelength finally resulted in $\lambda_{\rm{quantum}}=1.7-1.8$ {\AA},
which is even shorter than 
 2.2 {\AA} at 0.3 K (LQDA) shown in Fig.~\ref{fig:lamT}.

\begin{figure}
\includegraphics[width=8cm]{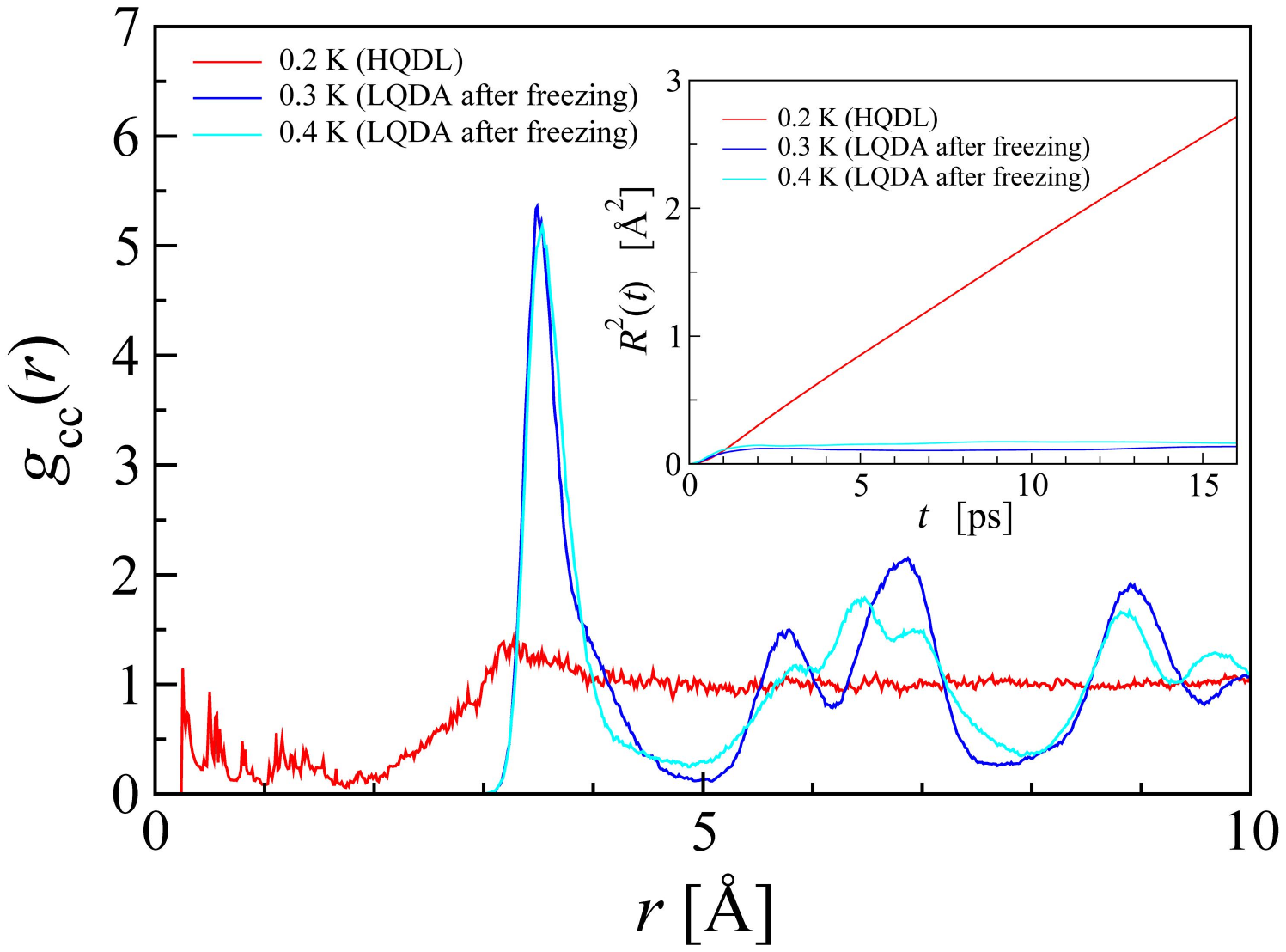}
\caption{\label{fig:heattest} 
The centroid-centroid radial distribution function $g_{\rm{cc}}$ and the 
mean square displacement $R^2(t)$ for the states  before and after the isobaric heating test 
at 40 bar.  The $R^2(t)$  is shown in the inset.}
\end{figure}

 The $g_{\rm{cc}}$ before and after heating are shown in 
  Fig.~\ref{fig:heattest}, while the MSDs for the states before and after the temperature switching are shown in the inset. The non-zero distribution characteristic of  the $g_{\rm{cc}}$ of  HQDL vanishes  
  at 0.3 and 0.4 K.
 The apparent self-diffusion coefficient estimated from the MSD at 0.2 K and 40 bar is  $D=2.9\times10^{-6}$ cm$^2$s$^{-1}$, while $D$ at 0.3 or 0.4 K is zero. 
  Thus, judging from  $g_{\rm{cc}}$ and $D$, we conclude that 
  the transition from HQDL  to LQDA, i.e., {\it{glassification by heating}},
  occurred between 0.2 and 0.3 K.  
The transition process is visualized in Fig. SC-1 in the Supplementary Material, 
which  shows the  snapshot change with MD time. 
The final state attained at 0.3 K is a partially crystallized LQDA.

The average thermodynamic properties  before and after the temperature switching are listed
  in Table SC-1 in the
  Supplementary Material. 
  If we assume the HQDL$\rightarrow$LQDA transition temperature as 
  $T_{\rm{tr}}=0.25$ K, the pseudo-Clapeyron equation,
  Eq. (\ref{eq:pclapeyron}),
  provides  a negative slope of the transition line, $dP/dT=-1.14$ kbarK$^{-1}$.

  The increase of kinetic energy $K$ in Table SC-1 in the Supplementary Material
  is explained by the
  shrinking of necklaces 
  by the HQDL$\rightarrow$LQDA transition. 
  Since  there is almost no 
concentration of  negative  $K_i$ in the HQDL at 0.2 K before heating
(Sec.~\ref{sec:tunneling}), 
the HQDL$\rightarrow$LQDA freezing transition is not a transition 
 from tunneling to trapped regimes.  This is unlike
the situation occurring in compressed $^4$He \cite{kinugawa2021}.

This test simulation showed  that the inverse freezing (endothermic freezing) 
of HQDL was actually caused by isobaric heating  
from 0.2 to 0.3-0.4 K at 40 bar.  This transition $P$-$T$ agrees with the HQDL-LQDA 
boundary line drawn in the state diagram (Fig.~\ref{fig:statediagram}).
Thus, the present test yielded the same conclusion as we saw in Sec.~\ref{sec:onstatediagram}.
Although the $P$-$T$ condition of the state boundary in  Fig.~\ref{fig:statediagram} cannot  completely be free from possible  hysteresis of production process,
the present agreement ensures that the HQDL existing in  
the upward convex region in  Fig.~\ref{fig:statediagram}
certainly freezes on the occasion of crossing over the drawn HQDL-LQDA boundary line.
This line almost overlaps  with the Kim-Chan boundary, as described in  Appendix \ref{sec:comp2}.

 A similar type of  inverse melting  was  observed for the classical systems 
  such as water (the transition from low-density amorphous solid  (LDA)
  to low-density amorphous liquid  (LDL)) \cite{mishima3961998} 
and classical polymers \cite{rastogi1991,mortensen1992}.
The higher entropy of the partially crystallized LQDA  than HQDL  can possibly 
be attributed to the existence of transverse phonon mode   absent in liquid phase 
 \cite{stillinger2001}.
We note that  the  inverse freezing of this system
is accompanied by the change of spatial extension of atomic necklaces, i.e., 
the atomic quantum dispersion.  The structural change of individual necklaces
resembles the coil-globule transition in classical polymer systems 
\cite{lifshitz1978,maffi2012,wu1998,podewitz2019,ma1995,simmons2008,simmons2010,sherman2006,matsuyama1991}.

\begin{figure*}
\includegraphics[width=12cm]{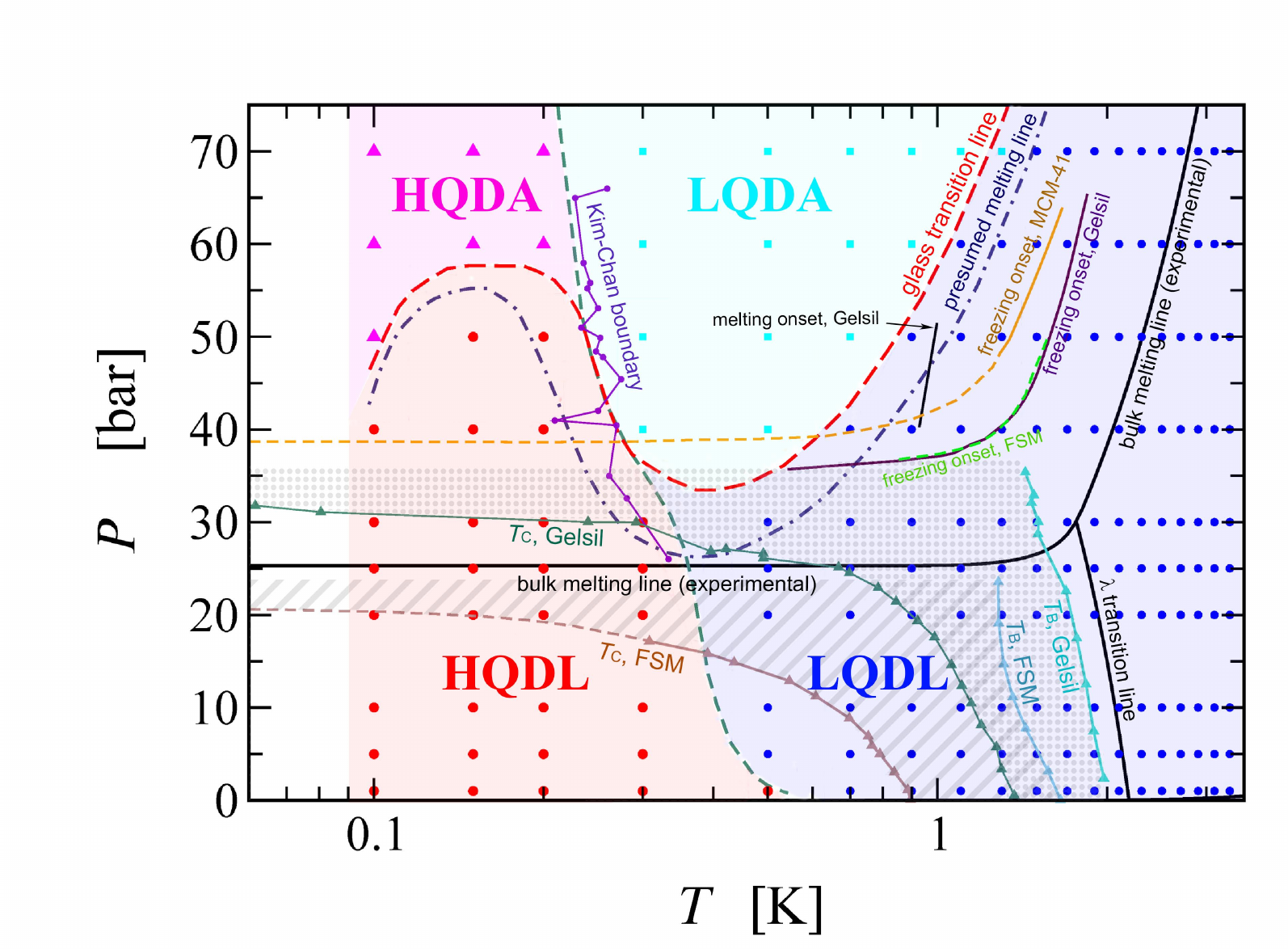}
\caption{\label{fig:confined} 
The  comparison of 
state diagram of distinguishable $^4$He 
and
confined real $^4$He systems.
 LQDL (blue): low quantum dispersion liquid;
 HQDL (red): high quantum dispersion liquid;
 LQDA (cyan): low quantum dispersion amorphous solid (metastable);
 HQDA (magenta): high quantum dispersion amorphous solid (metastable).
The Kim-Chan phase boundary is cited from Ref.~\onlinecite{kim2004}.  
 The phase boundary lines
of real $^4$He doped in 
Gelsil (25 \AA) \cite{shirahama2008ltp},
folded sheets meso-porous materials (FSM) (28 \AA) \cite{taniguchi2011}, 
and MCM-41 (core diameter 47 \AA) \cite{bossy2012},
are drawn together.  
``$T_{\rm{C}}$, FSM'': superfluid onset temperature of the FSM system; 
``$T_{\rm{B}}$, FSM'': rapid growth temperature of the superfluid fraction of the FSM system \cite{taniguchi2011};
``$T_{\rm{C}}$, Gelsil'': superfluid transition temperature of the Gelsil system;
``$T_{\rm{B}}$, Gelsil'': localized BEC (LBEC) temperature of the Gelsil system \cite{shirahama2008ltp}.
Experimentally, the BEC-like low-entropy state exists in gray slant stripe region \cite{taniguchi2011}, while 
the LBEC exists in gray polka-dot region \cite{shirahama2008ltp}.
}
\end{figure*}

\section{\label{sec:comexp} STATE DIAGRAM COMPARISON  WITH
NON-SUPERFLUID STATES OF REAL SYSTEMS}

In Appendix \ref{sec:comexp}, withholding speculative interpretation,
the present state diagram is compared with that of two types of real systems,
 (1) several confined $^4$He systems and  (2) reported {\it{supersolid}}.

\subsection{\label{sec:comp1} Comparison with confined $^4$He}
Recent experiments targeting  confined 
$^4$He revealed that the non-superfluid states,
unlike He-I and He-II, 
exist at temperatures lower than the ${\rm{\lambda}}$-transition line of the bulk 
system \cite{yamamoto2004,yamamoto2008,yamamoto2008jpsj,shirahama2008ltp,shirahama2008,bossy2008,taniguchi2010,taniguchi2011,taniguchi2013,bossy2012,bossy2019}.  
Under such confined conditions, the 
off-diagonal long-range Bosonic correlation is  hindered  
 by the  wall of glass substrate pores, so that Bosonic permutation effect
 should  partially be suppressed \cite{shirahama2008,shirahama2008ltp}. 
Figure \ref{fig:confined} shows the comparison of our state diagram with
that of confined $^4$He in three kinds of substrates.
Shirahama and co-workers revealed that the localized BEC  (LBEC) state 
of $^4$He confined in Gelsil exists in the region between the boundaries of 
``$T_{\rm{B}}$, Gelsil'' and ``$T_{\rm{C}}$, Gelsil'' \cite{yamamoto2004,yamamoto2008,yamamoto2008jpsj,shirahama2008ltp,shirahama2008}.  
For $^4$He confined in the folded sheets meso-porous 
materials (FSM) (28 \AA),
Taniguchi et al. reported that  the state 
at  temperatures between ``$T_{\rm{B}}$, FSM'' and ``$T_{\rm{C}}$, FSM''
was not  He-I  nor He-II but  a BEC-like  state with low entropy \cite{taniguchi2011}.
In this figure, 
 the experimental state boundaries have  the same negative 
slopes as  the HQDL-LQDA boundary line.
Some part of the $P$-$T$ regions of the  LBEC and the low-entropy BEC-like states
overlaps with the upward convex region ($18\lesssim{P}\lesssim{35}$ bar and $T\lesssim{0.35}$ K)
of the HQDL causing  inverse freezing.

\subsection{\label{sec:comp2} Comparison with reported supersolid
$^4$He region}
In Figs.~\ref{fig:statediagram} and \ref{fig:confined}, 
the HQDL-LQDA boundary emanating from 
the point at $T=0.3$ K and $P=26$ bar 
traces well the Kim-Chan  boundary line~\cite{kim2004}.  
This boundary  was   interpreted as an evidence of  {\it{supersolid}} which might exist at lower-temperature 
side 
of the line, on the basis of the detection of non-classical moment of inertia by the torsional oscillator experiment \cite{kim2004}.
However, nowadays the existence of  defect-less crystalline 
supersolid  phase is almost denied, 
while  this anomaly is attributed to some disorder
due to dislocation, grain boundaries, glassy domains, and $^3$He impurities \cite{balibar2008}.
In such a narrow domain unlike a bulk, the long-range atomic permutation should partially 
be suppressed  
so that the BEC should not completely be fulfilled. 
The HQDL has the upward convex $P$-$T$ region at $T\lesssim0.3$ K and
$P\gtrsim25$ bar, 
while it  causes such inverse freezing as shown in
Appendix \ref{sec:freezing}. 
This $P$-$T$ region of HQDL  coincides with the supersolid region.

\nocite{*}
\bibliography{Kinugawa.bib}

\providecommand{\noopsort}[1]{}\providecommand{\singleletter}[1]{#1}%
\begin{thebibliography}{86}%
\makeatletter
\providecommand \@ifxundefined [1]{%
 \@ifx{#1\undefined}
}%
\providecommand \@ifnum [1]{%
 \ifnum #1\expandafter \@firstoftwo
 \else \expandafter \@secondoftwo
 \fi
}%
\providecommand \@ifx [1]{%
 \ifx #1\expandafter \@firstoftwo
 \else \expandafter \@secondoftwo
 \fi
}%
\providecommand \natexlab [1]{#1}%
\providecommand \enquote  [1]{``#1''}%
\providecommand \bibnamefont  [1]{#1}%
\providecommand \bibfnamefont [1]{#1}%
\providecommand \citenamefont [1]{#1}%
\providecommand \href@noop [0]{\@secondoftwo}%
\providecommand \href [0]{\begingroup \@sanitize@url \@href}%
\providecommand \@href[1]{\@@startlink{#1}\@@href}%
\providecommand \@@href[1]{\endgroup#1\@@endlink}%
\providecommand \@sanitize@url [0]{\catcode `\\12\catcode `\$12\catcode `\&12\catcode `\#12\catcode `\^12\catcode `\_12\catcode `\%12\relax}%
\providecommand \@@startlink[1]{}%
\providecommand \@@endlink[0]{}%
\providecommand \url  [0]{\begingroup\@sanitize@url \@url }%
\providecommand \@url [1]{\endgroup\@href {#1}{\urlprefix }}%
\providecommand \urlprefix  [0]{URL }%
\providecommand \Eprint [0]{\href }%
\providecommand \doibase [0]{https://doi.org/}%
\providecommand \selectlanguage [0]{\@gobble}%
\providecommand \bibinfo  [0]{\@secondoftwo}%
\providecommand \bibfield  [0]{\@secondoftwo}%
\providecommand \translation [1]{[#1]}%
\providecommand \BibitemOpen [0]{}%
\providecommand \bibitemStop [0]{}%
\providecommand \bibitemNoStop [0]{.\EOS\space}%
\providecommand \EOS [0]{\spacefactor3000\relax}%
\providecommand \BibitemShut  [1]{\csname bibitem#1\endcsname}%
\let\auto@bib@innerbib\@empty
\bibitem [{\citenamefont {Feynman}(1972)}]{feynman1972}%
  \BibitemOpen
  \bibfield  {author} {\bibinfo {author} {\bibfnamefont {R.~P.}\ \bibnamefont {Feynman}},\ }\href@noop {} {\emph {\bibinfo {title} {Statistical Mechanics}}}\ (\bibinfo  {publisher} {Addison-Wesley},\ \bibinfo {address} {Reading, MA},\ \bibinfo {year} {1972})\BibitemShut {NoStop}%
\bibitem [{\citenamefont {London}(1938{\natexlab{a}})}]{london1938}%
  \BibitemOpen
  \bibfield  {author} {\bibinfo {author} {\bibfnamefont {F.}~\bibnamefont {London}},\ }\href@noop {} {\bibfield  {journal} {\bibinfo  {journal} {Nature}\ }\textbf {\bibinfo {volume} {141}},\ \bibinfo {pages} {643} (\bibinfo {year} {1938}{\natexlab{a}})}\BibitemShut {NoStop}%
\bibitem [{\citenamefont {London}(1938{\natexlab{b}})}]{london1938pr}%
  \BibitemOpen
  \bibfield  {author} {\bibinfo {author} {\bibfnamefont {F.}~\bibnamefont {London}},\ }\href@noop {} {\bibfield  {journal} {\bibinfo  {journal} {Phys. Rev.}\ }\textbf {\bibinfo {volume} {54}},\ \bibinfo {pages} {947} (\bibinfo {year} {1938}{\natexlab{b}})}\BibitemShut {NoStop}%
\bibitem [{\citenamefont {Pitaevskii}\ and\ \citenamefont {Stringari}(2016)}]{pitaevskii2016}%
  \BibitemOpen
  \bibfield  {author} {\bibinfo {author} {\bibfnamefont {L.}~\bibnamefont {Pitaevskii}}\ and\ \bibinfo {author} {\bibfnamefont {S.}~\bibnamefont {Stringari}},\ }\href@noop {} {\emph {\bibinfo {title} {Bose-Einstein Condensation and Superfluidity}}},\ \bibinfo {series} {International Series of Monographs on Physics}, Vol.\ \bibinfo {volume} {164}\ (\bibinfo  {publisher} {Oxford University Press},\ \bibinfo {address} {NY},\ \bibinfo {year} {2016})\BibitemShut {NoStop}%
\bibitem [{\citenamefont {Boninsegni}\ \emph {et~al.}(2012)\citenamefont {Boninsegni}, \citenamefont {Pollet}, \citenamefont {Prokof'ev},\ and\ \citenamefont {Svistunov}}]{boninsegni2012}%
  \BibitemOpen
  \bibfield  {author} {\bibinfo {author} {\bibfnamefont {M.}~\bibnamefont {Boninsegni}}, \bibinfo {author} {\bibfnamefont {L.}~\bibnamefont {Pollet}}, \bibinfo {author} {\bibfnamefont {N.}~\bibnamefont {Prokof'ev}},\ and\ \bibinfo {author} {\bibfnamefont {B.}~\bibnamefont {Svistunov}},\ }\href@noop {} {\bibfield  {journal} {\bibinfo  {journal} {Phys. Rev. Lett.}\ }\textbf {\bibinfo {volume} {109}},\ \bibinfo {pages} {025302} (\bibinfo {year} {2012})}\BibitemShut {NoStop}%
\bibitem [{\citenamefont {Huang}(1987)}]{huang1987}%
  \BibitemOpen
  \bibfield  {author} {\bibinfo {author} {\bibfnamefont {K.}~\bibnamefont {Huang}},\ }\href@noop {} {\emph {\bibinfo {title} {Statistical Mechanics}}},\ \bibinfo {edition} {2nd}\ ed.\ (\bibinfo  {publisher} {Wiley},\ \bibinfo {address} {NY},\ \bibinfo {year} {1987})\BibitemShut {NoStop}%
\bibitem [{\citenamefont {Bernu}\ \emph {et~al.}(1987)\citenamefont {Bernu}, \citenamefont {Hansen}, \citenamefont {Hiwatari},\ and\ \citenamefont {Pastore}}]{bernu1987}%
  \BibitemOpen
  \bibfield  {author} {\bibinfo {author} {\bibfnamefont {B.}~\bibnamefont {Bernu}}, \bibinfo {author} {\bibfnamefont {J.~P.}\ \bibnamefont {Hansen}}, \bibinfo {author} {\bibfnamefont {Y.}~\bibnamefont {Hiwatari}},\ and\ \bibinfo {author} {\bibfnamefont {G.}~\bibnamefont {Pastore}},\ }\href@noop {} {\bibfield  {journal} {\bibinfo  {journal} {Phys. Rev. A}\ }\textbf {\bibinfo {volume} {36}},\ \bibinfo {pages} {4891} (\bibinfo {year} {1987})}\BibitemShut {NoStop}%
\bibitem [{\citenamefont {Markland}\ \emph {et~al.}(2011)\citenamefont {Markland}, \citenamefont {Morrone}, \citenamefont {Berne}, \citenamefont {Miyazaki}, \citenamefont {Rabani},\ and\ \citenamefont {Reichman}}]{markland2011}%
  \BibitemOpen
  \bibfield  {author} {\bibinfo {author} {\bibfnamefont {T.~E.}\ \bibnamefont {Markland}}, \bibinfo {author} {\bibfnamefont {J.~A.}\ \bibnamefont {Morrone}}, \bibinfo {author} {\bibfnamefont {B.~J.}\ \bibnamefont {Berne}}, \bibinfo {author} {\bibfnamefont {K.}~\bibnamefont {Miyazaki}}, \bibinfo {author} {\bibfnamefont {E.}~\bibnamefont {Rabani}},\ and\ \bibinfo {author} {\bibfnamefont {D.~R.}\ \bibnamefont {Reichman}},\ }\href@noop {} {\bibfield  {journal} {\bibinfo  {journal} {Nat. Phys.}\ }\textbf {\bibinfo {volume} {7}},\ \bibinfo {pages} {134} (\bibinfo {year} {2011})}\BibitemShut {NoStop}%
\bibitem [{\citenamefont {Markland}\ \emph {et~al.}(2012)\citenamefont {Markland}, \citenamefont {Morrone}, \citenamefont {Miyazaki}, \citenamefont {Berne}, \citenamefont {Reichman},\ and\ \citenamefont {Rabani}}]{markland2012}%
  \BibitemOpen
  \bibfield  {author} {\bibinfo {author} {\bibfnamefont {T.~E.}\ \bibnamefont {Markland}}, \bibinfo {author} {\bibfnamefont {J.~A.}\ \bibnamefont {Morrone}}, \bibinfo {author} {\bibfnamefont {K.}~\bibnamefont {Miyazaki}}, \bibinfo {author} {\bibfnamefont {B.~J.}\ \bibnamefont {Berne}}, \bibinfo {author} {\bibfnamefont {D.~R.}\ \bibnamefont {Reichman}},\ and\ \bibinfo {author} {\bibfnamefont {E.}~\bibnamefont {Rabani}},\ }\href@noop {} {\bibfield  {journal} {\bibinfo  {journal} {J. Chem. Phys.}\ }\textbf {\bibinfo {volume} {136}},\ \bibinfo {pages} {074511} (\bibinfo {year} {2012})}\BibitemShut {NoStop}%
\bibitem [{\citenamefont {Anderson}\ \emph {et~al.}(1972)\citenamefont {Anderson}, \citenamefont {Halperin},\ and\ \citenamefont {Varma}}]{anderson1972}%
  \BibitemOpen
  \bibfield  {author} {\bibinfo {author} {\bibfnamefont {P.~W.}\ \bibnamefont {Anderson}}, \bibinfo {author} {\bibfnamefont {B.~I.}\ \bibnamefont {Halperin}},\ and\ \bibinfo {author} {\bibfnamefont {C.~M.}\ \bibnamefont {Varma}},\ }\href@noop {} {\bibfield  {journal} {\bibinfo  {journal} {Philos. Mag.}\ }\textbf {\bibinfo {volume} {25}},\ \bibinfo {pages} {1} (\bibinfo {year} {1972})}\BibitemShut {NoStop}%
\bibitem [{\citenamefont {Phillips}(1972)}]{phillips1972}%
  \BibitemOpen
  \bibfield  {author} {\bibinfo {author} {\bibfnamefont {W.~A.}\ \bibnamefont {Phillips}},\ }\href@noop {} {\bibfield  {journal} {\bibinfo  {journal} {J. Low Temp. Phys.}\ }\textbf {\bibinfo {volume} {7}},\ \bibinfo {pages} {351} (\bibinfo {year} {1972})}\BibitemShut {NoStop}%
\bibitem [{\citenamefont {Stanley}(2013)}]{stanley2013}%
  \BibitemOpen
  \bibfield  {author} {\bibinfo {author} {\bibfnamefont {H.~E.}\ \bibnamefont {Stanley}},\ }\href@noop {} {\emph {\bibinfo {title} {Liquid Polyamorphism}}},\ \bibinfo {series} {Advances in Chemical Physics}, Vol.\ \bibinfo {volume} {152}\ (\bibinfo  {publisher} {Wiley},\ \bibinfo {address} {Hoboken, NJ},\ \bibinfo {year} {2013})\BibitemShut {NoStop}%
\bibitem [{\citenamefont {Tanaka}(2020)}]{tanaka2020}%
  \BibitemOpen
  \bibfield  {author} {\bibinfo {author} {\bibfnamefont {H.}~\bibnamefont {Tanaka}},\ }\href@noop {} {\bibfield  {journal} {\bibinfo  {journal} {J. Chem. Phys.}\ }\textbf {\bibinfo {volume} {153}},\ \bibinfo {pages} {130901} (\bibinfo {year} {2020})}\BibitemShut {NoStop}%
\bibitem [{\citenamefont {Ediger}\ \emph {et~al.}(1996)\citenamefont {Ediger}, \citenamefont {Angell},\ and\ \citenamefont {Nagel}}]{ediger1996}%
  \BibitemOpen
  \bibfield  {author} {\bibinfo {author} {\bibfnamefont {M.~D.}\ \bibnamefont {Ediger}}, \bibinfo {author} {\bibfnamefont {C.~A.}\ \bibnamefont {Angell}},\ and\ \bibinfo {author} {\bibfnamefont {S.~R.}\ \bibnamefont {Nagel}},\ }\href@noop {} {\bibfield  {journal} {\bibinfo  {journal} {J. Phys. Chem.}\ }\textbf {\bibinfo {volume} {100}},\ \bibinfo {pages} {13200} (\bibinfo {year} {1996})}\BibitemShut {NoStop}%
\bibitem [{\citenamefont {Mishima}\ and\ \citenamefont {Stanley}(1998{\natexlab{a}})}]{mishima3921998}%
  \BibitemOpen
  \bibfield  {author} {\bibinfo {author} {\bibfnamefont {O.}~\bibnamefont {Mishima}}\ and\ \bibinfo {author} {\bibfnamefont {H.~G.}\ \bibnamefont {Stanley}},\ }\href@noop {} {\bibfield  {journal} {\bibinfo  {journal} {Nature}\ }\textbf {\bibinfo {volume} {392}},\ \bibinfo {pages} {164} (\bibinfo {year} {1998}{\natexlab{a}})}\BibitemShut {NoStop}%
\bibitem [{\citenamefont {Mishima}\ and\ \citenamefont {Stanley}(1998{\natexlab{b}})}]{mishima3961998}%
  \BibitemOpen
  \bibfield  {author} {\bibinfo {author} {\bibfnamefont {O.}~\bibnamefont {Mishima}}\ and\ \bibinfo {author} {\bibfnamefont {H.~G.}\ \bibnamefont {Stanley}},\ }\href@noop {} {\bibfield  {journal} {\bibinfo  {journal} {Nature}\ }\textbf {\bibinfo {volume} {396}},\ \bibinfo {pages} {329} (\bibinfo {year} {1998}{\natexlab{b}})}\BibitemShut {NoStop}%
\bibitem [{\citenamefont {Katayama}\ \emph {et~al.}(2004)\citenamefont {Katayama}, \citenamefont {Inamura}, \citenamefont {Mizutani}, \citenamefont {Yamakata}, \citenamefont {Utsumi},\ and\ \citenamefont {Shimomura}}]{katayama2004}%
  \BibitemOpen
  \bibfield  {author} {\bibinfo {author} {\bibfnamefont {Y.}~\bibnamefont {Katayama}}, \bibinfo {author} {\bibfnamefont {Y.}~\bibnamefont {Inamura}}, \bibinfo {author} {\bibfnamefont {T.}~\bibnamefont {Mizutani}}, \bibinfo {author} {\bibfnamefont {M.}~\bibnamefont {Yamakata}}, \bibinfo {author} {\bibfnamefont {W.}~\bibnamefont {Utsumi}},\ and\ \bibinfo {author} {\bibfnamefont {O.}~\bibnamefont {Shimomura}},\ }\href@noop {} {\bibfield  {journal} {\bibinfo  {journal} {Science}\ }\textbf {\bibinfo {volume} {306}},\ \bibinfo {pages} {848} (\bibinfo {year} {2004})}\BibitemShut {NoStop}%
\bibitem [{\citenamefont {Morales}\ \emph {et~al.}(2010)\citenamefont {Morales}, \citenamefont {Pierleoni}, \citenamefont {Schwegler},\ and\ \citenamefont {Ceperley}}]{morales2010}%
  \BibitemOpen
  \bibfield  {author} {\bibinfo {author} {\bibfnamefont {M.~A.}\ \bibnamefont {Morales}}, \bibinfo {author} {\bibfnamefont {C.}~\bibnamefont {Pierleoni}}, \bibinfo {author} {\bibfnamefont {E.}~\bibnamefont {Schwegler}},\ and\ \bibinfo {author} {\bibfnamefont {D.~M.}\ \bibnamefont {Ceperley}},\ }\href@noop {} {\bibfield  {journal} {\bibinfo  {journal} {Proc. Natl. Acad. Sci. U. S. A.}\ }\textbf {\bibinfo {volume} {107}},\ \bibinfo {pages} {12799} (\bibinfo {year} {2010})}\BibitemShut {NoStop}%
\bibitem [{\citenamefont {Nguyen}\ \emph {et~al.}(2018)\citenamefont {Nguyen}, \citenamefont {Lopez},\ and\ \citenamefont {Giovambattista}}]{nguyen2018}%
  \BibitemOpen
  \bibfield  {author} {\bibinfo {author} {\bibfnamefont {B.}~\bibnamefont {Nguyen}}, \bibinfo {author} {\bibfnamefont {G.~E.}\ \bibnamefont {Lopez}},\ and\ \bibinfo {author} {\bibfnamefont {N.}~\bibnamefont {Giovambattista}},\ }\href@noop {} {\bibfield  {journal} {\bibinfo  {journal} {Phys. Chem. Chem. Phys.}\ }\textbf {\bibinfo {volume} {20}},\ \bibinfo {pages} {8210} (\bibinfo {year} {2018})}\BibitemShut {NoStop}%
\bibitem [{\citenamefont {Eltareb}\ \emph {et~al.}(2022)\citenamefont {Eltareb}, \citenamefont {Lopez},\ and\ \citenamefont {Giovambattista}}]{eltareb2022}%
  \BibitemOpen
  \bibfield  {author} {\bibinfo {author} {\bibfnamefont {A.}~\bibnamefont {Eltareb}}, \bibinfo {author} {\bibfnamefont {G.~E.}\ \bibnamefont {Lopez}},\ and\ \bibinfo {author} {\bibfnamefont {N.}~\bibnamefont {Giovambattista}},\ }\href@noop {} {\bibfield  {journal} {\bibinfo  {journal} {J. Chem. Phys.}\ }\textbf {\bibinfo {volume} {156}},\ \bibinfo {pages} {204502} (\bibinfo {year} {2022})}\BibitemShut {NoStop}%
\bibitem [{\citenamefont {Kinugawa}\ and\ \citenamefont {Takemoto}(2021)}]{kinugawa2021}%
  \BibitemOpen
  \bibfield  {author} {\bibinfo {author} {\bibfnamefont {K.}~\bibnamefont {Kinugawa}}\ and\ \bibinfo {author} {\bibfnamefont {A.}~\bibnamefont {Takemoto}},\ }\href@noop {} {\bibfield  {journal} {\bibinfo  {journal} {J. Chem. Phys.}\ }\textbf {\bibinfo {volume} {154}},\ \bibinfo {pages} {224503} (\bibinfo {year} {2021})}\BibitemShut {NoStop}%
\bibitem [{Note1()}]{Note1}%
  \BibitemOpen
  \bibinfo {note} {Here we should mention the abbreviation of these names. In Ref.~\protect \rev@citealp {kinugawa2021}, we abbreviated them as ``LDA'' and ``HDA'', respectively. However, since these may be confused with the low and high density amorphous state of classical systems (e.g., water \cite {mishima3921998,mishima3961998}), we presently call them LQDA and HQDA instead.}\BibitemShut {Stop}%
\bibitem [{\citenamefont {Lifshitz}\ \emph {et~al.}(1978)\citenamefont {Lifshitz}, \citenamefont {Grosberg},\ and\ \citenamefont {Khokhlov}}]{lifshitz1978}%
  \BibitemOpen
  \bibfield  {author} {\bibinfo {author} {\bibfnamefont {I.~M.}\ \bibnamefont {Lifshitz}}, \bibinfo {author} {\bibfnamefont {A.~Y.}\ \bibnamefont {Grosberg}},\ and\ \bibinfo {author} {\bibfnamefont {A.~R.}\ \bibnamefont {Khokhlov}},\ }\href@noop {} {\bibfield  {journal} {\bibinfo  {journal} {Rev. Mod. Phys.}\ }\textbf {\bibinfo {volume} {50}},\ \bibinfo {pages} {683} (\bibinfo {year} {1978})}\BibitemShut {NoStop}%
\bibitem [{\citenamefont {Maffi}\ \emph {et~al.}(2012)\citenamefont {Maffi}, \citenamefont {Baiesi}, \citenamefont {Casetti}, \citenamefont {Piazza},\ and\ \citenamefont {Rios}}]{maffi2012}%
  \BibitemOpen
  \bibfield  {author} {\bibinfo {author} {\bibfnamefont {C.}~\bibnamefont {Maffi}}, \bibinfo {author} {\bibfnamefont {M.}~\bibnamefont {Baiesi}}, \bibinfo {author} {\bibfnamefont {L.}~\bibnamefont {Casetti}}, \bibinfo {author} {\bibfnamefont {F.}~\bibnamefont {Piazza}},\ and\ \bibinfo {author} {\bibfnamefont {P.~D.~L.}\ \bibnamefont {Rios}},\ }\href@noop {} {\bibfield  {journal} {\bibinfo  {journal} {Nat. Commun.}\ }\textbf {\bibinfo {volume} {3}},\ \bibinfo {pages} {1065} (\bibinfo {year} {2012})}\BibitemShut {NoStop}%
\bibitem [{\citenamefont {Wu}\ and\ \citenamefont {Wang}(1998)}]{wu1998}%
  \BibitemOpen
  \bibfield  {author} {\bibinfo {author} {\bibfnamefont {C.}~\bibnamefont {Wu}}\ and\ \bibinfo {author} {\bibfnamefont {X.}~\bibnamefont {Wang}},\ }\href@noop {} {\bibfield  {journal} {\bibinfo  {journal} {Phys. Rev. Lett.}\ }\textbf {\bibinfo {volume} {80}},\ \bibinfo {pages} {4092} (\bibinfo {year} {1998})}\BibitemShut {NoStop}%
\bibitem [{\citenamefont {Podewitz}\ \emph {et~al.}(2019)\citenamefont {Podewitz}, \citenamefont {Wang}, \citenamefont {Quoika}, \citenamefont {Loeffler}, \citenamefont {Schauperl},\ and\ \citenamefont {Liedl}}]{podewitz2019}%
  \BibitemOpen
  \bibfield  {author} {\bibinfo {author} {\bibfnamefont {M.}~\bibnamefont {Podewitz}}, \bibinfo {author} {\bibfnamefont {Y.}~\bibnamefont {Wang}}, \bibinfo {author} {\bibfnamefont {P.~K.}\ \bibnamefont {Quoika}}, \bibinfo {author} {\bibfnamefont {J.~R.}\ \bibnamefont {Loeffler}}, \bibinfo {author} {\bibfnamefont {M.}~\bibnamefont {Schauperl}},\ and\ \bibinfo {author} {\bibfnamefont {K.~R.}\ \bibnamefont {Liedl}},\ }\href@noop {} {\bibfield  {journal} {\bibinfo  {journal} {J. Phys. Chem. B}\ }\textbf {\bibinfo {volume} {123}},\ \bibinfo {pages} {8838} (\bibinfo {year} {2019})}\BibitemShut {NoStop}%
\bibitem [{\citenamefont {Ma}\ \emph {et~al.}(1995)\citenamefont {Ma}, \citenamefont {Straub},\ and\ \citenamefont {Shakhnovich}}]{ma1995}%
  \BibitemOpen
  \bibfield  {author} {\bibinfo {author} {\bibfnamefont {J.}~\bibnamefont {Ma}}, \bibinfo {author} {\bibfnamefont {J.~E.}\ \bibnamefont {Straub}},\ and\ \bibinfo {author} {\bibfnamefont {E.~I.}\ \bibnamefont {Shakhnovich}},\ }\href@noop {} {\bibfield  {journal} {\bibinfo  {journal} {J. Chem. Phys.}\ }\textbf {\bibinfo {volume} {103}},\ \bibinfo {pages} {2615} (\bibinfo {year} {1995})}\BibitemShut {NoStop}%
\bibitem [{\citenamefont {Simmons}\ and\ \citenamefont {Sanchez}(2008)}]{simmons2008}%
  \BibitemOpen
  \bibfield  {author} {\bibinfo {author} {\bibfnamefont {D.~S.}\ \bibnamefont {Simmons}}\ and\ \bibinfo {author} {\bibfnamefont {I.~C.}\ \bibnamefont {Sanchez}},\ }\href@noop {} {\bibfield  {journal} {\bibinfo  {journal} {Macromolecules}\ }\textbf {\bibinfo {volume} {41}},\ \bibinfo {pages} {5885} (\bibinfo {year} {2008})}\BibitemShut {NoStop}%
\bibitem [{\citenamefont {Simmons}\ and\ \citenamefont {Sanchez}(2010)}]{simmons2010}%
  \BibitemOpen
  \bibfield  {author} {\bibinfo {author} {\bibfnamefont {D.~S.}\ \bibnamefont {Simmons}}\ and\ \bibinfo {author} {\bibfnamefont {I.~C.}\ \bibnamefont {Sanchez}},\ }\href@noop {} {\bibfield  {journal} {\bibinfo  {journal} {Macromolecules}\ }\textbf {\bibinfo {volume} {43}},\ \bibinfo {pages} {1571} (\bibinfo {year} {2010})}\BibitemShut {NoStop}%
\bibitem [{\citenamefont {Sherman}\ and\ \citenamefont {Haran}(2006)}]{sherman2006}%
  \BibitemOpen
  \bibfield  {author} {\bibinfo {author} {\bibfnamefont {E.}~\bibnamefont {Sherman}}\ and\ \bibinfo {author} {\bibfnamefont {G.}~\bibnamefont {Haran}},\ }\href@noop {} {\bibfield  {journal} {\bibinfo  {journal} {Proc. Natl. Acad. Sci. U. S. A.}\ }\textbf {\bibinfo {volume} {103}},\ \bibinfo {pages} {11539} (\bibinfo {year} {2006})}\BibitemShut {NoStop}%
\bibitem [{\citenamefont {Matsuyama}\ and\ \citenamefont {Tanaka}(1991)}]{matsuyama1991}%
  \BibitemOpen
  \bibfield  {author} {\bibinfo {author} {\bibfnamefont {A.}~\bibnamefont {Matsuyama}}\ and\ \bibinfo {author} {\bibfnamefont {F.}~\bibnamefont {Tanaka}},\ }\href@noop {} {\bibfield  {journal} {\bibinfo  {journal} {J. Chem. Phys.}\ }\textbf {\bibinfo {volume} {94}},\ \bibinfo {pages} {781} (\bibinfo {year} {1991})}\BibitemShut {NoStop}%
\bibitem [{\citenamefont {Kim}\ and\ \citenamefont {Chan}(2004)}]{kim2004}%
  \BibitemOpen
  \bibfield  {author} {\bibinfo {author} {\bibfnamefont {E.}~\bibnamefont {Kim}}\ and\ \bibinfo {author} {\bibfnamefont {M.~H.~W.}\ \bibnamefont {Chan}},\ }\href@noop {} {\bibfield  {journal} {\bibinfo  {journal} {Science}\ }\textbf {\bibinfo {volume} {305}},\ \bibinfo {pages} {1941} (\bibinfo {year} {2004})}\BibitemShut {NoStop}%
\bibitem [{\citenamefont {Hallock}(2015)}]{hallock2015}%
  \BibitemOpen
  \bibfield  {author} {\bibinfo {author} {\bibfnamefont {R.}~\bibnamefont {Hallock}},\ }\href@noop {} {\bibfield  {journal} {\bibinfo  {journal} {Phys. Today}\ }\textbf {\bibinfo {volume} {68}},\ \bibinfo {pages} {30} (\bibinfo {year} {2015})}\BibitemShut {NoStop}%
\bibitem [{\citenamefont {Wilks}(1970)}]{wilks1970}%
  \BibitemOpen
  \bibfield  {author} {\bibinfo {author} {\bibfnamefont {J.}~\bibnamefont {Wilks}},\ }\href@noop {} {\emph {\bibinfo {title} {An Introduction to Liquid Helium}}},\ Oxford Library of the Physical Sciences\ (\bibinfo  {publisher} {Clarendon},\ \bibinfo {address} {Oxford, UK},\ \bibinfo {year} {1970})\BibitemShut {NoStop}%
\bibitem [{\citenamefont {Rapoport}(1967)}]{rapoport1967}%
  \BibitemOpen
  \bibfield  {author} {\bibinfo {author} {\bibfnamefont {E.}~\bibnamefont {Rapoport}},\ }\href@noop {} {\bibfield  {journal} {\bibinfo  {journal} {J. Chem. Phys.}\ }\textbf {\bibinfo {volume} {46}},\ \bibinfo {pages} {2891} (\bibinfo {year} {1967})}\BibitemShut {NoStop}%
\bibitem [{\citenamefont {Stillinger}\ \emph {et~al.}(2001)\citenamefont {Stillinger}, \citenamefont {Debenedetti},\ and\ \citenamefont {Truskett}}]{stillinger2001}%
  \BibitemOpen
  \bibfield  {author} {\bibinfo {author} {\bibfnamefont {F.~H.}\ \bibnamefont {Stillinger}}, \bibinfo {author} {\bibfnamefont {P.~G.}\ \bibnamefont {Debenedetti}},\ and\ \bibinfo {author} {\bibfnamefont {T.~M.}\ \bibnamefont {Truskett}},\ }\href@noop {} {\bibfield  {journal} {\bibinfo  {journal} {J. Phys. Chem. B}\ }\textbf {\bibinfo {volume} {105}},\ \bibinfo {pages} {11809} (\bibinfo {year} {2001})}\BibitemShut {NoStop}%
\bibitem [{\citenamefont {Stillinger}\ and\ \citenamefont {Debenedetti}(2003)}]{stillinger2003}%
  \BibitemOpen
  \bibfield  {author} {\bibinfo {author} {\bibfnamefont {F.~H.}\ \bibnamefont {Stillinger}}\ and\ \bibinfo {author} {\bibfnamefont {P.~G.}\ \bibnamefont {Debenedetti}},\ }\href@noop {} {\bibfield  {journal} {\bibinfo  {journal} {Biophys. Chem.}\ }\textbf {\bibinfo {volume} {105}},\ \bibinfo {pages} {211} (\bibinfo {year} {2003})}\BibitemShut {NoStop}%
\bibitem [{\citenamefont {Feeney}\ \emph {et~al.}(2003)\citenamefont {Feeney}, \citenamefont {Debenedetti},\ and\ \citenamefont {Stillinger}}]{feeney2003}%
  \BibitemOpen
  \bibfield  {author} {\bibinfo {author} {\bibfnamefont {M.~R.}\ \bibnamefont {Feeney}}, \bibinfo {author} {\bibfnamefont {P.~G.}\ \bibnamefont {Debenedetti}},\ and\ \bibinfo {author} {\bibfnamefont {F.~H.}\ \bibnamefont {Stillinger}},\ }\href@noop {} {\bibfield  {journal} {\bibinfo  {journal} {J. Chem. Phys.}\ }\textbf {\bibinfo {volume} {119}},\ \bibinfo {pages} {4582} (\bibinfo {year} {2003})}\BibitemShut {NoStop}%
\bibitem [{\citenamefont {Schupper}\ and\ \citenamefont {Shnerb}(2005)}]{schupper2005}%
  \BibitemOpen
  \bibfield  {author} {\bibinfo {author} {\bibfnamefont {N.}~\bibnamefont {Schupper}}\ and\ \bibinfo {author} {\bibfnamefont {N.~M.}\ \bibnamefont {Shnerb}},\ }\href@noop {} {\bibfield  {journal} {\bibinfo  {journal} {Phys. Rev. E}\ }\textbf {\bibinfo {volume} {72}},\ \bibinfo {pages} {046107} (\bibinfo {year} {2005})}\BibitemShut {NoStop}%
\bibitem [{\citenamefont {Prestipino}(2007)}]{prestipino2007}%
  \BibitemOpen
  \bibfield  {author} {\bibinfo {author} {\bibfnamefont {S.}~\bibnamefont {Prestipino}},\ }\href@noop {} {\bibfield  {journal} {\bibinfo  {journal} {Phys. Rev. E}\ }\textbf {\bibinfo {volume} {75}},\ \bibinfo {pages} {011107} (\bibinfo {year} {2007})}\BibitemShut {NoStop}%
\bibitem [{\citenamefont {Greer}(2000)}]{greer2000}%
  \BibitemOpen
  \bibfield  {author} {\bibinfo {author} {\bibfnamefont {A.~L.}\ \bibnamefont {Greer}},\ }\href@noop {} {\bibfield  {journal} {\bibinfo  {journal} {Nature}\ }\textbf {\bibinfo {volume} {404}},\ \bibinfo {pages} {134} (\bibinfo {year} {2000})}\BibitemShut {NoStop}%
\bibitem [{\citenamefont {Landau}\ and\ \citenamefont {Lifshitz}(1980)}]{landau1980}%
  \BibitemOpen
  \bibfield  {author} {\bibinfo {author} {\bibfnamefont {L.~D.}\ \bibnamefont {Landau}}\ and\ \bibinfo {author} {\bibfnamefont {E.~M.}\ \bibnamefont {Lifshitz}},\ }\href@noop {} {\emph {\bibinfo {title} {Statistical Physics}}},\ \bibinfo {edition} {3rd}\ ed.\ (\bibinfo  {publisher} {Pergamon},\ \bibinfo {address} {Oxford, UK},\ \bibinfo {year} {1980})\ \bibinfo {note} {revised and enlarged by E. M. Lifshitz and L. P. Pitaevskii}\BibitemShut {NoStop}%
\bibitem [{\citenamefont {Chandler}\ and\ \citenamefont {Wolynes}(1981)}]{chandler1981}%
  \BibitemOpen
  \bibfield  {author} {\bibinfo {author} {\bibfnamefont {D.}~\bibnamefont {Chandler}}\ and\ \bibinfo {author} {\bibfnamefont {P.~G.}\ \bibnamefont {Wolynes}},\ }\href@noop {} {\bibfield  {journal} {\bibinfo  {journal} {J. Chem. Phys.}\ }\textbf {\bibinfo {volume} {74}},\ \bibinfo {pages} {4078} (\bibinfo {year} {1981})}\BibitemShut {NoStop}%
\bibitem [{\citenamefont {Frenkel}\ and\ \citenamefont {Smit}(2002)}]{frenkel2002}%
  \BibitemOpen
  \bibfield  {author} {\bibinfo {author} {\bibfnamefont {D.}~\bibnamefont {Frenkel}}\ and\ \bibinfo {author} {\bibfnamefont {B.}~\bibnamefont {Smit}},\ }\href@noop {} {\emph {\bibinfo {title} {Understanding Molecular Simulation: From Algorithm to Applications}}}\ (\bibinfo  {publisher} {Academic Press},\ \bibinfo {address} {NY},\ \bibinfo {year} {2002})\BibitemShut {NoStop}%
\bibitem [{\citenamefont {Aziz}\ \emph {et~al.}(1995)\citenamefont {Aziz}, \citenamefont {Janzen},\ and\ \citenamefont {Moldover}}]{aziz1995}%
  \BibitemOpen
  \bibfield  {author} {\bibinfo {author} {\bibfnamefont {R.~A.}\ \bibnamefont {Aziz}}, \bibinfo {author} {\bibfnamefont {A.~R.}\ \bibnamefont {Janzen}},\ and\ \bibinfo {author} {\bibfnamefont {M.~R.}\ \bibnamefont {Moldover}},\ }\href@noop {} {\bibfield  {journal} {\bibinfo  {journal} {Phys. Rev. Lett.}\ }\textbf {\bibinfo {volume} {74}},\ \bibinfo {pages} {1586} (\bibinfo {year} {1995})}\BibitemShut {NoStop}%
\bibitem [{\citenamefont {Takemoto}\ and\ \citenamefont {Kinugawa}(2018)}]{takemoto2018}%
  \BibitemOpen
  \bibfield  {author} {\bibinfo {author} {\bibfnamefont {A.}~\bibnamefont {Takemoto}}\ and\ \bibinfo {author} {\bibfnamefont {K.}~\bibnamefont {Kinugawa}},\ }\href@noop {} {\bibfield  {journal} {\bibinfo  {journal} {J. Chem. Phys.}\ }\textbf {\bibinfo {volume} {149}},\ \bibinfo {pages} {204504} (\bibinfo {year} {2018})}\BibitemShut {NoStop}%
\bibitem [{\citenamefont {Martyna}\ \emph {et~al.}(1999)\citenamefont {Martyna}, \citenamefont {Hughes},\ and\ \citenamefont {Tuckerman}}]{martyna1999}%
  \BibitemOpen
  \bibfield  {author} {\bibinfo {author} {\bibfnamefont {G.~J.}\ \bibnamefont {Martyna}}, \bibinfo {author} {\bibfnamefont {A.}~\bibnamefont {Hughes}},\ and\ \bibinfo {author} {\bibfnamefont {M.~E.}\ \bibnamefont {Tuckerman}},\ }\href@noop {} {\bibfield  {journal} {\bibinfo  {journal} {J. Chem. Phys.}\ }\textbf {\bibinfo {volume} {110}},\ \bibinfo {pages} {3275} (\bibinfo {year} {1999})}\BibitemShut {NoStop}%
\bibitem [{\citenamefont {Yamamoto}(2005)}]{yamamoto2005}%
  \BibitemOpen
  \bibfield  {author} {\bibinfo {author} {\bibfnamefont {T.~M.}\ \bibnamefont {Yamamoto}},\ }\href@noop {} {\bibfield  {journal} {\bibinfo  {journal} {J. Chem. Phys.}\ }\textbf {\bibinfo {volume} {123}},\ \bibinfo {pages} {104101} (\bibinfo {year} {2005})}\BibitemShut {NoStop}%
\bibitem [{\citenamefont {Glaesemann}\ and\ \citenamefont {Fried}(2002)}]{glaesemann2002}%
  \BibitemOpen
  \bibfield  {author} {\bibinfo {author} {\bibfnamefont {K.~R.}\ \bibnamefont {Glaesemann}}\ and\ \bibinfo {author} {\bibfnamefont {L.~E.}\ \bibnamefont {Fried}},\ }\href@noop {} {\bibfield  {journal} {\bibinfo  {journal} {J. Chem. Phys.}\ }\textbf {\bibinfo {volume} {117}},\ \bibinfo {pages} {3020} (\bibinfo {year} {2002})}\BibitemShut {NoStop}%
\bibitem [{\citenamefont {Tuckerman}\ \emph {et~al.}(1993)\citenamefont {Tuckerman}, \citenamefont {Berne}, \citenamefont {Martyna},\ and\ \citenamefont {Klein}}]{tuckerman1993}%
  \BibitemOpen
  \bibfield  {author} {\bibinfo {author} {\bibfnamefont {M.~E.}\ \bibnamefont {Tuckerman}}, \bibinfo {author} {\bibfnamefont {B.~J.}\ \bibnamefont {Berne}}, \bibinfo {author} {\bibfnamefont {G.~J.}\ \bibnamefont {Martyna}},\ and\ \bibinfo {author} {\bibfnamefont {M.~L.}\ \bibnamefont {Klein}},\ }\href@noop {} {\bibfield  {journal} {\bibinfo  {journal} {J. Chem. Phys.}\ }\textbf {\bibinfo {volume} {99}},\ \bibinfo {pages} {2796} (\bibinfo {year} {1993})}\BibitemShut {NoStop}%
\bibitem [{\citenamefont {Glyde}\ \emph {et~al.}(2011)\citenamefont {Glyde}, \citenamefont {Diallo}, \citenamefont {Azuah}, \citenamefont {Kirichek},\ and\ \citenamefont {Taylor}}]{glyde2011}%
  \BibitemOpen
  \bibfield  {author} {\bibinfo {author} {\bibfnamefont {H.~R.}\ \bibnamefont {Glyde}}, \bibinfo {author} {\bibfnamefont {S.~O.}\ \bibnamefont {Diallo}}, \bibinfo {author} {\bibfnamefont {R.~T.}\ \bibnamefont {Azuah}}, \bibinfo {author} {\bibfnamefont {O.}~\bibnamefont {Kirichek}},\ and\ \bibinfo {author} {\bibfnamefont {J.~W.}\ \bibnamefont {Taylor}},\ }\href@noop {} {\bibfield  {journal} {\bibinfo  {journal} {Phys. Rev. B}\ }\textbf {\bibinfo {volume} {84}},\ \bibinfo {pages} {184506} (\bibinfo {year} {2011})}\BibitemShut {NoStop}%
\bibitem [{\citenamefont {Goto}\ \emph {et~al.}(2023)\citenamefont {Goto}, \citenamefont {Kim},\ and\ \citenamefont {Matsubayashi}}]{goto2023}%
  \BibitemOpen
  \bibfield  {author} {\bibinfo {author} {\bibfnamefont {S.}~\bibnamefont {Goto}}, \bibinfo {author} {\bibfnamefont {K.}~\bibnamefont {Kim}},\ and\ \bibinfo {author} {\bibfnamefont {N.}~\bibnamefont {Matsubayashi}},\ }\href@noop {} {\bibfield  {journal} {\bibinfo  {journal} {ACS Polymer}\ }\textbf {\bibinfo {volume} {3}},\ \bibinfo {pages} {437} (\bibinfo {year} {2023})}\BibitemShut {NoStop}%
\bibitem [{\citenamefont {Cai}\ \emph {et~al.}(2022)\citenamefont {Cai}, \citenamefont {Liang}, \citenamefont {Liu},\ and\ \citenamefont {Zhang}}]{cai2022}%
  \BibitemOpen
  \bibfield  {author} {\bibinfo {author} {\bibfnamefont {X.}~\bibnamefont {Cai}}, \bibinfo {author} {\bibfnamefont {C.}~\bibnamefont {Liang}}, \bibinfo {author} {\bibfnamefont {H.}~\bibnamefont {Liu}},\ and\ \bibinfo {author} {\bibfnamefont {G.}~\bibnamefont {Zhang}},\ }\href@noop {} {\bibfield  {journal} {\bibinfo  {journal} {Polymer}\ }\textbf {\bibinfo {volume} {253}},\ \bibinfo {pages} {124953} (\bibinfo {year} {2022})}\BibitemShut {NoStop}%
\bibitem [{\citenamefont {Nichols}\ \emph {et~al.}(1984)\citenamefont {Nichols}, \citenamefont {Chandler}, \citenamefont {Singh},\ and\ \citenamefont {Richardson}}]{nichols1984}%
  \BibitemOpen
  \bibfield  {author} {\bibinfo {author} {\bibfnamefont {A.~L.}\ \bibnamefont {Nichols}}, \bibinfo {author} {\bibfnamefont {D.}~\bibnamefont {Chandler}}, \bibinfo {author} {\bibfnamefont {Y.}~\bibnamefont {Singh}},\ and\ \bibinfo {author} {\bibfnamefont {D.~M.}\ \bibnamefont {Richardson}},\ }\href@noop {} {\bibfield  {journal} {\bibinfo  {journal} {J. Chem. Phys.}\ }\textbf {\bibinfo {volume} {81}},\ \bibinfo {pages} {5109} (\bibinfo {year} {1984})}\BibitemShut {NoStop}%
\bibitem [{\citenamefont {Rubinstein}\ and\ \citenamefont {Colby}(2003)}]{rubinstein2003}%
  \BibitemOpen
  \bibfield  {author} {\bibinfo {author} {\bibfnamefont {M.}~\bibnamefont {Rubinstein}}\ and\ \bibinfo {author} {\bibfnamefont {R.~H.}\ \bibnamefont {Colby}},\ }\href@noop {} {\emph {\bibinfo {title} {Polymer Physics}}}\ (\bibinfo  {publisher} {Oxford University Press},\ \bibinfo {address} {NY},\ \bibinfo {year} {2003})\BibitemShut {NoStop}%
\bibitem [{\citenamefont {Brazahkin}\ \emph {et~al.}(2013)\citenamefont {Brazahkin}, \citenamefont {Fomin}, \citenamefont {Lyapin}, \citenamefont {Ryzhov}, \citenamefont {Tsiok},\ and\ \citenamefont {Trachenko}}]{brazhkin2013}%
  \BibitemOpen
  \bibfield  {author} {\bibinfo {author} {\bibfnamefont {V.~V.}\ \bibnamefont {Brazahkin}}, \bibinfo {author} {\bibfnamefont {Y.~D.}\ \bibnamefont {Fomin}}, \bibinfo {author} {\bibfnamefont {A.~G.}\ \bibnamefont {Lyapin}}, \bibinfo {author} {\bibfnamefont {V.~N.}\ \bibnamefont {Ryzhov}}, \bibinfo {author} {\bibfnamefont {E.~N.}\ \bibnamefont {Tsiok}},\ and\ \bibinfo {author} {\bibfnamefont {K.}~\bibnamefont {Trachenko}},\ }\href@noop {} {\bibfield  {journal} {\bibinfo  {journal} {Phys. Rev. Lett.}\ }\textbf {\bibinfo {volume} {111}},\ \bibinfo {pages} {145901} (\bibinfo {year} {2013})}\BibitemShut {NoStop}%
\bibitem [{\citenamefont {Caupin}\ \emph {et~al.}(2003)\citenamefont {Caupin}, \citenamefont {Balibar},\ and\ \citenamefont {Maris}}]{caupin2003}%
  \BibitemOpen
  \bibfield  {author} {\bibinfo {author} {\bibfnamefont {F.}~\bibnamefont {Caupin}}, \bibinfo {author} {\bibfnamefont {S.}~\bibnamefont {Balibar}},\ and\ \bibinfo {author} {\bibfnamefont {H.~J.}\ \bibnamefont {Maris}},\ }\href@noop {} {\bibfield  {journal} {\bibinfo  {journal} {Physica B}\ }\textbf {\bibinfo {volume} {329-333}},\ \bibinfo {pages} {356} (\bibinfo {year} {2003})}\BibitemShut {NoStop}%
\bibitem [{\citenamefont {Werner}\ \emph {et~al.}(2004)\citenamefont {Werner}, \citenamefont {Beaume}, \citenamefont {Hobeika}, \citenamefont {Nascimbene}, \citenamefont {Herrmann}, \citenamefont {Caupin},\ and\ \citenamefont {Balibar}}]{werner2004}%
  \BibitemOpen
  \bibfield  {author} {\bibinfo {author} {\bibfnamefont {F.}~\bibnamefont {Werner}}, \bibinfo {author} {\bibfnamefont {G.}~\bibnamefont {Beaume}}, \bibinfo {author} {\bibfnamefont {A.}~\bibnamefont {Hobeika}}, \bibinfo {author} {\bibfnamefont {S.}~\bibnamefont {Nascimbene}}, \bibinfo {author} {\bibfnamefont {C.}~\bibnamefont {Herrmann}}, \bibinfo {author} {\bibfnamefont {F.}~\bibnamefont {Caupin}},\ and\ \bibinfo {author} {\bibfnamefont {S.}~\bibnamefont {Balibar}},\ }\href@noop {} {\bibfield  {journal} {\bibinfo  {journal} {J. Low Temp. Phys.}\ }\textbf {\bibinfo {volume} {136}},\ \bibinfo {pages} {93} (\bibinfo {year} {2004})}\BibitemShut {NoStop}%
\bibitem [{\citenamefont {Tombari}\ \emph {et~al.}(2005)\citenamefont {Tombari}, \citenamefont {Ferrari}, \citenamefont {Salvetti},\ and\ \citenamefont {Johari}}]{tombari2005}%
  \BibitemOpen
  \bibfield  {author} {\bibinfo {author} {\bibfnamefont {E.}~\bibnamefont {Tombari}}, \bibinfo {author} {\bibfnamefont {C.}~\bibnamefont {Ferrari}}, \bibinfo {author} {\bibfnamefont {G.}~\bibnamefont {Salvetti}},\ and\ \bibinfo {author} {\bibfnamefont {G.~P.}\ \bibnamefont {Johari}},\ }\href@noop {} {\bibfield  {journal} {\bibinfo  {journal} {J. Chem. Phys.}\ }\textbf {\bibinfo {volume} {123}},\ \bibinfo {pages} {051104} (\bibinfo {year} {2005})}\BibitemShut {NoStop}%
\bibitem [{\citenamefont {Hoffer}\ \emph {et~al.}(1976)\citenamefont {Hoffer}, \citenamefont {Gardner}, \citenamefont {Waterfield},\ and\ \citenamefont {Phillips}}]{hoffer1976}%
  \BibitemOpen
  \bibfield  {author} {\bibinfo {author} {\bibfnamefont {J.~K.}\ \bibnamefont {Hoffer}}, \bibinfo {author} {\bibfnamefont {W.~R.}\ \bibnamefont {Gardner}}, \bibinfo {author} {\bibfnamefont {C.~G.}\ \bibnamefont {Waterfield}},\ and\ \bibinfo {author} {\bibfnamefont {N.~E.}\ \bibnamefont {Phillips}},\ }\href@noop {} {\bibfield  {journal} {\bibinfo  {journal} {J. Low Temp. Phys.}\ }\textbf {\bibinfo {volume} {23}},\ \bibinfo {pages} {63} (\bibinfo {year} {1976})}\BibitemShut {NoStop}%
\bibitem [{\citenamefont {Goldstein}(1961)}]{goldstein1961}%
  \BibitemOpen
  \bibfield  {author} {\bibinfo {author} {\bibfnamefont {L.}~\bibnamefont {Goldstein}},\ }\href@noop {} {\bibfield  {journal} {\bibinfo  {journal} {Phys. Rev.}\ }\textbf {\bibinfo {volume} {122}},\ \bibinfo {pages} {726} (\bibinfo {year} {1961})}\BibitemShut {NoStop}%
\bibitem [{\citenamefont {Lima}\ \emph {et~al.}(2018)\citenamefont {Lima}, \citenamefont {Faria}, \citenamefont {Paschoal},\ and\ \citenamefont {Ribeiro}}]{lima2018}%
  \BibitemOpen
  \bibfield  {author} {\bibinfo {author} {\bibfnamefont {T.~A.}\ \bibnamefont {Lima}}, \bibinfo {author} {\bibfnamefont {L.~F.~O.}\ \bibnamefont {Faria}}, \bibinfo {author} {\bibfnamefont {V.~H.}\ \bibnamefont {Paschoal}},\ and\ \bibinfo {author} {\bibfnamefont {M.~C.~C.}\ \bibnamefont {Ribeiro}},\ }\href@noop {} {\bibfield  {journal} {\bibinfo  {journal} {J. Chem. Phys.}\ }\textbf {\bibinfo {volume} {148}},\ \bibinfo {pages} {171101} (\bibinfo {year} {2018})}\BibitemShut {NoStop}%
\bibitem [{\citenamefont {Johari}(2001)}]{johari2001}%
  \BibitemOpen
  \bibfield  {author} {\bibinfo {author} {\bibfnamefont {G.~P.}\ \bibnamefont {Johari}},\ }\href@noop {} {\bibfield  {journal} {\bibinfo  {journal} {Phys. Chem. Chem. Phys.}\ }\textbf {\bibinfo {volume} {3}},\ \bibinfo {pages} {2483} (\bibinfo {year} {2001})}\BibitemShut {NoStop}%
\bibitem [{\citenamefont {Imaoka}\ and\ \citenamefont {Kinugawa}(2017)}]{imaoka2017}%
  \BibitemOpen
  \bibfield  {author} {\bibinfo {author} {\bibfnamefont {H.}~\bibnamefont {Imaoka}}\ and\ \bibinfo {author} {\bibfnamefont {K.}~\bibnamefont {Kinugawa}},\ }\href@noop {} {\bibfield  {journal} {\bibinfo  {journal} {Chem. Phys. Lett.}\ }\textbf {\bibinfo {volume} {671}},\ \bibinfo {pages} {174} (\bibinfo {year} {2017})}\BibitemShut {NoStop}%
\bibitem [{\citenamefont {Markham}\ \emph {et~al.}(2020)\citenamefont {Markham}, \citenamefont {Ji},\ and\ \citenamefont {Held}}]{markham2020}%
  \BibitemOpen
  \bibfield  {author} {\bibinfo {author} {\bibfnamefont {T.}~\bibnamefont {Markham}}, \bibinfo {author} {\bibfnamefont {J.-Y.}\ \bibnamefont {Ji}},\ and\ \bibinfo {author} {\bibfnamefont {E.~D.}\ \bibnamefont {Held}},\ }\href@noop {} {\bibfield  {journal} {\bibinfo  {journal} {J. Stat. Mech.}\ }\textbf {\bibinfo {volume} {2020}},\ \bibinfo {pages} {103103} (\bibinfo {year} {2020})}\BibitemShut {NoStop}%
\bibitem [{\citenamefont {Glaum}\ \emph {et~al.}(2007)\citenamefont {Glaum}, \citenamefont {Kleinert},\ and\ \citenamefont {Pelster}}]{glaum2007}%
  \BibitemOpen
  \bibfield  {author} {\bibinfo {author} {\bibfnamefont {K.}~\bibnamefont {Glaum}}, \bibinfo {author} {\bibfnamefont {H.}~\bibnamefont {Kleinert}},\ and\ \bibinfo {author} {\bibfnamefont {A.}~\bibnamefont {Pelster}},\ }\href@noop {} {\bibfield  {journal} {\bibinfo  {journal} {Phys. Rev. A}\ }\textbf {\bibinfo {volume} {76}},\ \bibinfo {pages} {063604} (\bibinfo {year} {2007})}\BibitemShut {NoStop}%
\bibitem [{\citenamefont {Yamamoto}\ \emph {et~al.}(2008{\natexlab{a}})\citenamefont {Yamamoto}, \citenamefont {Shibayama},\ and\ \citenamefont {Shirahama}}]{yamamoto2008}%
  \BibitemOpen
  \bibfield  {author} {\bibinfo {author} {\bibfnamefont {K.}~\bibnamefont {Yamamoto}}, \bibinfo {author} {\bibfnamefont {Y.}~\bibnamefont {Shibayama}},\ and\ \bibinfo {author} {\bibfnamefont {K.}~\bibnamefont {Shirahama}},\ }\href@noop {} {\bibfield  {journal} {\bibinfo  {journal} {Phys. Rev. Lett.}\ }\textbf {\bibinfo {volume} {100}},\ \bibinfo {pages} {195301} (\bibinfo {year} {2008}{\natexlab{a}})}\BibitemShut {NoStop}%
\bibitem [{\citenamefont {Tani}\ \emph {et~al.}(2022)\citenamefont {Tani}, \citenamefont {Nago}, \citenamefont {Murakawa},\ and\ \citenamefont {Shirahama}}]{tani2022}%
  \BibitemOpen
  \bibfield  {author} {\bibinfo {author} {\bibfnamefont {T.}~\bibnamefont {Tani}}, \bibinfo {author} {\bibfnamefont {Y.}~\bibnamefont {Nago}}, \bibinfo {author} {\bibfnamefont {S.}~\bibnamefont {Murakawa}},\ and\ \bibinfo {author} {\bibfnamefont {K.}~\bibnamefont {Shirahama}},\ }\href@noop {} {\bibfield  {journal} {\bibinfo  {journal} {J. Phys. Soc. Jpn.}\ }\textbf {\bibinfo {volume} {91}},\ \bibinfo {pages} {014603} (\bibinfo {year} {2022})}\BibitemShut {NoStop}%
\bibitem [{\citenamefont {Gasparini}\ \emph {et~al.}(2008)\citenamefont {Gasparini}, \citenamefont {Kimball}, \citenamefont {Mooney},\ and\ \citenamefont {Diaz-Avila}}]{gasparini2008}%
  \BibitemOpen
  \bibfield  {author} {\bibinfo {author} {\bibfnamefont {F.~M.}\ \bibnamefont {Gasparini}}, \bibinfo {author} {\bibfnamefont {M.~O.}\ \bibnamefont {Kimball}}, \bibinfo {author} {\bibfnamefont {K.~P.}\ \bibnamefont {Mooney}},\ and\ \bibinfo {author} {\bibfnamefont {M.}~\bibnamefont {Diaz-Avila}},\ }\href@noop {} {\bibfield  {journal} {\bibinfo  {journal} {Rev. Mod. Phys.}\ }\textbf {\bibinfo {volume} {80}},\ \bibinfo {pages} {1009} (\bibinfo {year} {2008})}\BibitemShut {NoStop}%
\bibitem [{Note2()}]{Note2}%
  \BibitemOpen
  \bibinfo {note} {M. Tsujimoto and K. Kinugawa, to be submitted.}\BibitemShut {Stop}%
\bibitem [{\citenamefont {Tanaka}(2006)}]{ftanaka2006}%
  \BibitemOpen
  \bibfield  {author} {\bibinfo {author} {\bibfnamefont {F.}~\bibnamefont {Tanaka}},\ }\href@noop {} {\bibfield  {journal} {\bibinfo  {journal} {Phys. Rev. E}\ }\textbf {\bibinfo {volume} {73}},\ \bibinfo {pages} {061405} (\bibinfo {year} {2006})}\BibitemShut {NoStop}%
\bibitem [{\citenamefont {Nakayama}\ and\ \citenamefont {Makri}(2005)}]{nakayama2005}%
  \BibitemOpen
  \bibfield  {author} {\bibinfo {author} {\bibfnamefont {A.}~\bibnamefont {Nakayama}}\ and\ \bibinfo {author} {\bibfnamefont {N.}~\bibnamefont {Makri}},\ }\href@noop {} {\bibfield  {journal} {\bibinfo  {journal} {Proc. Natl. Acad. Sci. U. S. A.}\ }\textbf {\bibinfo {volume} {102}},\ \bibinfo {pages} {4230} (\bibinfo {year} {2005})}\BibitemShut {NoStop}%
\bibitem [{\citenamefont {Boninsegni}\ \emph {et~al.}(2006)\citenamefont {Boninsegni}, \citenamefont {Prokof'ev},\ and\ \citenamefont {Svistunov}}]{boninsegni2006}%
  \BibitemOpen
  \bibfield  {author} {\bibinfo {author} {\bibfnamefont {M.}~\bibnamefont {Boninsegni}}, \bibinfo {author} {\bibfnamefont {N.}~\bibnamefont {Prokof'ev}},\ and\ \bibinfo {author} {\bibfnamefont {B.}~\bibnamefont {Svistunov}},\ }\href@noop {} {\bibfield  {journal} {\bibinfo  {journal} {Phys. Rev. Lett.}\ }\textbf {\bibinfo {volume} {96}},\ \bibinfo {pages} {105301} (\bibinfo {year} {2006})}\BibitemShut {NoStop}%
\bibitem [{\citenamefont {Rastogi}\ \emph {et~al.}(1991)\citenamefont {Rastogi}, \citenamefont {Newman},\ and\ \citenamefont {Keller}}]{rastogi1991}%
  \BibitemOpen
  \bibfield  {author} {\bibinfo {author} {\bibfnamefont {S.}~\bibnamefont {Rastogi}}, \bibinfo {author} {\bibfnamefont {M.}~\bibnamefont {Newman}},\ and\ \bibinfo {author} {\bibfnamefont {A.}~\bibnamefont {Keller}},\ }\href@noop {} {\bibfield  {journal} {\bibinfo  {journal} {Nature}\ }\textbf {\bibinfo {volume} {353}},\ \bibinfo {pages} {55} (\bibinfo {year} {1991})}\BibitemShut {NoStop}%
\bibitem [{\citenamefont {Mortensen}\ \emph {et~al.}(1992)\citenamefont {Mortensen}, \citenamefont {Brown},\ and\ \citenamefont {Norden}}]{mortensen1992}%
  \BibitemOpen
  \bibfield  {author} {\bibinfo {author} {\bibfnamefont {K.}~\bibnamefont {Mortensen}}, \bibinfo {author} {\bibfnamefont {W.}~\bibnamefont {Brown}},\ and\ \bibinfo {author} {\bibfnamefont {B.}~\bibnamefont {Norden}},\ }\href@noop {} {\bibfield  {journal} {\bibinfo  {journal} {Phys. Rev. Lett.}\ }\textbf {\bibinfo {volume} {68}},\ \bibinfo {pages} {2340} (\bibinfo {year} {1992})}\BibitemShut {NoStop}%
\bibitem [{\citenamefont {Shirahama}\ \emph {et~al.}(2008{\natexlab{a}})\citenamefont {Shirahama}, \citenamefont {Yamamoto},\ and\ \citenamefont {Shibayama}}]{shirahama2008ltp}%
  \BibitemOpen
  \bibfield  {author} {\bibinfo {author} {\bibfnamefont {K.}~\bibnamefont {Shirahama}}, \bibinfo {author} {\bibfnamefont {K.}~\bibnamefont {Yamamoto}},\ and\ \bibinfo {author} {\bibfnamefont {Y.}~\bibnamefont {Shibayama}},\ }\href@noop {} {\bibfield  {journal} {\bibinfo  {journal} {Low Temp. Phys.}\ }\textbf {\bibinfo {volume} {34}},\ \bibinfo {pages} {273} (\bibinfo {year} {2008}{\natexlab{a}})}\BibitemShut {NoStop}%
\bibitem [{\citenamefont {Taniguchi}\ \emph {et~al.}(2011)\citenamefont {Taniguchi}, \citenamefont {Fujii},\ and\ \citenamefont {Suzuki}}]{taniguchi2011}%
  \BibitemOpen
  \bibfield  {author} {\bibinfo {author} {\bibfnamefont {J.}~\bibnamefont {Taniguchi}}, \bibinfo {author} {\bibfnamefont {R.}~\bibnamefont {Fujii}},\ and\ \bibinfo {author} {\bibfnamefont {M.}~\bibnamefont {Suzuki}},\ }\href@noop {} {\bibfield  {journal} {\bibinfo  {journal} {Phys. Rev. B}\ }\textbf {\bibinfo {volume} {84}},\ \bibinfo {pages} {134511} (\bibinfo {year} {2011})}\BibitemShut {NoStop}%
\bibitem [{\citenamefont {Bossy}\ \emph {et~al.}(2012)\citenamefont {Bossy}, \citenamefont {Ollivier}, \citenamefont {Schober},\ and\ \citenamefont {Glyde}}]{bossy2012}%
  \BibitemOpen
  \bibfield  {author} {\bibinfo {author} {\bibfnamefont {J.}~\bibnamefont {Bossy}}, \bibinfo {author} {\bibfnamefont {J.}~\bibnamefont {Ollivier}}, \bibinfo {author} {\bibfnamefont {H.}~\bibnamefont {Schober}},\ and\ \bibinfo {author} {\bibfnamefont {H.~R.}\ \bibnamefont {Glyde}},\ }\href@noop {} {\bibfield  {journal} {\bibinfo  {journal} {Phys. Rev. B}\ }\textbf {\bibinfo {volume} {86}},\ \bibinfo {pages} {224503} (\bibinfo {year} {2012})}\BibitemShut {NoStop}%
\bibitem [{\citenamefont {Yamamoto}\ \emph {et~al.}(2004)\citenamefont {Yamamoto}, \citenamefont {Nakashima}, \citenamefont {Shibayama},\ and\ \citenamefont {Shirahama}}]{yamamoto2004}%
  \BibitemOpen
  \bibfield  {author} {\bibinfo {author} {\bibfnamefont {K.}~\bibnamefont {Yamamoto}}, \bibinfo {author} {\bibfnamefont {H.}~\bibnamefont {Nakashima}}, \bibinfo {author} {\bibfnamefont {Y.}~\bibnamefont {Shibayama}},\ and\ \bibinfo {author} {\bibfnamefont {K.}~\bibnamefont {Shirahama}},\ }\href@noop {} {\bibfield  {journal} {\bibinfo  {journal} {Phys. Rev. Lett.}\ }\textbf {\bibinfo {volume} {93}},\ \bibinfo {pages} {075302} (\bibinfo {year} {2004})}\BibitemShut {NoStop}%
\bibitem [{\citenamefont {Yamamoto}\ \emph {et~al.}(2008{\natexlab{b}})\citenamefont {Yamamoto}, \citenamefont {Shibayama},\ and\ \citenamefont {Shirahama}}]{yamamoto2008jpsj}%
  \BibitemOpen
  \bibfield  {author} {\bibinfo {author} {\bibfnamefont {K.}~\bibnamefont {Yamamoto}}, \bibinfo {author} {\bibfnamefont {Y.}~\bibnamefont {Shibayama}},\ and\ \bibinfo {author} {\bibfnamefont {K.}~\bibnamefont {Shirahama}},\ }\href@noop {} {\bibfield  {journal} {\bibinfo  {journal} {J. Phys. Soc. Jpn.}\ }\textbf {\bibinfo {volume} {77}},\ \bibinfo {pages} {013601} (\bibinfo {year} {2008}{\natexlab{b}})}\BibitemShut {NoStop}%
\bibitem [{\citenamefont {Shirahama}\ \emph {et~al.}(2008{\natexlab{b}})\citenamefont {Shirahama}, \citenamefont {Yamamoto},\ and\ \citenamefont {Shibayama}}]{shirahama2008}%
  \BibitemOpen
  \bibfield  {author} {\bibinfo {author} {\bibfnamefont {K.}~\bibnamefont {Shirahama}}, \bibinfo {author} {\bibfnamefont {K.}~\bibnamefont {Yamamoto}},\ and\ \bibinfo {author} {\bibfnamefont {Y.}~\bibnamefont {Shibayama}},\ }\href@noop {} {\bibfield  {journal} {\bibinfo  {journal} {J. Phys. Soc. Jpn.}\ }\textbf {\bibinfo {volume} {77}},\ \bibinfo {pages} {111011} (\bibinfo {year} {2008}{\natexlab{b}})}\BibitemShut {NoStop}%
\bibitem [{\citenamefont {Bossy}\ \emph {et~al.}(2008)\citenamefont {Bossy}, \citenamefont {Pearce}, \citenamefont {Schober},\ and\ \citenamefont {Glyde}}]{bossy2008}%
  \BibitemOpen
  \bibfield  {author} {\bibinfo {author} {\bibfnamefont {J.}~\bibnamefont {Bossy}}, \bibinfo {author} {\bibfnamefont {J.~V.}\ \bibnamefont {Pearce}}, \bibinfo {author} {\bibfnamefont {H.}~\bibnamefont {Schober}},\ and\ \bibinfo {author} {\bibfnamefont {H.~R.}\ \bibnamefont {Glyde}},\ }\href@noop {} {\bibfield  {journal} {\bibinfo  {journal} {Phys. Rev. Lett}\ }\textbf {\bibinfo {volume} {101}},\ \bibinfo {pages} {025301} (\bibinfo {year} {2008})}\BibitemShut {NoStop}%
\bibitem [{\citenamefont {Taniguchi}\ \emph {et~al.}(2010)\citenamefont {Taniguchi}, \citenamefont {Aoki},\ and\ \citenamefont {Suzuki}}]{taniguchi2010}%
  \BibitemOpen
  \bibfield  {author} {\bibinfo {author} {\bibfnamefont {J.}~\bibnamefont {Taniguchi}}, \bibinfo {author} {\bibfnamefont {Y.}~\bibnamefont {Aoki}},\ and\ \bibinfo {author} {\bibfnamefont {M.}~\bibnamefont {Suzuki}},\ }\href@noop {} {\bibfield  {journal} {\bibinfo  {journal} {Phys. Rev. B}\ }\textbf {\bibinfo {volume} {82}},\ \bibinfo {pages} {104509} (\bibinfo {year} {2010})}\BibitemShut {NoStop}%
\bibitem [{\citenamefont {Taniguchi}\ \emph {et~al.}(2013)\citenamefont {Taniguchi}, \citenamefont {Demura},\ and\ \citenamefont {Suzuki}}]{taniguchi2013}%
  \BibitemOpen
  \bibfield  {author} {\bibinfo {author} {\bibfnamefont {J.}~\bibnamefont {Taniguchi}}, \bibinfo {author} {\bibfnamefont {K.}~\bibnamefont {Demura}},\ and\ \bibinfo {author} {\bibfnamefont {M.}~\bibnamefont {Suzuki}},\ }\href@noop {} {\bibfield  {journal} {\bibinfo  {journal} {Phys. Rev. B}\ }\textbf {\bibinfo {volume} {88}},\ \bibinfo {pages} {014502} (\bibinfo {year} {2013})}\BibitemShut {NoStop}%
\bibitem [{\citenamefont {Bossy}\ \emph {et~al.}(2019)\citenamefont {Bossy}, \citenamefont {Ollivier},\ and\ \citenamefont {Glyde}}]{bossy2019}%
  \BibitemOpen
  \bibfield  {author} {\bibinfo {author} {\bibfnamefont {J.}~\bibnamefont {Bossy}}, \bibinfo {author} {\bibfnamefont {J.}~\bibnamefont {Ollivier}},\ and\ \bibinfo {author} {\bibfnamefont {H.~R.}\ \bibnamefont {Glyde}},\ }\href@noop {} {\bibfield  {journal} {\bibinfo  {journal} {Phys. Rev. B}\ }\textbf {\bibinfo {volume} {99}},\ \bibinfo {pages} {165425} (\bibinfo {year} {2019})}\BibitemShut {NoStop}%
\bibitem [{\citenamefont {Balibar}\ and\ \citenamefont {Caupin}(2008)}]{balibar2008}%
  \BibitemOpen
  \bibfield  {author} {\bibinfo {author} {\bibfnamefont {S.}~\bibnamefont {Balibar}}\ and\ \bibinfo {author} {\bibfnamefont {F.}~\bibnamefont {Caupin}},\ }\href@noop {} {\bibfield  {journal} {\bibinfo  {journal} {J. Phys. Condens. Matter}\ }\textbf {\bibinfo {volume} {20}},\ \bibinfo {pages} {173201} (\bibinfo {year} {2008})}\BibitemShut {NoStop}%
\end{thebibliography}%

\newpage 
\widetext
\newpage { \center \bf \large  SUPPLEMENTARY MATERIAL
\\of  
\\ Two liquid states of distinguishable helium-4: the
existence of another non-superfluid frozen  by heating
\vspace*{0.1cm}\\  \vspace*{0.0cm} } 

\begin{center} Momoko Tsujimoto and Kenichi Kinugawa \\ 
\vspace*{0.15cm}
\small{\textit{Department of Chemistry, Graduate School of Humanities and Sciences, Nara Women's University, Nara 630-8506, Japan}} \\
\vspace*{0.25cm} 
\end{center}

\setcounter{equation}{0}
\setcounter{section}{0}
\setcounter{figure}{0}
\setcounter{table}{0}
\setcounter{page}{1}
\makeatletter
\renewcommand{\theequation}{S\arabic{equation}}
\renewcommand{\bibnumfmt}[1]{[S#1]}
\renewcommand{\citenumfont}[1]{S#1}

\section*{Part A. Supplementary figures}
\renewcommand{\thefigure}{SA-\arabic{figure}}

\begin{midpage}
\begin{figure}[H]
\centering
\includegraphics[width=16cm]{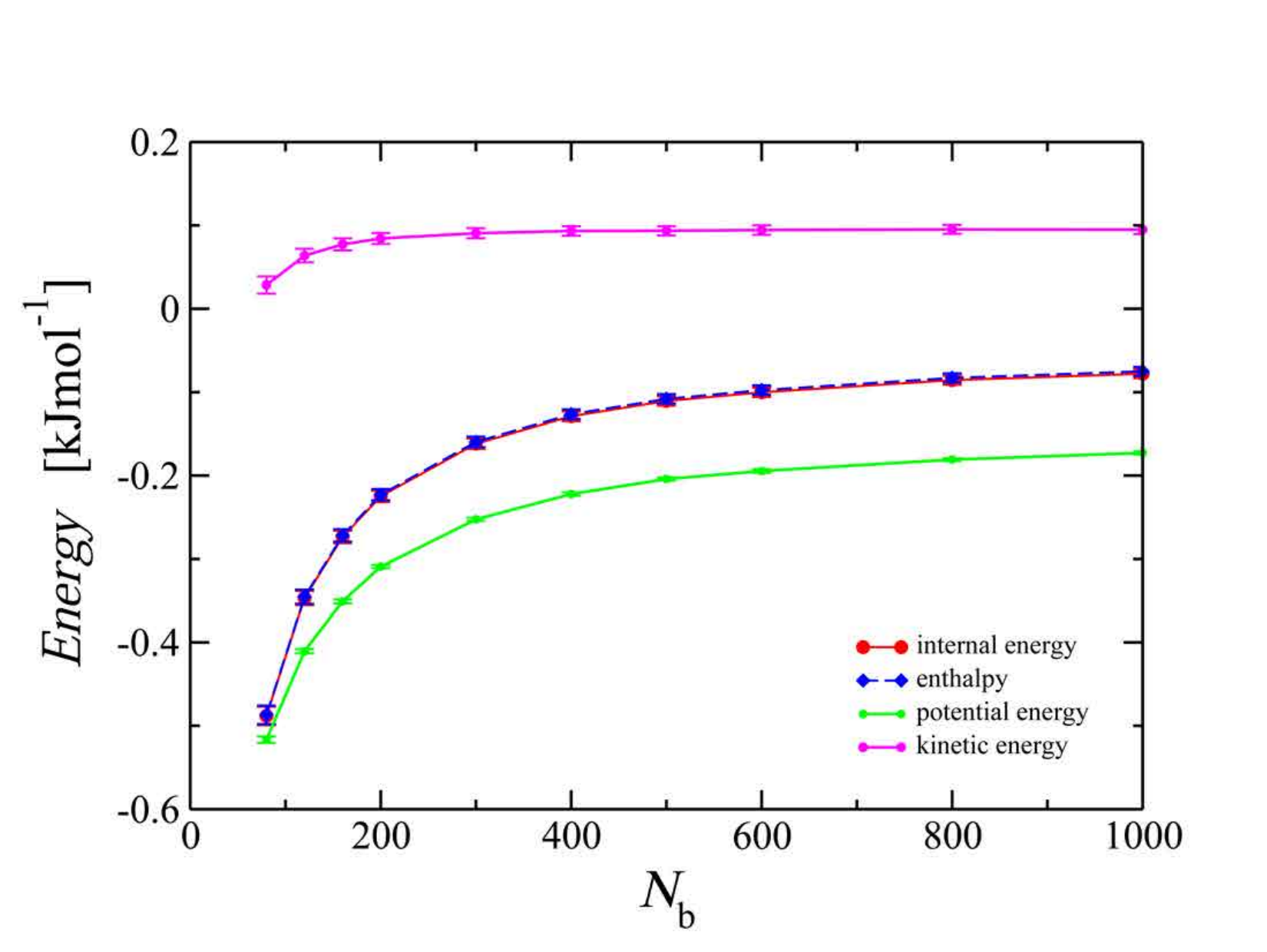}
\caption{\label{fig:FigSA-1} 
The Trotter number dependence of energetic properties.  This test was carried out for the canonical ensemble at 0.1 K and SVP density (0.145 gcm$^{-3}$) spanning 100,000 steps.}
\end{figure}  
\end{midpage}

\pagebreak

\begin{midpage}
\begin{figure}[H]
\centering
\includegraphics[width=18cm]{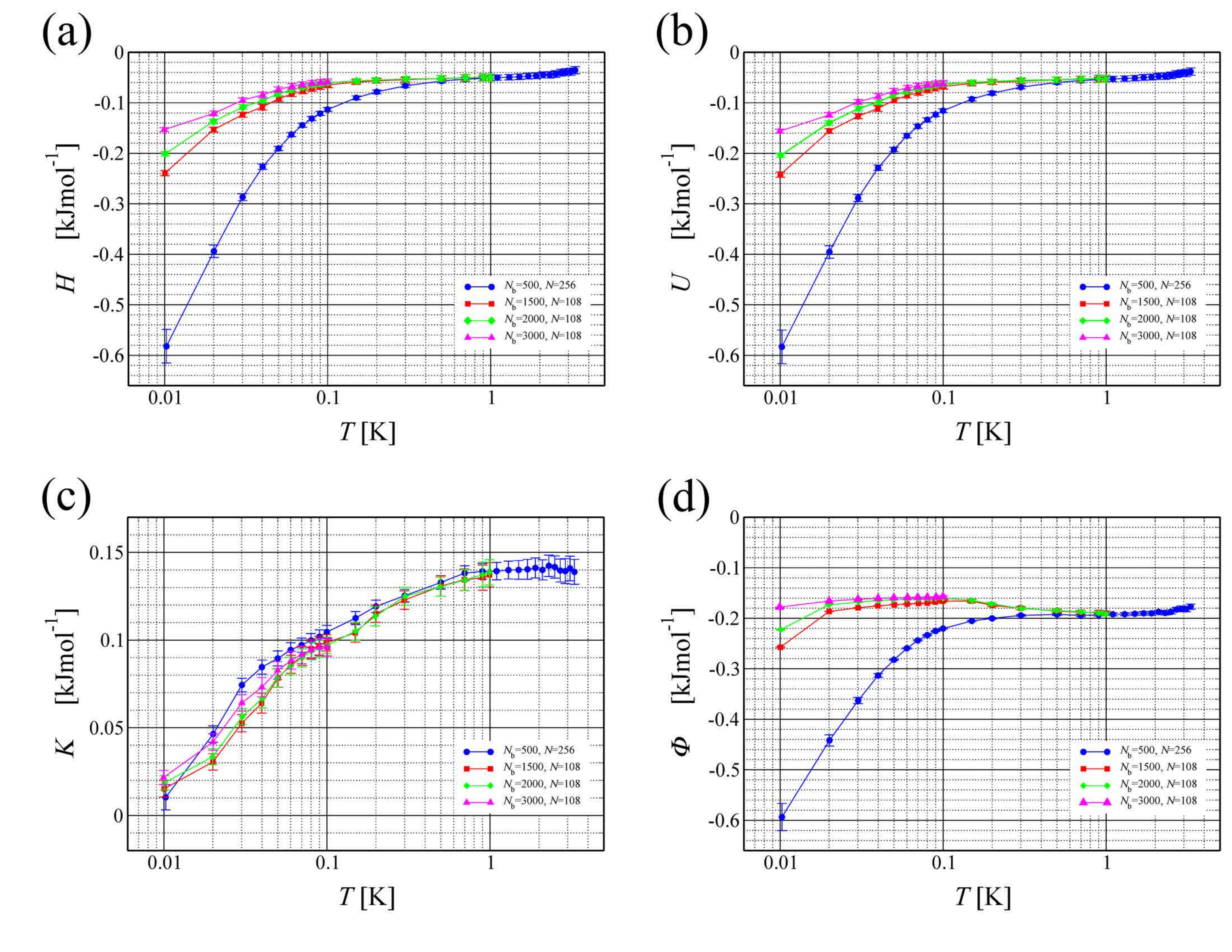}
\caption{\label{fig:FigSA-2} 
 The temperature dependence of the system energetic properties with different Trotter numbers. (a) enthalpy; (b) internal energy; (c) kinetic energy; (d) potential energy; (e) enlarged plot of enthalpy and kinetic energy for 0.01-0.2 K.  All the results were obtained from the isobaric-isothermal CMD simulations at 1 bar and each temperature.  In (e), we can see slight inflection points of the curves for $N_{\rm{b}}=3000$ at about 0.02 K.}
\end{figure}   
\end{midpage}

\pagebreak

\setcounter{figure}{1}

\begin{midpage}
\begin{figure}[H]
\centering
\includegraphics[width=12cm]{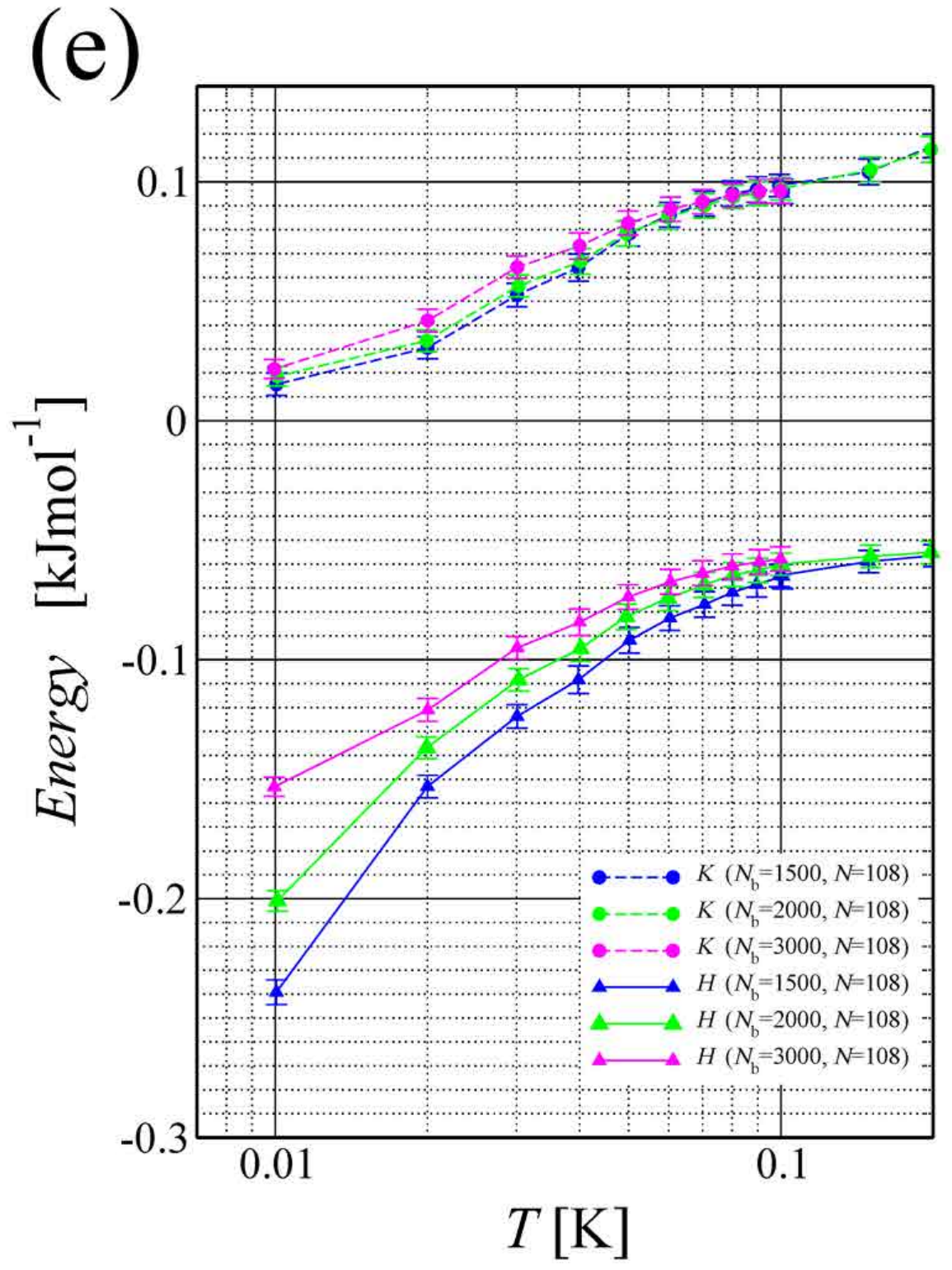}
\caption{\label{fig:FigSA-2-e} 
(continued) The temperature dependence of the system energetic properties with different Trotter numbers.}
\end{figure}   
\end{midpage}

\pagebreak

\begin{midpage}
\begin{figure}[H]
\centering
\includegraphics[width=12cm]{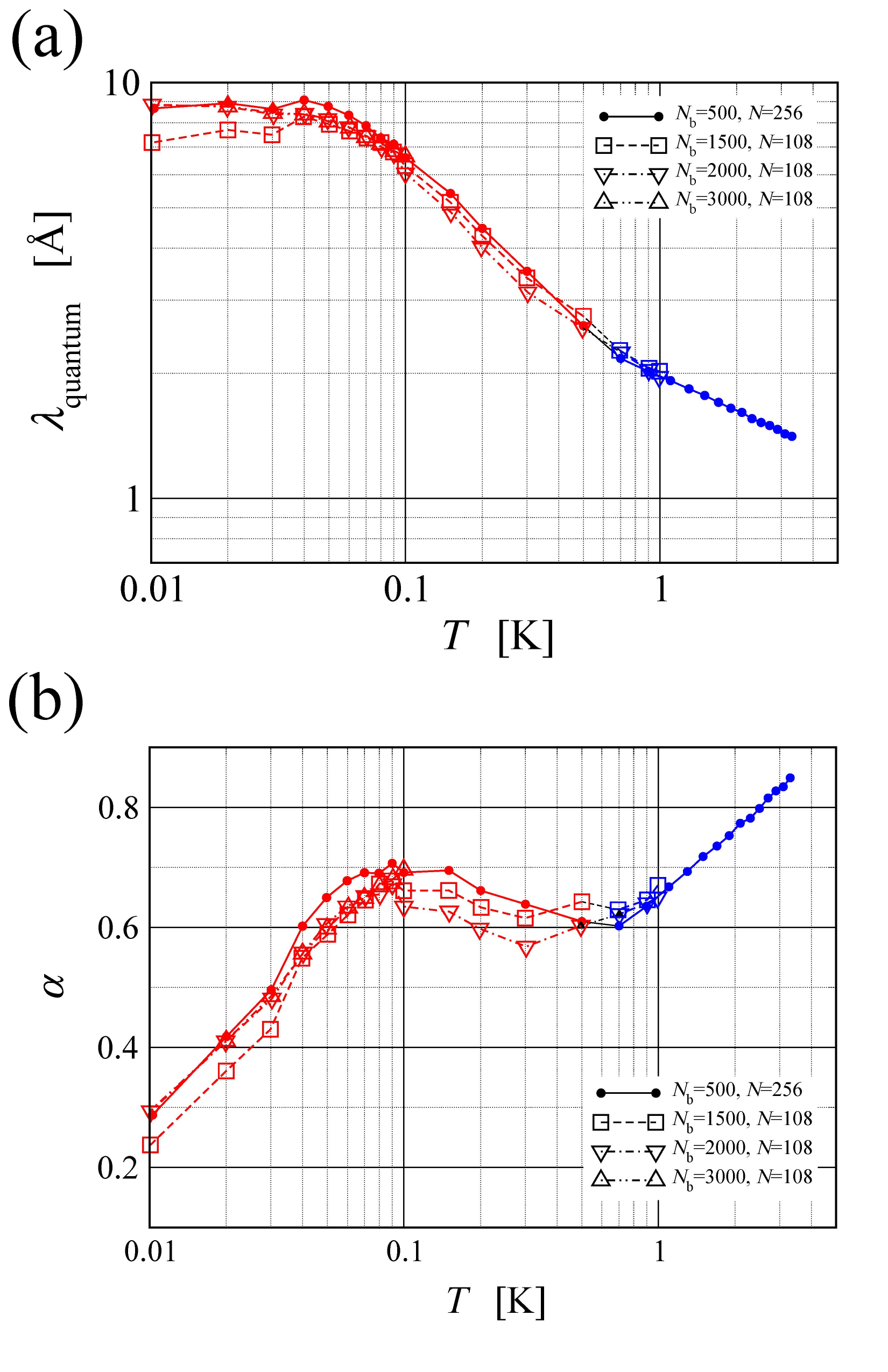}
\caption{\label{fig:FigSA-3} 
The temperature dependence of quantum wavelength and expansion factor with different Trotter numbers at 1 bar.  (a) quantum wavelength; (b) expansion factor.  The color of the symbols is the same as that displayed in FIG. 7 and FIG. 8 in the main text: LQDL (blue) and HQDL (red).  The definition of these properties is given in Sec. III D.  The plotted results of $N_{\rm{b}}=500$ and $N=256$ involve the 1 bar data shown in FIG. 7 and FIG. 8(a) in the main text.
}
\end{figure}     
\end{midpage}

\pagebreak

\begin{midpage}
\begin{figure}[H]
\centering
\includegraphics[width=12cm]{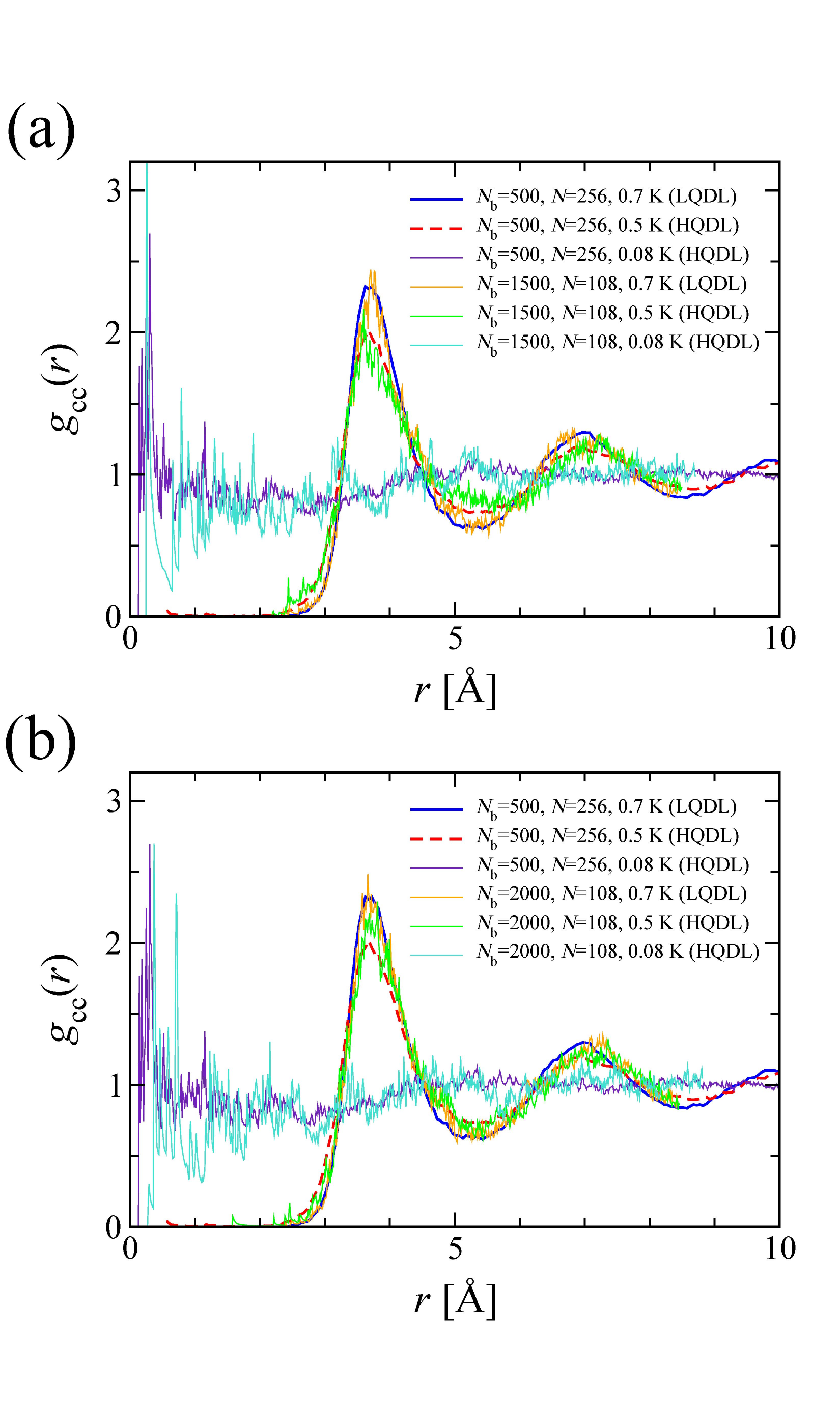}
\caption{\label{fig:FigSA-4} 
The bead-bead radial distribution fucntion with different Trotter numbers at 1 bar.  (a) comparison between $N_{\rm{b}}=500$ and $N_{\rm{b}}=1500$. (b) comparison between $N_{\rm{b}}=500$ and $N_{\rm{b}}=2000$.  The plotted results of $N_{\rm{b}}=500$ and $N=256$ involve the 1 bar data shown in FIG. 6(b) in the main text.  We can see that the transition between the LQDL and HQDL occurs between 0.7 and 0.5 K for all $N_{\rm{b}}$ conditions.
}
\end{figure}     
\end{midpage}

\pagebreak

\begin{midpage}
\begin{figure}[H]
\centering
\includegraphics[width=16cm]{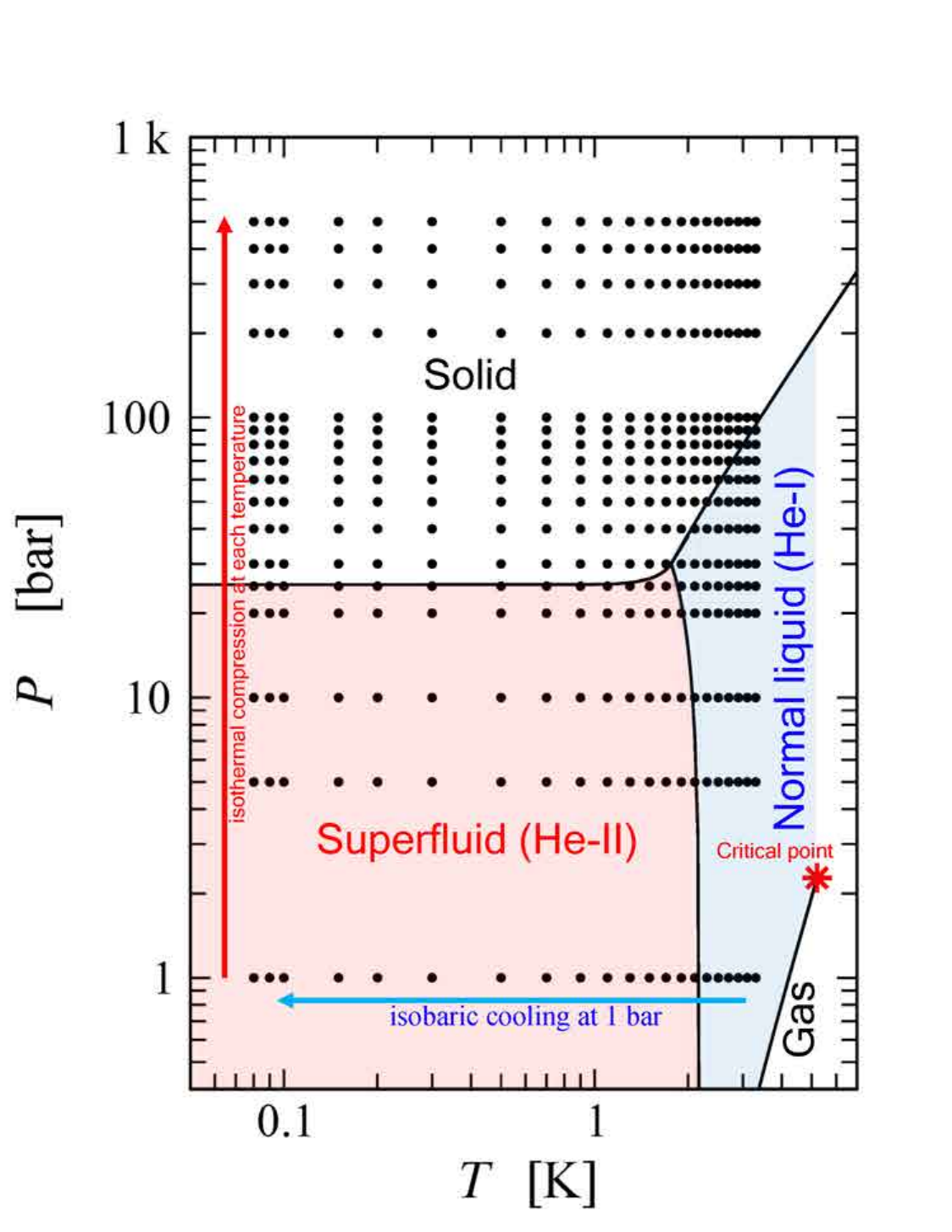}
\caption{\label{fig:FigSA-5} 
The $P$-$T$ conditions set for the CMD simulations.  The black dots denote 357 conditions 
(21 temperatures times 17 pressures).  The shown phases are experimental ones.  Before conducting isothermal 
compression simulations, the isobaric cooling was conducted at 1 bar to generate the equilibrium state at each 
temperature at 1 bar.
}
\end{figure}     
\end{midpage}

\pagebreak

\begin{midpage}
\begin{figure}[H]
\centering
\includegraphics[width=14cm]{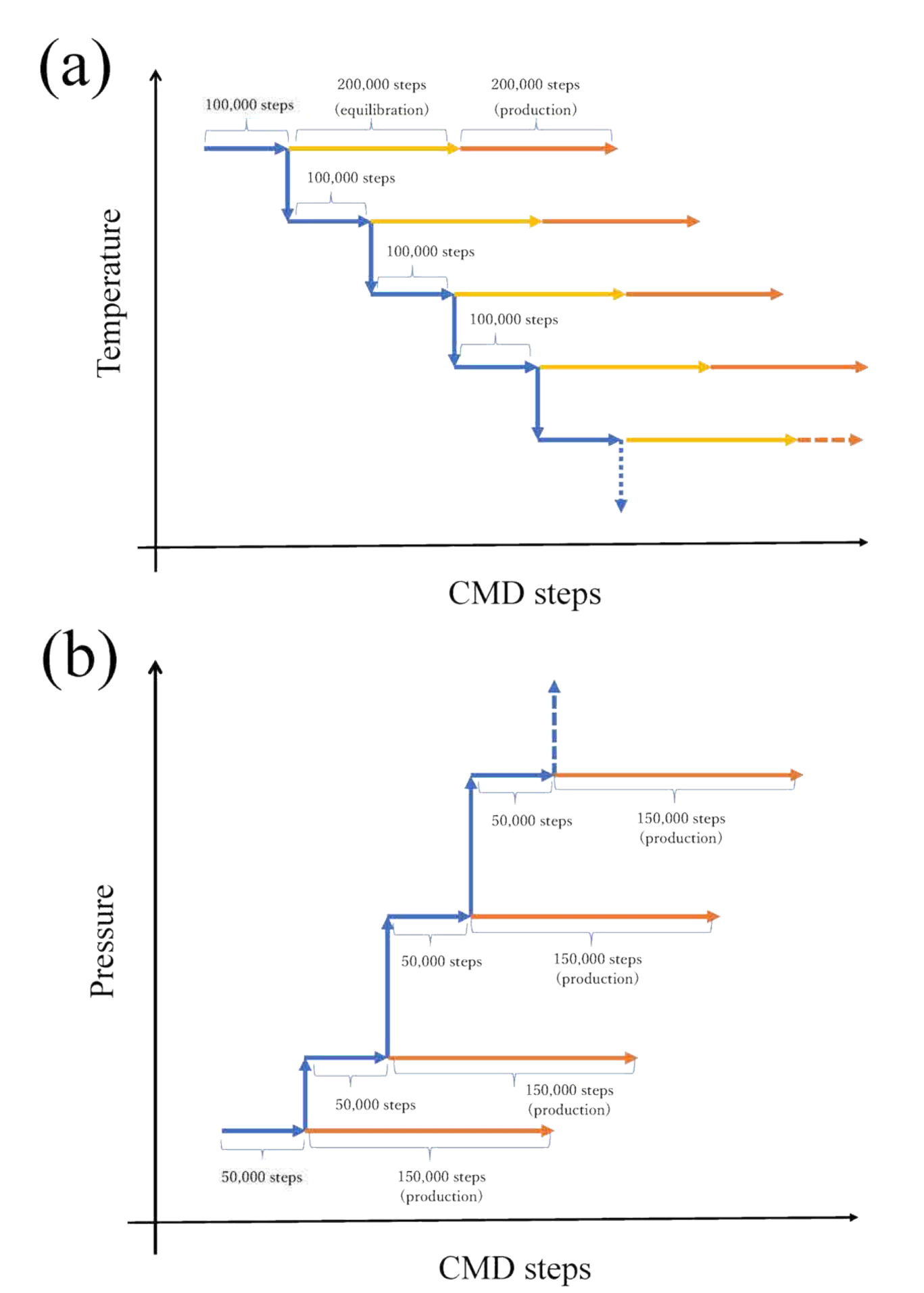}
\caption{\label{fig:FigSA-6} 
Schematic representation of CMD procedure.  (a) Isobaric cooling run at 1 bar conducted prior to starting the isothermal compression; (b) Isothermal compression run started from each temperature point at 1 bar.
}
\end{figure}     
\end{midpage}

\pagebreak

\begin{midpage}
\begin{figure}[H]
\centering
\includegraphics[width=10cm]{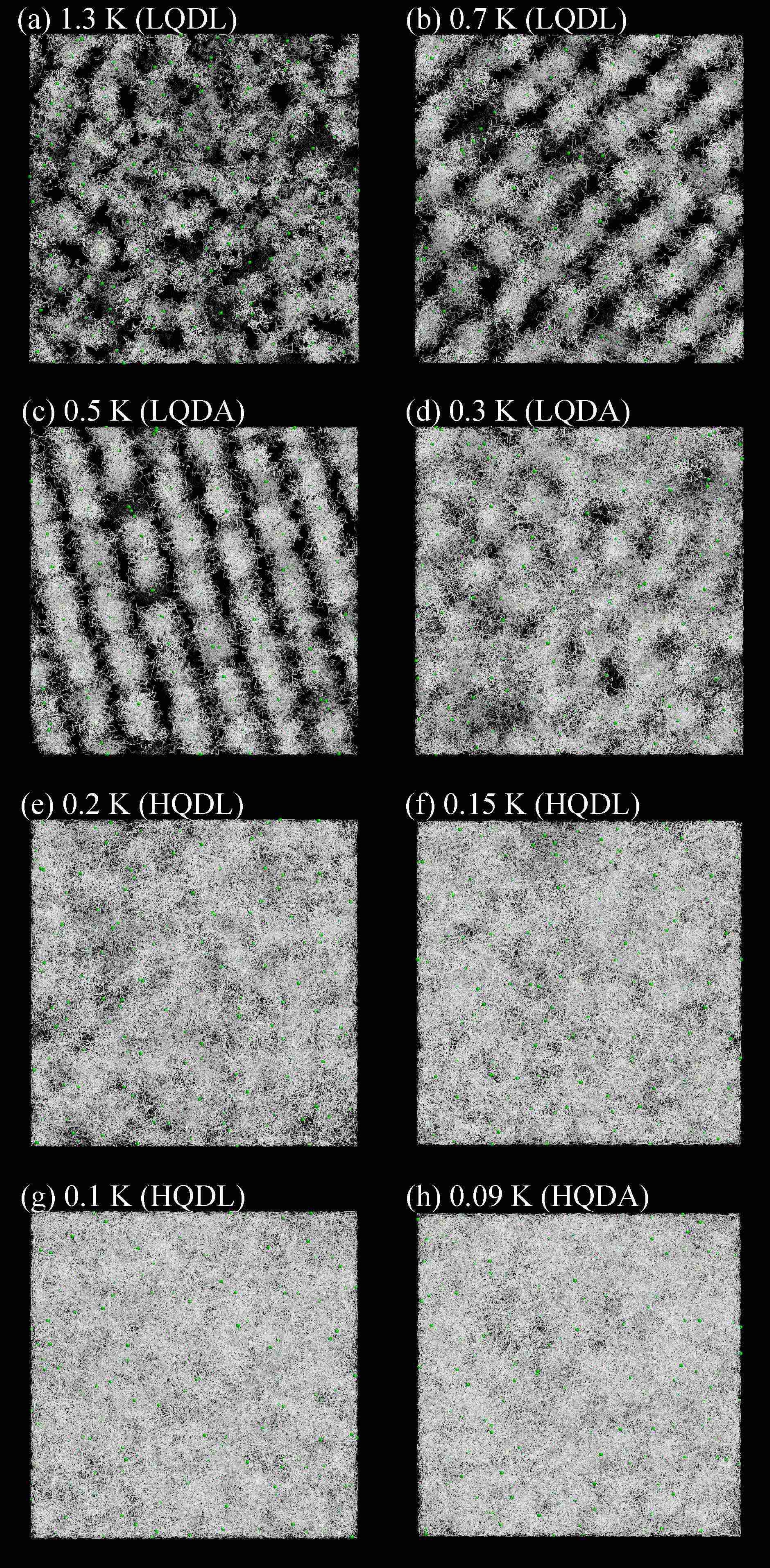}
\caption{\label{fig:FigSA-7} 
The $xy$-projected snapshots of the configurations of the $^4$He atomic necklaces and centroids at 40 bar.  Green spheres and white ones denote the centroids and the beads, respectively.  The drawn scales of the centroids and the beads are arbitrary.
}
\end{figure}     
\end{midpage}  

\pagebreak

\begin{midpage}
\begin{figure}[H]
\centering
\includegraphics[width=16cm]{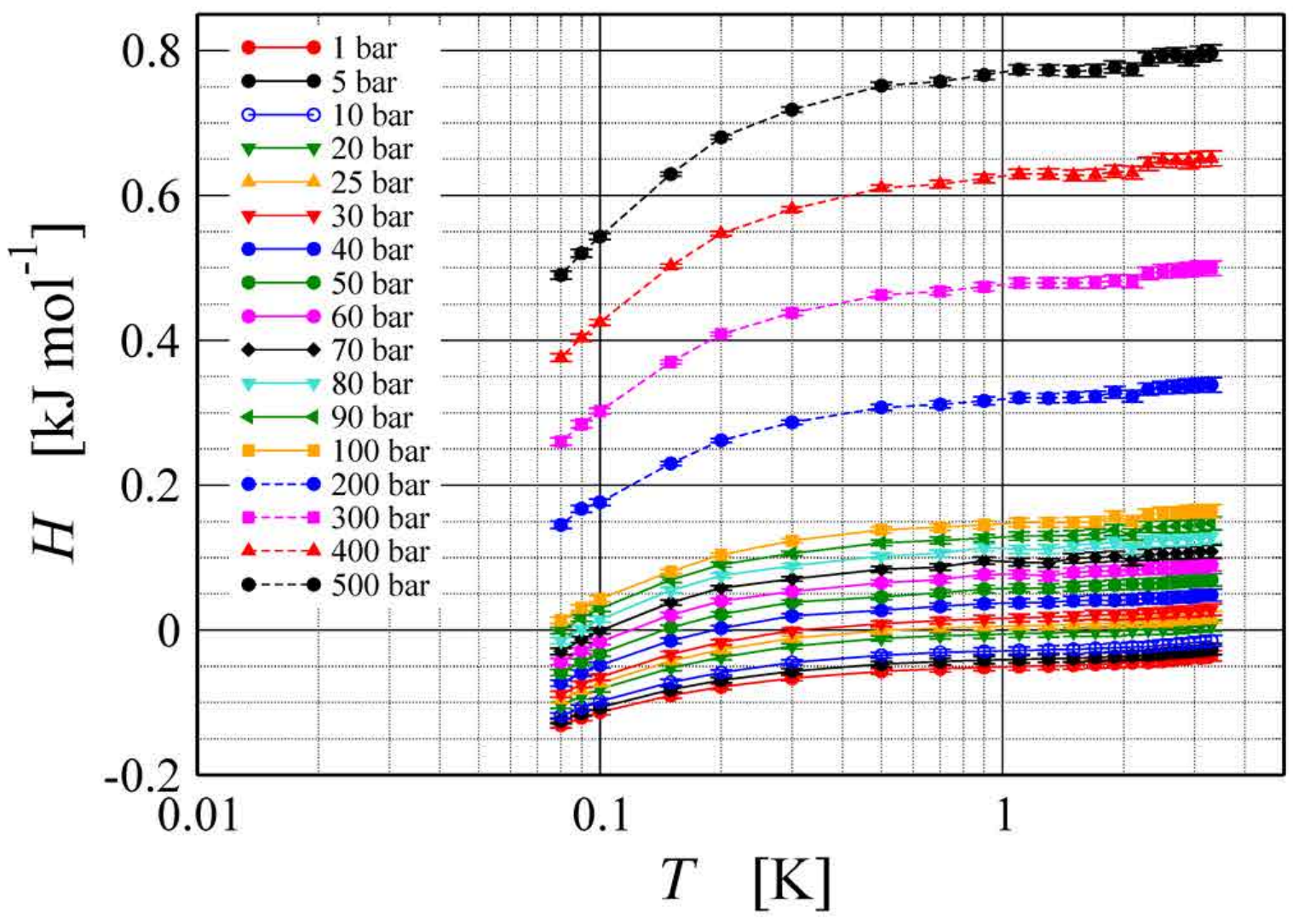}
\caption{\label{fig:FigSA-8} 
The plot of enthalpy versus temperature over the whole pressure range.
}
\end{figure}     
\end{midpage}

\pagebreak

\begin{midpage}
\begin{figure}[H]
\centering
\includegraphics[width=16cm]{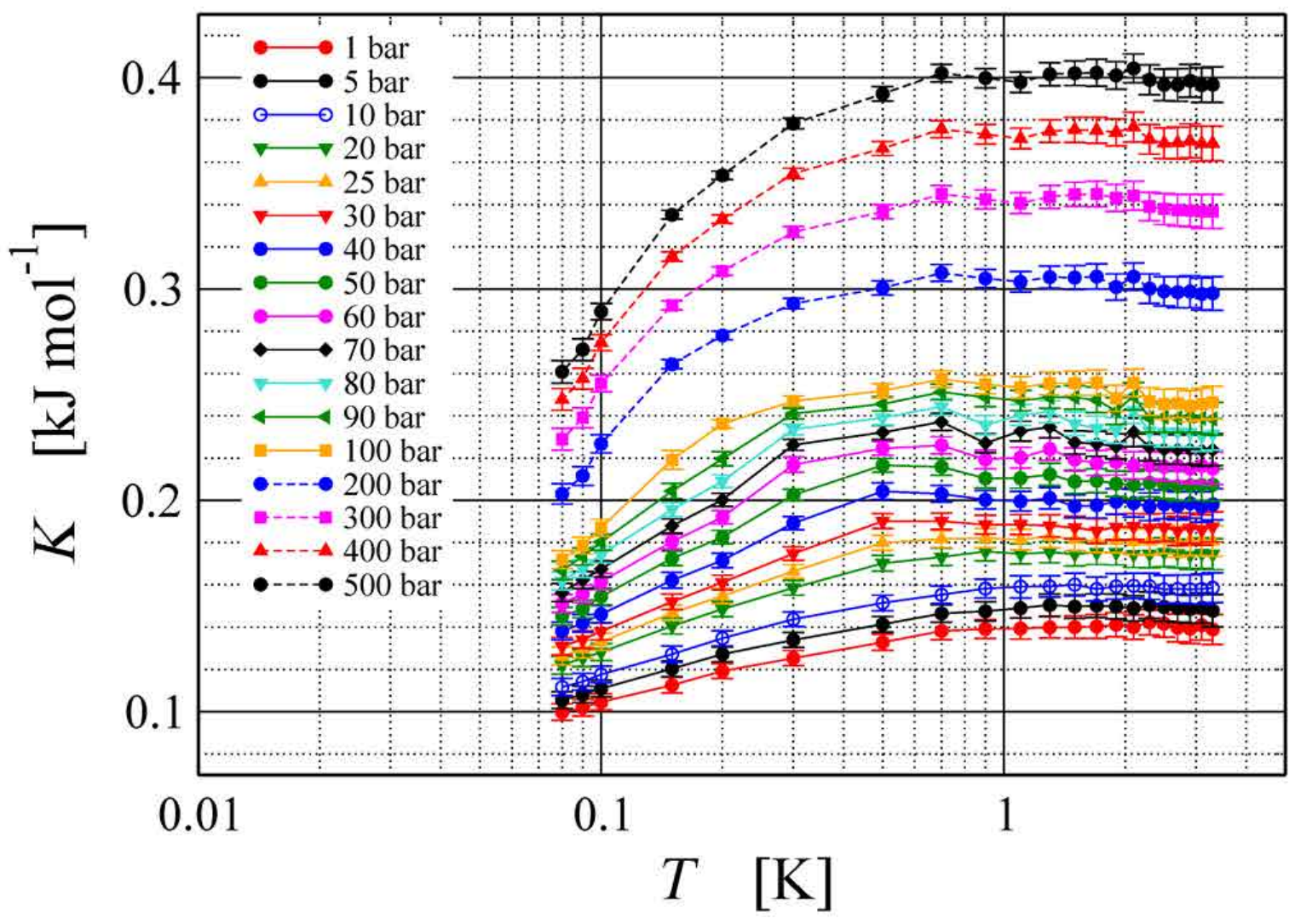}
\caption{\label{fig:FigSA-9} 
The plot of kinetic energy versus temperature.
}
\end{figure}   
\end{midpage}

\pagebreak

\begin{midpage}
\begin{figure}[H]
\centering
\includegraphics[width=16cm]{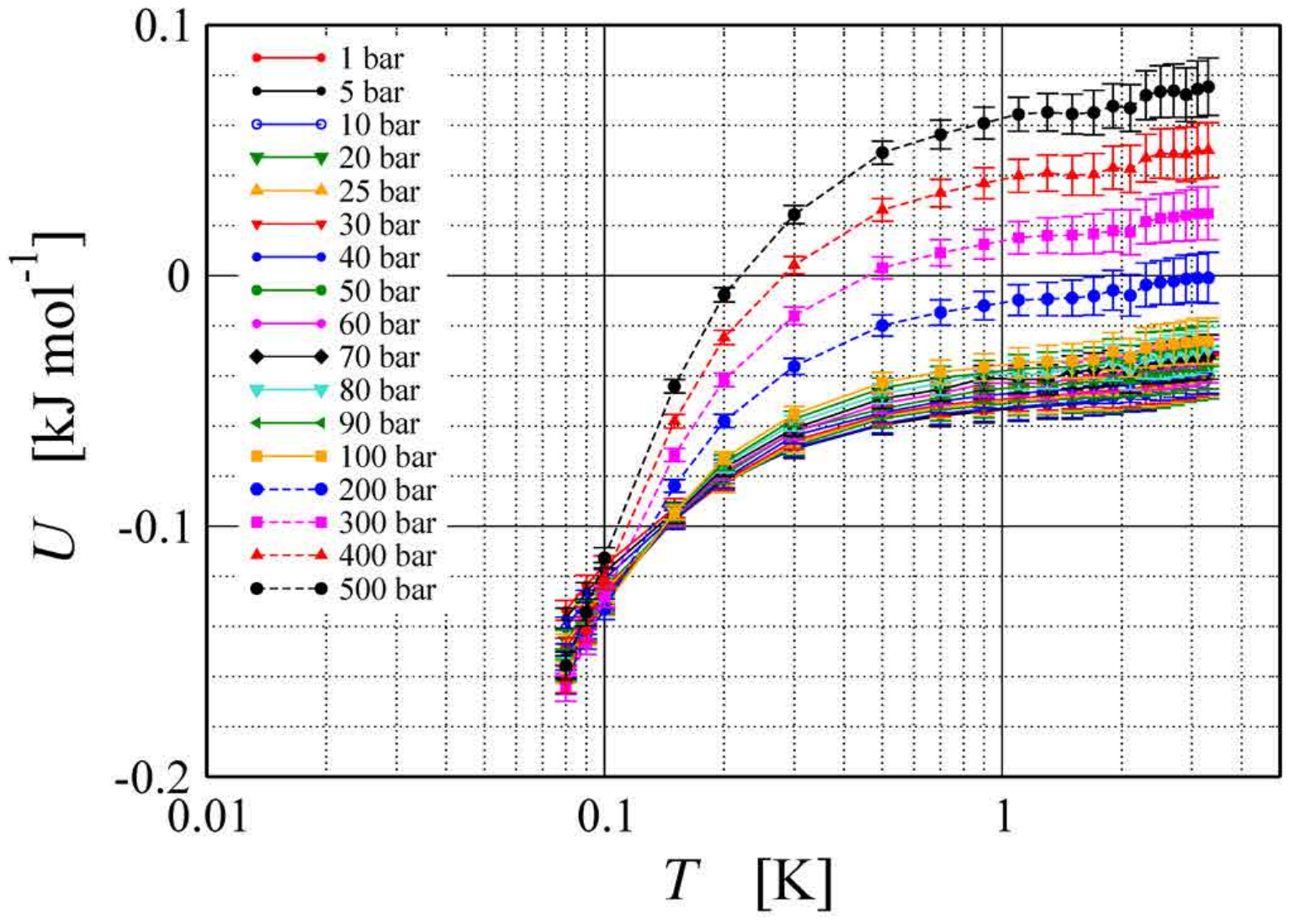}
\caption{\label{fig:FigSA-10} 
The plot of internal energy versus temperature.
}
\end{figure}   
\end{midpage}

\pagebreak

\begin{midpage}
\begin{figure}[H]
\centering
\includegraphics[width=16cm]{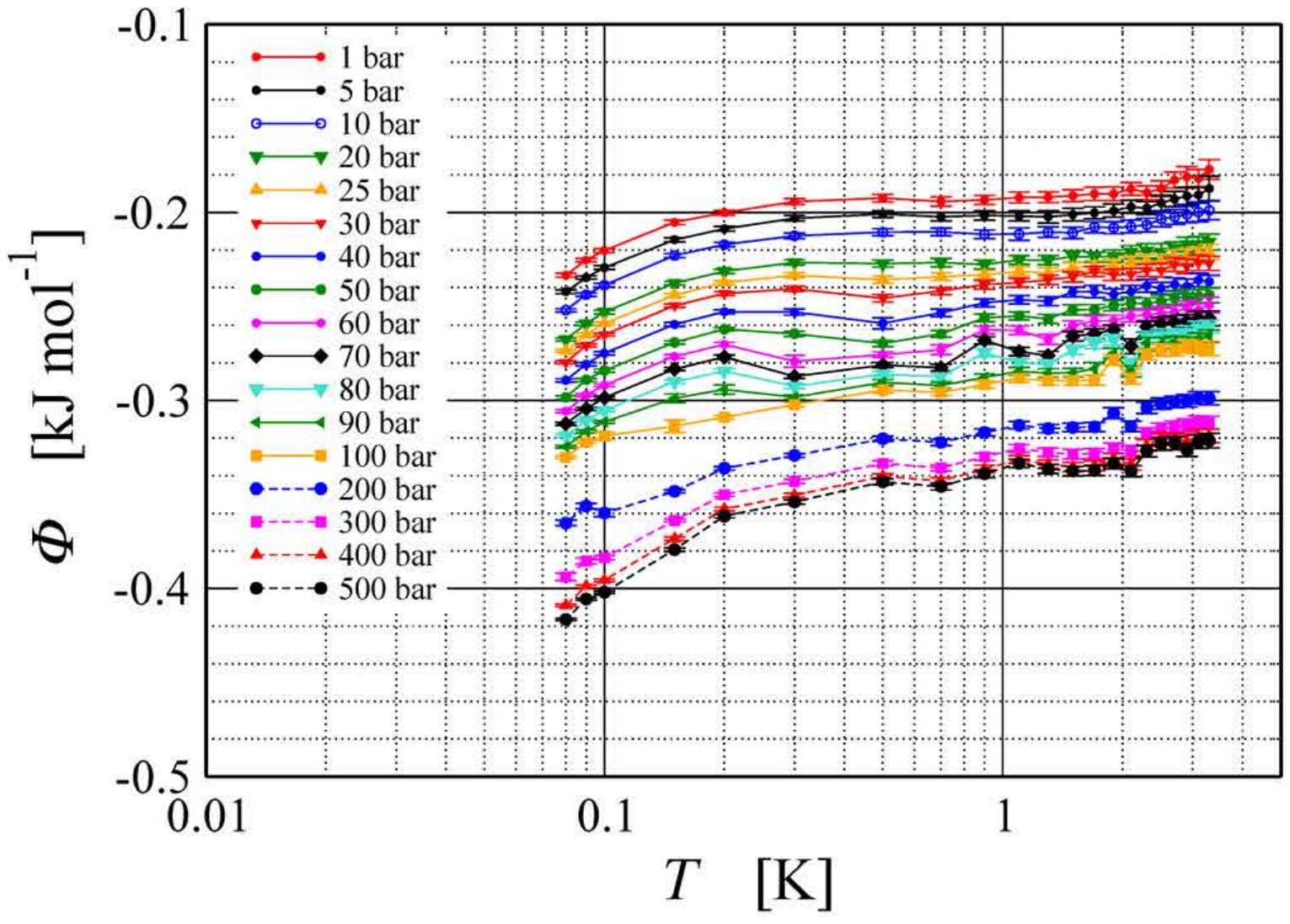}
\caption{\label{fig:FigSA-11} 
The plot of potential energy versus temperature.
}
\end{figure}   
\end{midpage}

\pagebreak

\begin{midpage}
\begin{figure}[H]
\centering
\includegraphics[width=16cm]{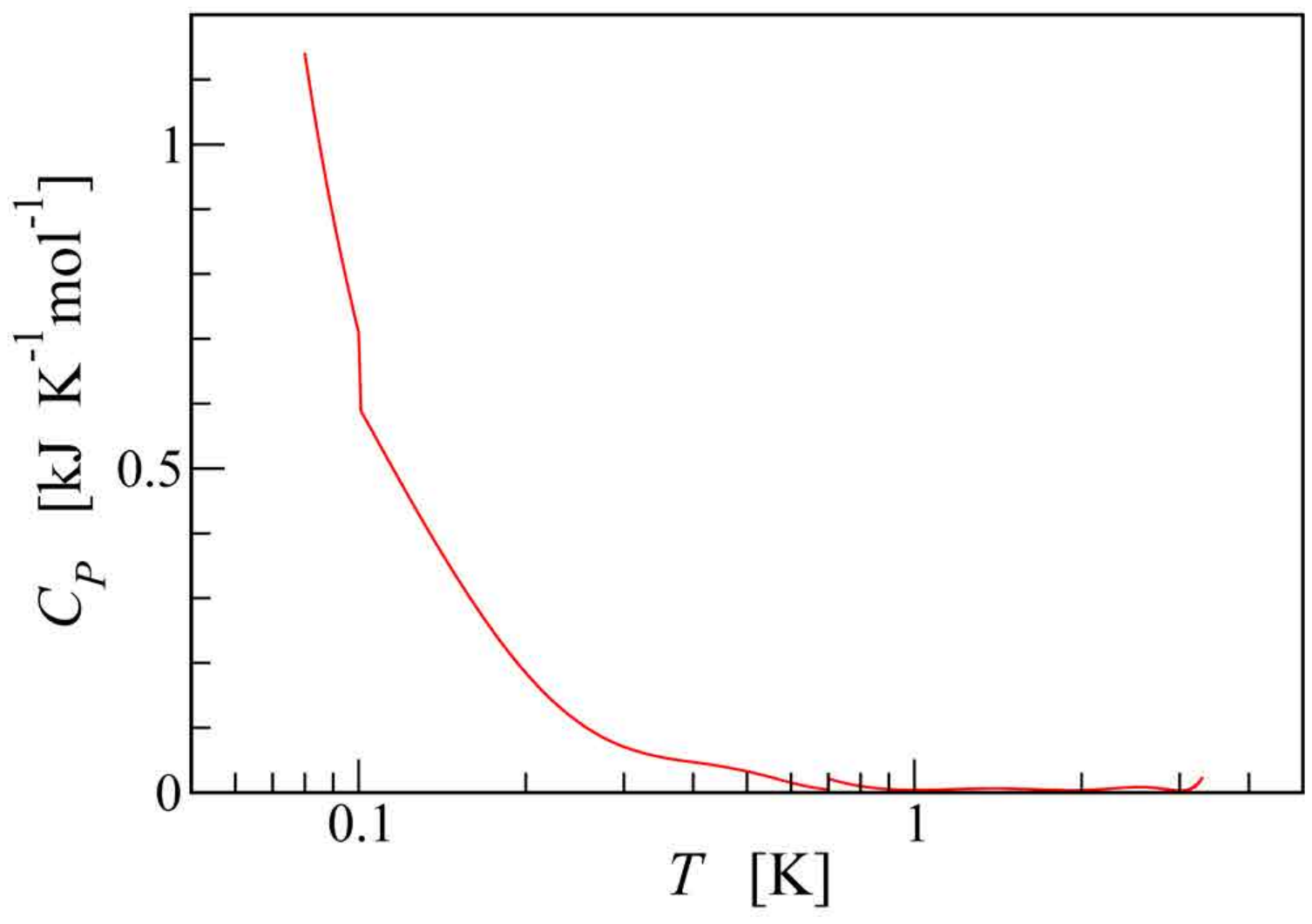}
\caption{\label{fig:FigSA-12} 
The isobaric heat capacity obtained from the polynomial curve fitting of enthalpy-temperature relation at 1 bar.
}
\end{figure}   
\end{midpage}  

\pagebreak

\begin{midpage}
\begin{figure}[H]
\centering
\includegraphics[width=16cm]{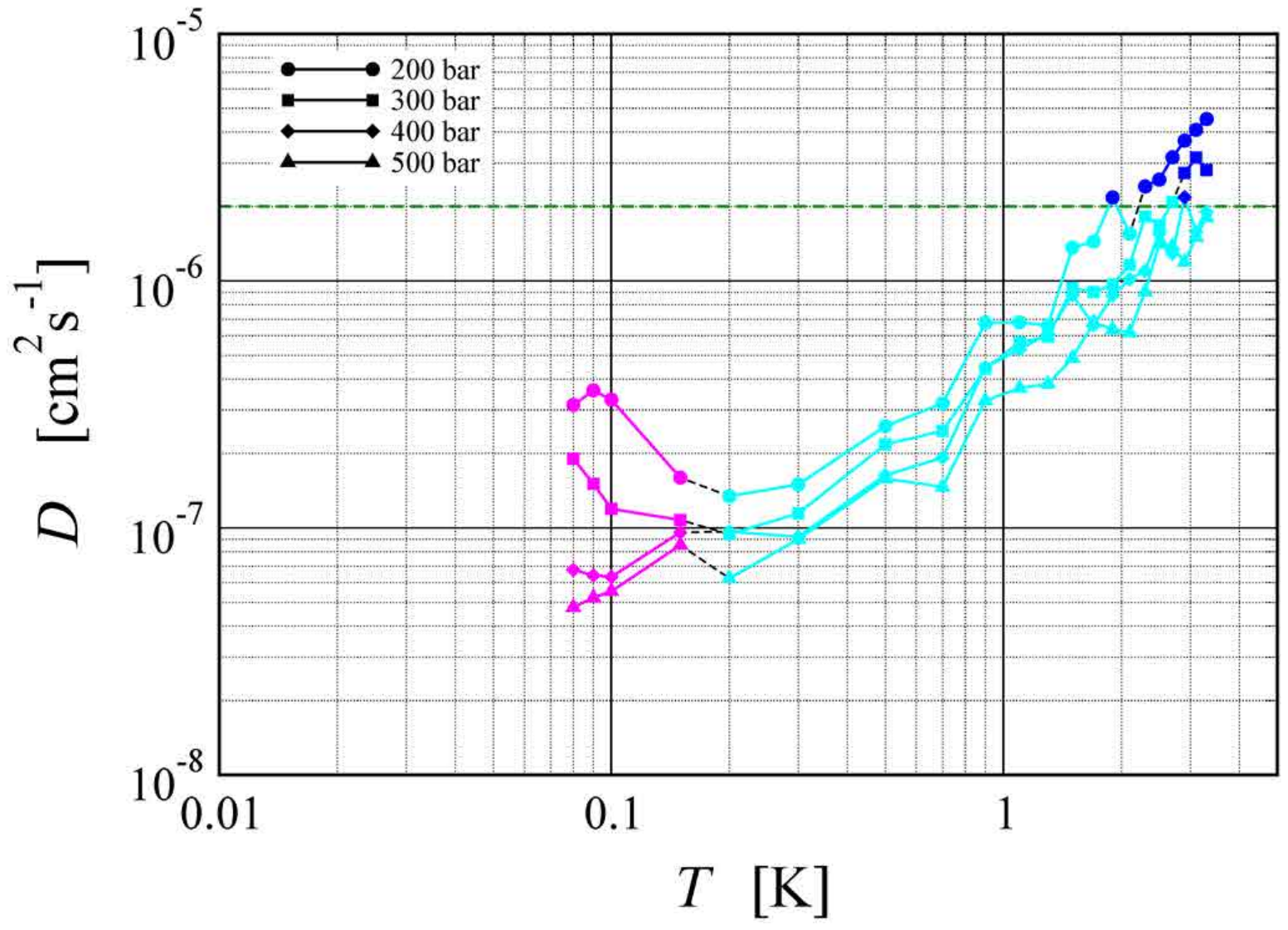}
\caption{\label{fig:FigSA-13} 
 The temperature dependence of apparent self-diffusion coefficient above 200 bar.
The colors of the symbols are the same as those displayed in Fig. 1 in the main text: LQDL (blue), 
LQDA (cyan), HQDL (red), and HQDA (magenta).  Green horizontal dashed line denotes the
threshold between liquids and glasses.
}
\end{figure}   
\end{midpage}

\pagebreak

\begin{midpage}
\begin{figure}[H]
\centering
\includegraphics[width=16cm]{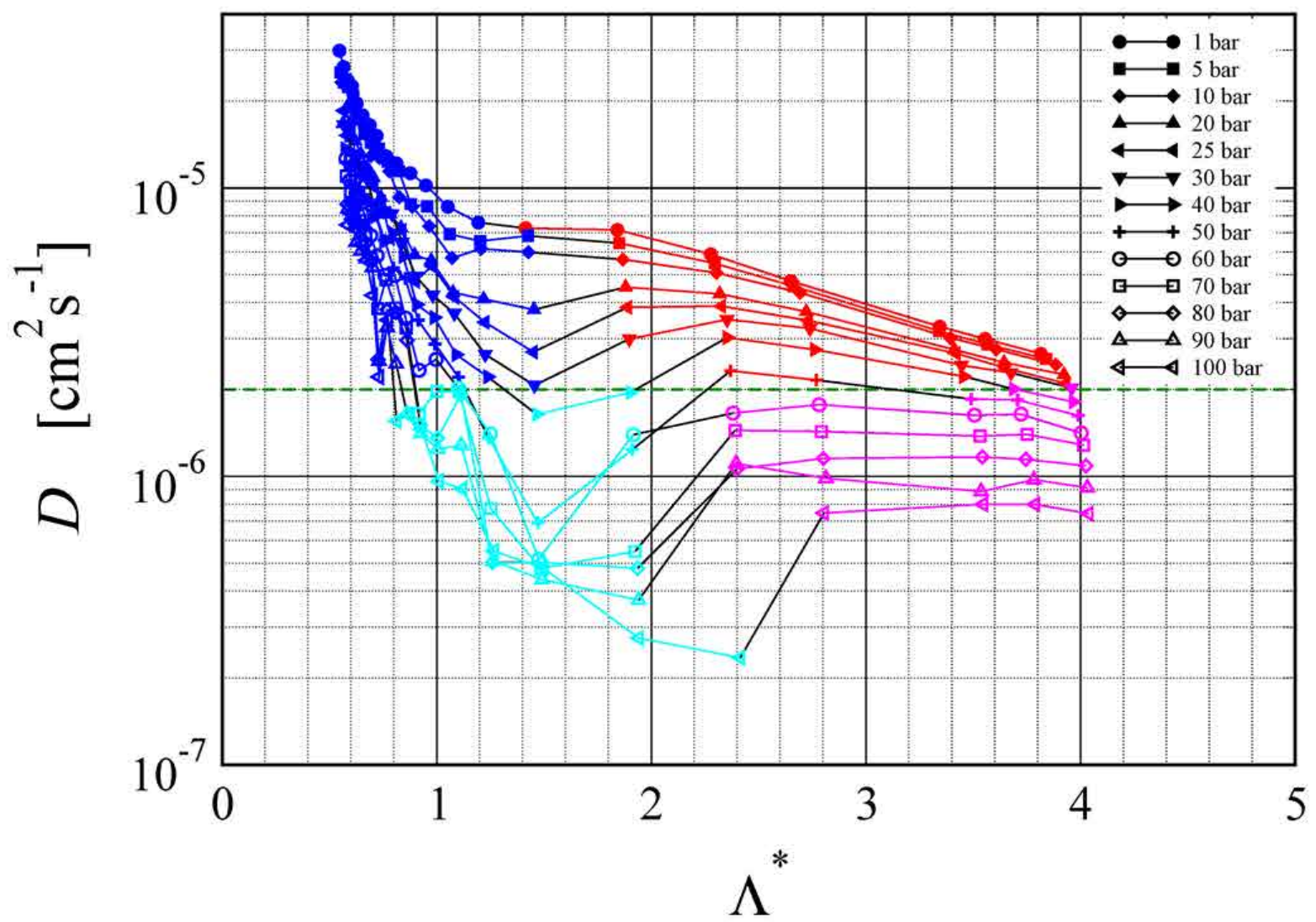}
\caption{\label{fig:FigSA-14} 
 The quantumness parameter dependence of apparent self-diffusion coefficient below 100 
bar.  The colors of the symbols are the same as those displayed in Fig. 1 in the main text: LQDL (blue), 
LQDA (cyan), HQDL (red), and HQDA (magenta).  Green horizontal dashed line denotes the
threshold between liquids and glasses.
}
\end{figure}   
\end{midpage}

\newpage

\section*{Part B. Complete set of radial distribution functions and mean square displacements}
\setcounter{figure}{0}
\renewcommand{\thefigure}{SB-\arabic{figure}}

\begin{midpage}
\begin{figure}[H]
\centering
\includegraphics[width=18cm]{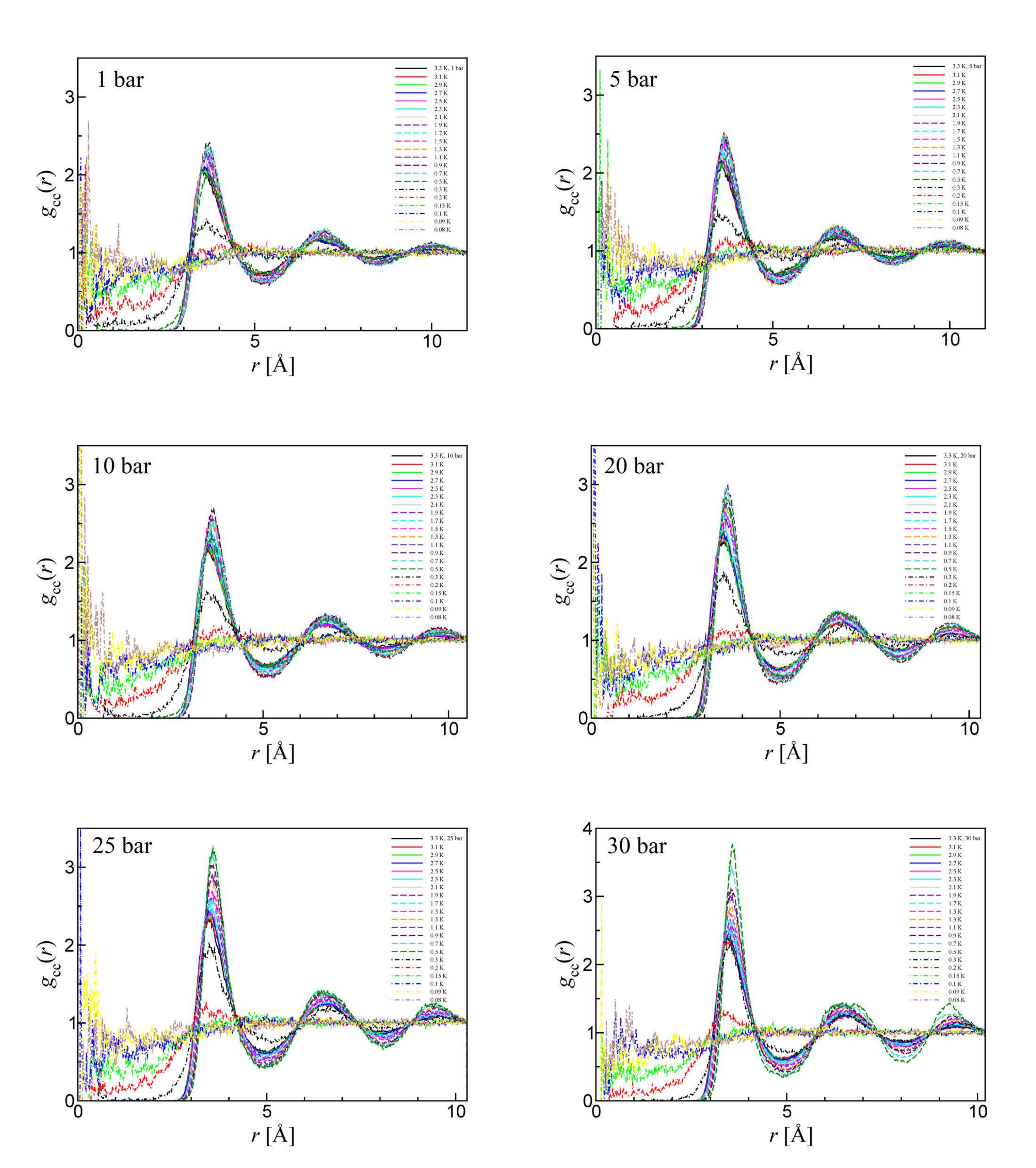}
\caption{\label{fig:FigSB-1-1} 
 The centroid-centroid radial distribution functions.
 }
\end{figure}    
\end{midpage}

\pagebreak

\setcounter{figure}{0}

\begin{midpage}
\begin{figure}[H]
\centering
\includegraphics[width=18cm]{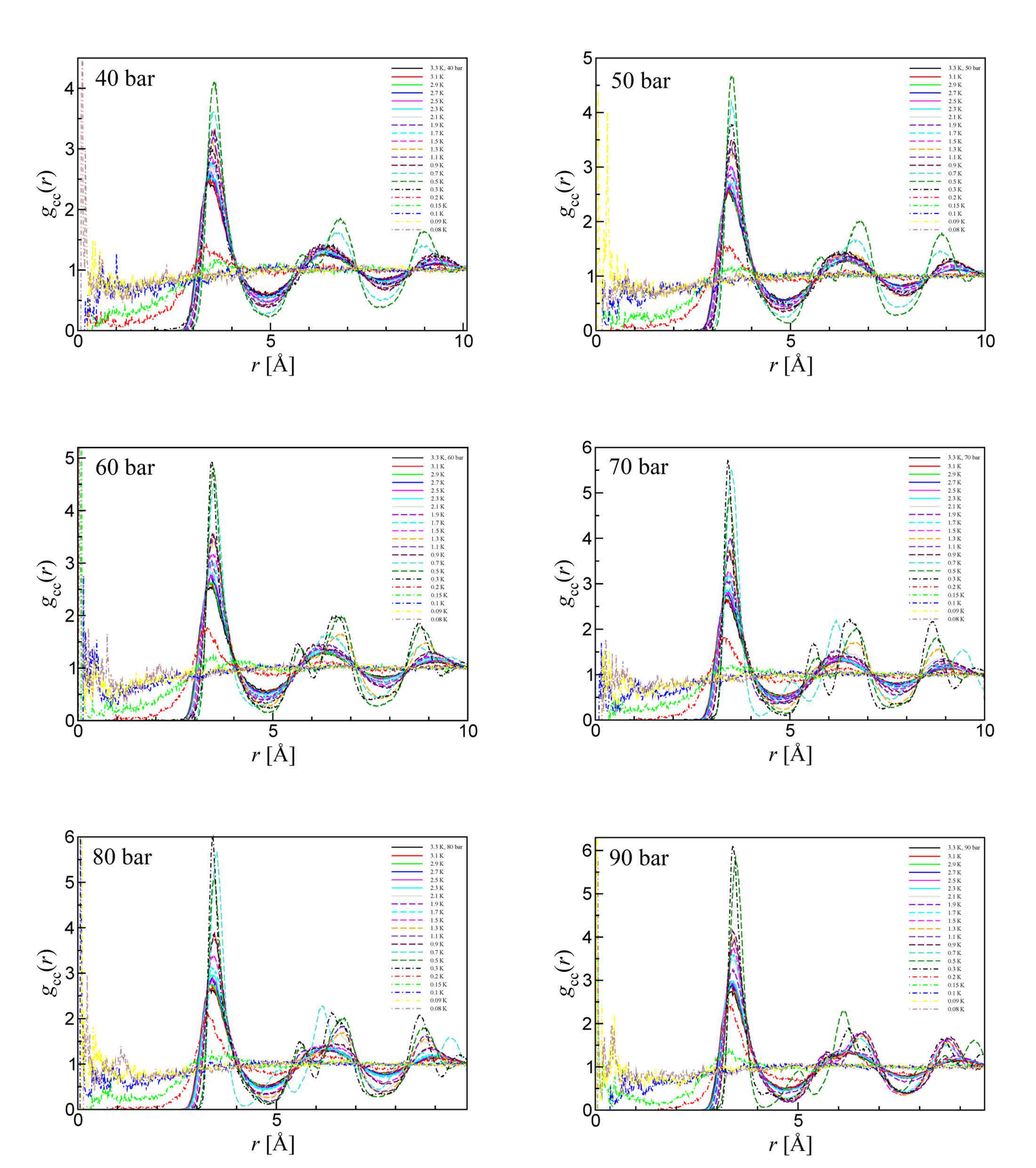}
\caption{\label{fig:FigSB-1-2} 
(continued) The centroid-centroid radial distribution functions.
}
\end{figure}    
\end{midpage}

\pagebreak

\setcounter{figure}{0}

\begin{midpage}
\begin{figure}[H]
\centering
\includegraphics[width=18cm]{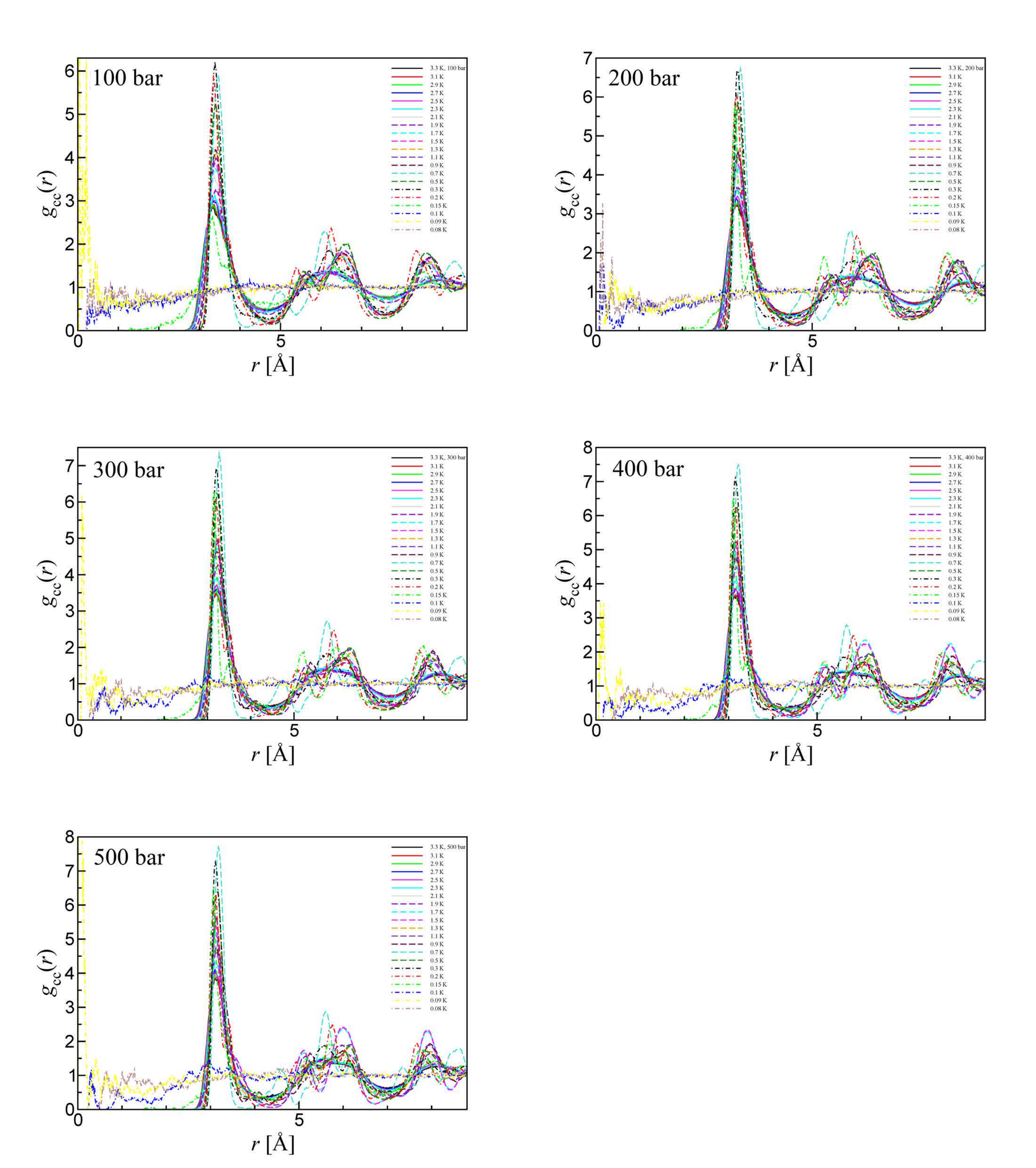}
\caption{\label{fig:FigSB-1-3} 
(continued) The centroid-centroid radial distribution functions.
}
\end{figure}   
\end{midpage}

\pagebreak

\begin{midpage}  
\begin{figure}[H]
\centering
\includegraphics[width=18cm]{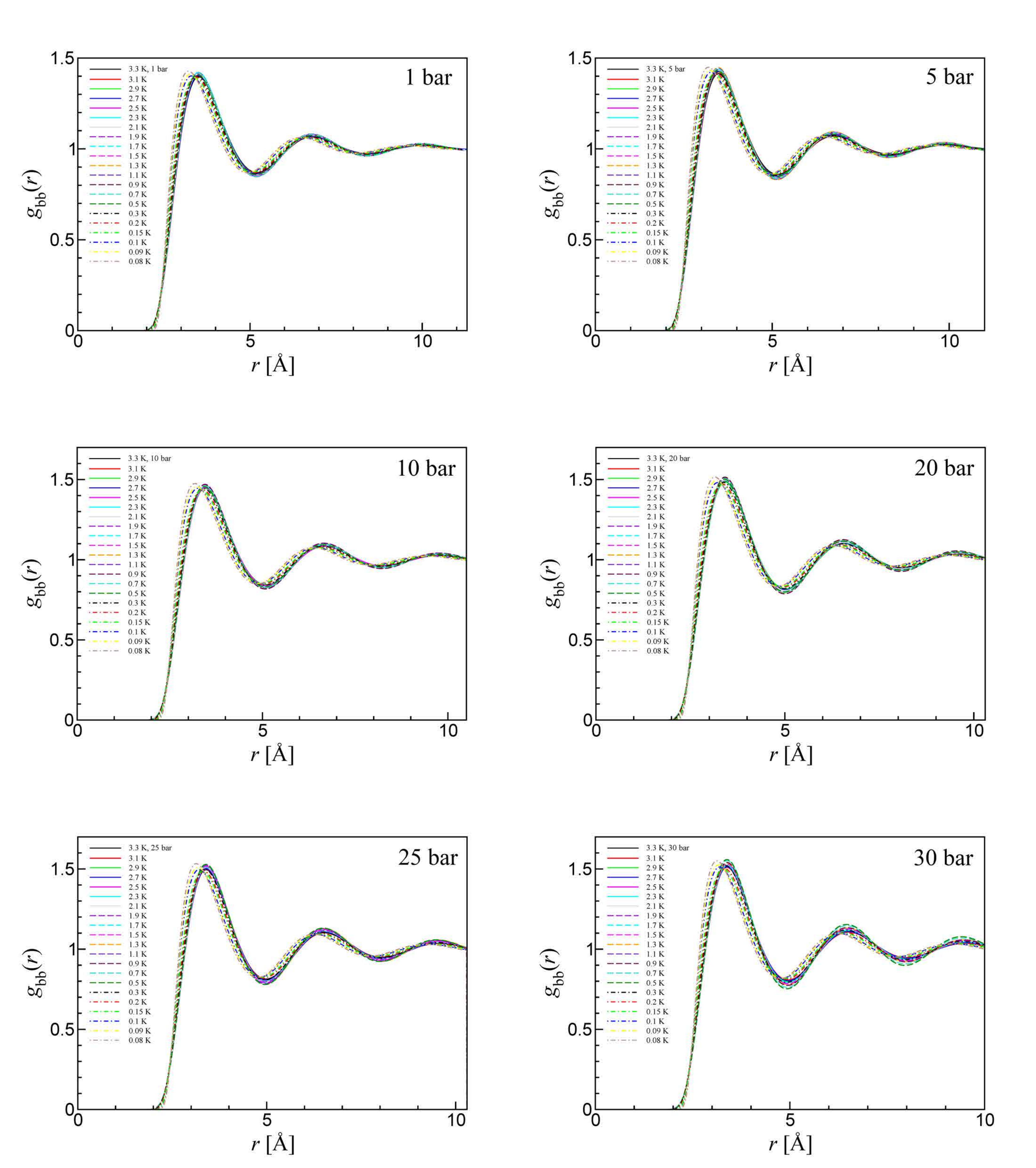}
\caption{\label{fig:FigSB-2-1} 
 The bead-bead radial distribution functions.
 }
\end{figure}  
\end{midpage}

\pagebreak

\setcounter{figure}{1}

\begin{midpage}
\begin{figure}[H]
\centering
\includegraphics[width=18cm]{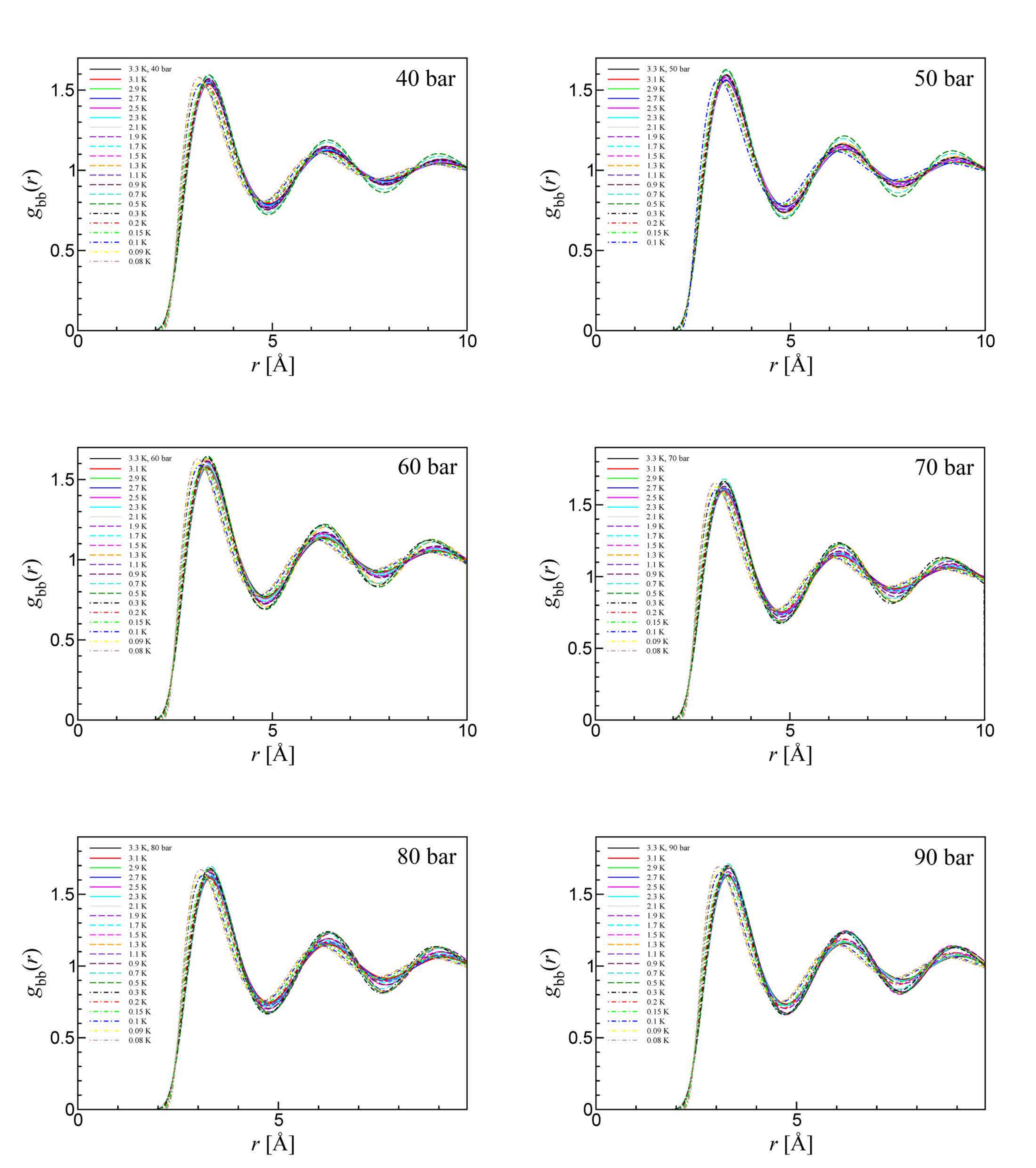}
\caption{\label{fig:FigSB-2-2} 
(continued) The bead-bead radial distribution functions.
}
\end{figure}   
\end{midpage}  

\pagebreak

\setcounter{figure}{1}

\begin{midpage}
\begin{figure}[H]
\centering
\includegraphics[width=18cm]{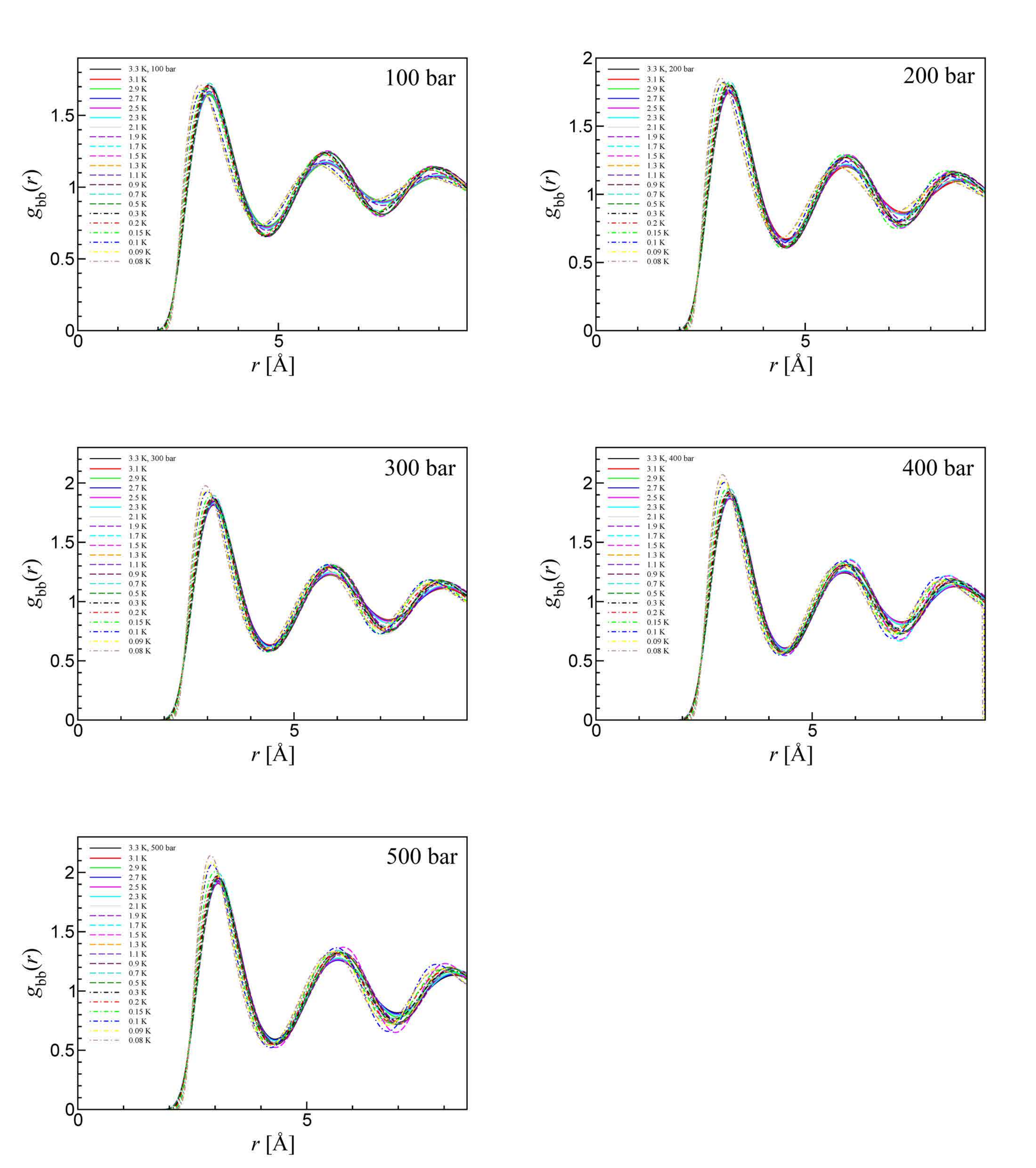}
\caption{\label{fig:FigSB-2-3} 
(continued) The bead-bead radial distribution functions.
}
\end{figure}   
\end{midpage}

\pagebreak

\begin{midpage}
\begin{figure}[H]
\centering
\includegraphics[width=18cm]{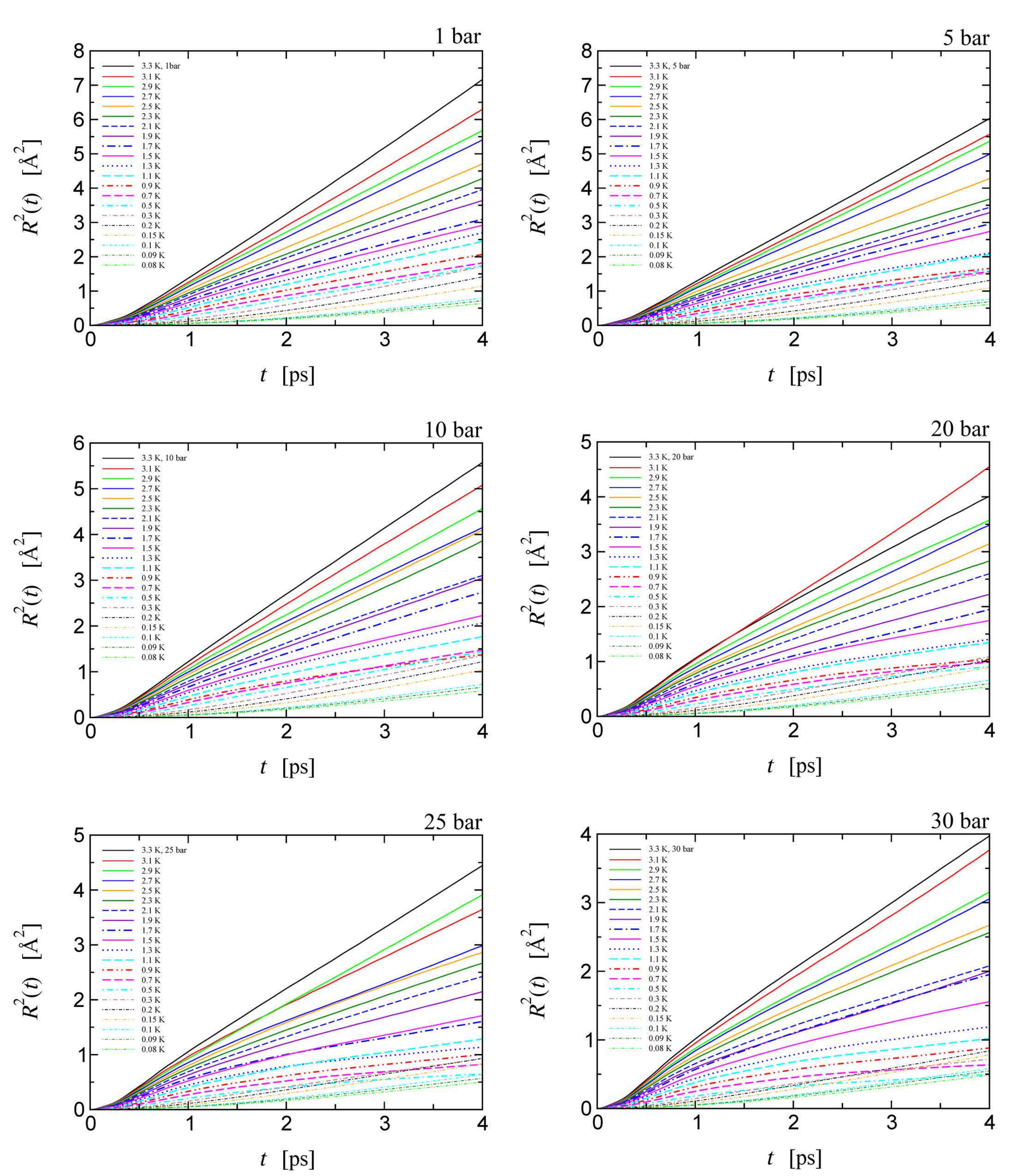}
\caption{\label{fig:FigSB-3-1} 
 Mean square dispacement of atomic centroids.}
\end{figure}    
\end{midpage}

\pagebreak

\setcounter{figure}{2}

\begin{midpage}
\begin{figure}[H]
\centering
\includegraphics[width=18cm]{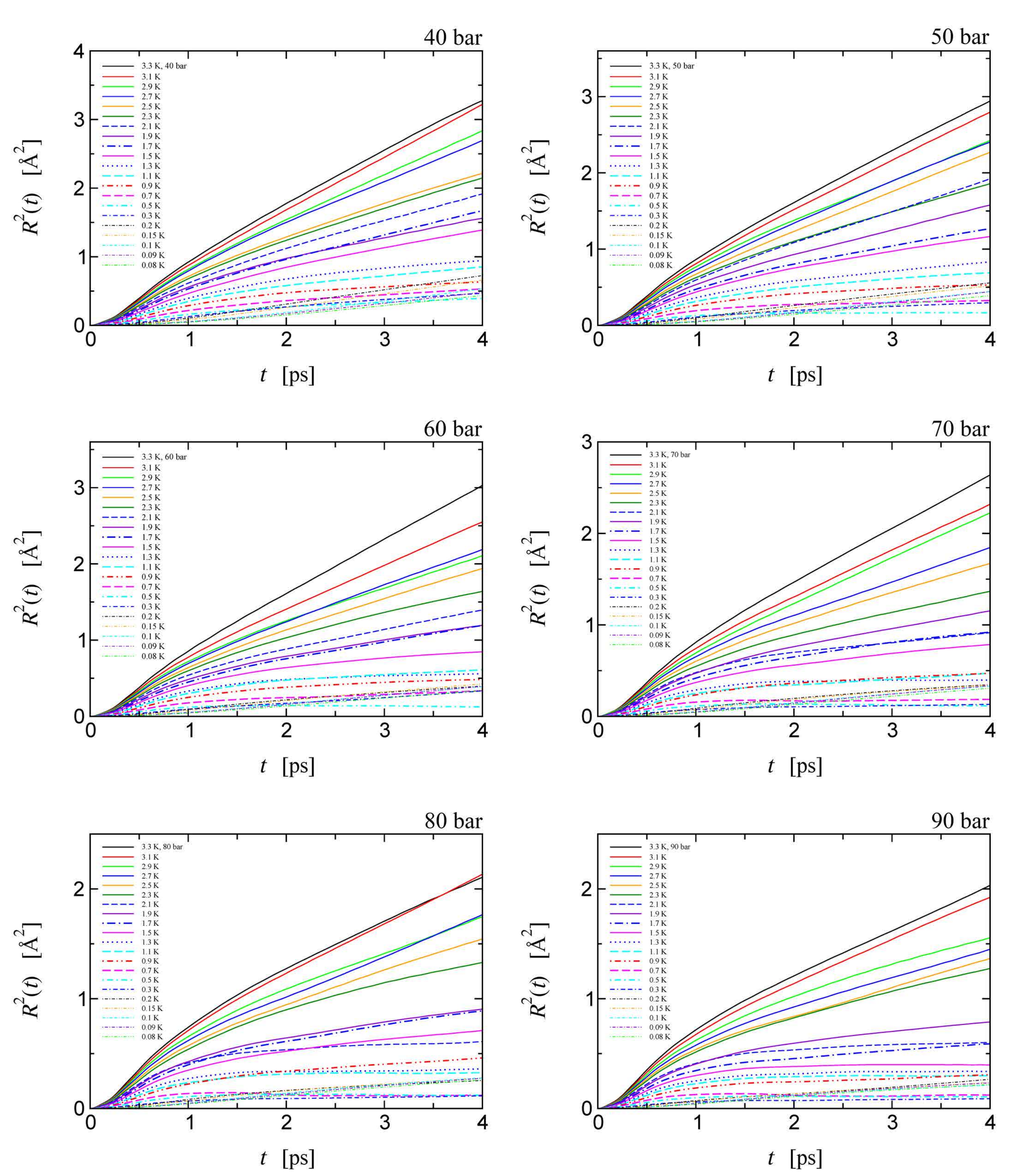}
\caption{\label{fig:FigSB-3-2} 
(continued) Mean square dispacement of atomic centroids.
}
\end{figure}     
\end{midpage}  

\pagebreak

\setcounter{figure}{2}

\begin{midpage}
\begin{figure}[H]
\centering
\includegraphics[width=18cm]{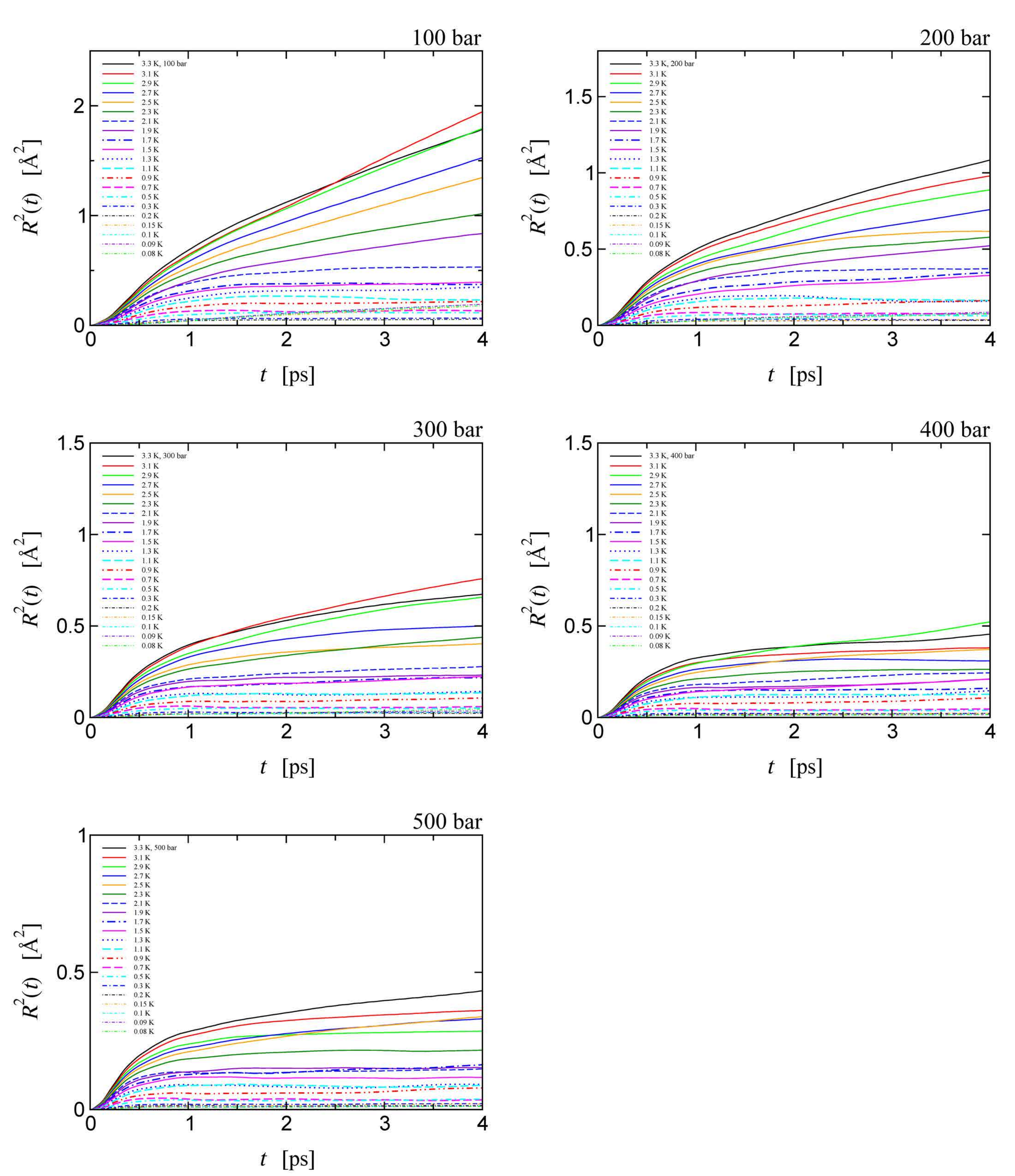}
\caption{\label{fig:FigSB-3-3} 
(continued) Mean square dispacement of atomic centroids.
}
\end{figure}   
\end{midpage}

\pagebreak

\setcounter{figure}{3}

\begin{midpage}  
\begin{figure}[H]
\centering
\includegraphics[width=18cm]{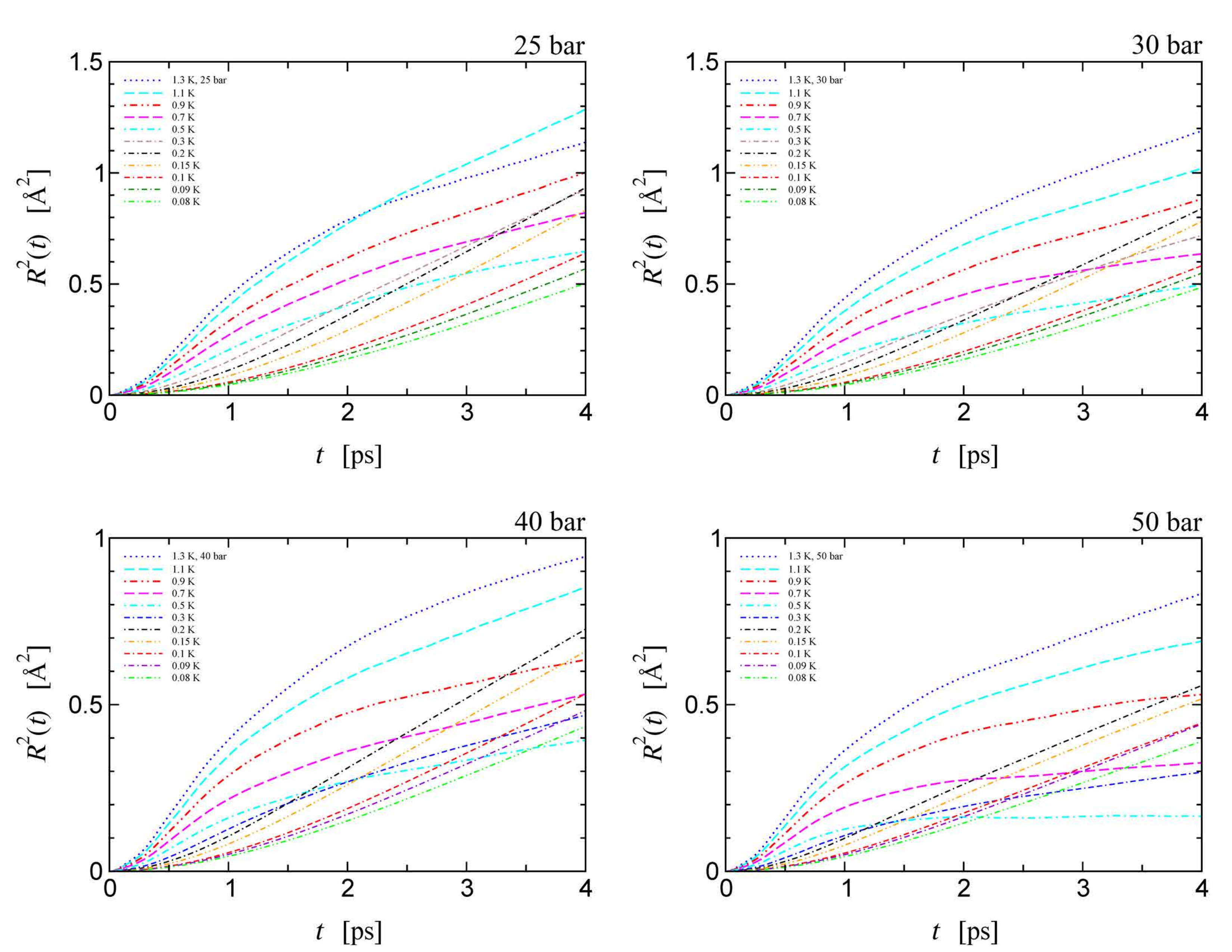}
\caption{\label{fig:FigSB-4} 
Enlarged plot of mean square dispacement of atomic centroids.
}
\end{figure}   
\end{midpage}

\newpage

\section*{Part C. Supplementary data of the test simulation of inverse freezing}
\setcounter{figure}{0}
\setcounter{table}{0}
\renewcommand{\thefigure}{SC-\arabic{figure}}

\renewcommand{\tablename}{TABLE SC-}

\begin{table}[H]
\caption{\label{tab:tableSC1}The averaged properties before and after the state transitions by isobaric heating test
at 40 bar.}
\begin{ruledtabular}
\begin{tabular}{ccccc}
Property          &  0.2 K   &  0.3 K                         &   0.4 K                         &  Increment $\Delta{X}$ \\
 $X$              &  (HQDL)  & (LQDA, partially crystallized) & (LQDA, partially crystallized)  & from 0.2 K to 0.3 K    \\
\hline
$U$ [Jmol$^{-1}$] &  -80.8   &   -64.6                        & -57.6                           &  16.2                  \\
$K$ [Jmol$^{-1}$] & 173.0    &   197.0                        &     200.3                       &  24.0                  \\
$\Phi$ [Jmol$^{-1}$] & -253.8  &   -261.6                     &    -257.9                       &  -7.8                  \\
$H$ [Jmol$^{-1}$] & 2.8     &    17.0                        &      24.6                       &  14.2                  \\
$V$ [cm$^{3}$mol$^{-1}$] & 20.89  &   20.40                  &      20.56                      &  -0.49                 \\
$\lambda_{\rm{quantum}}$ [\AA] & 3.47 &   1.84               &      1.68                      &  -1.63                 \\
$\alpha$                & 0.514  &   0.334                  &       0.353                      &  -0.180               \\
\end{tabular}
\end{ruledtabular}
\end{table}

\pagebreak

\begin{midpage}
\begin{figure}[H]
\centering
\includegraphics[width=8cm]{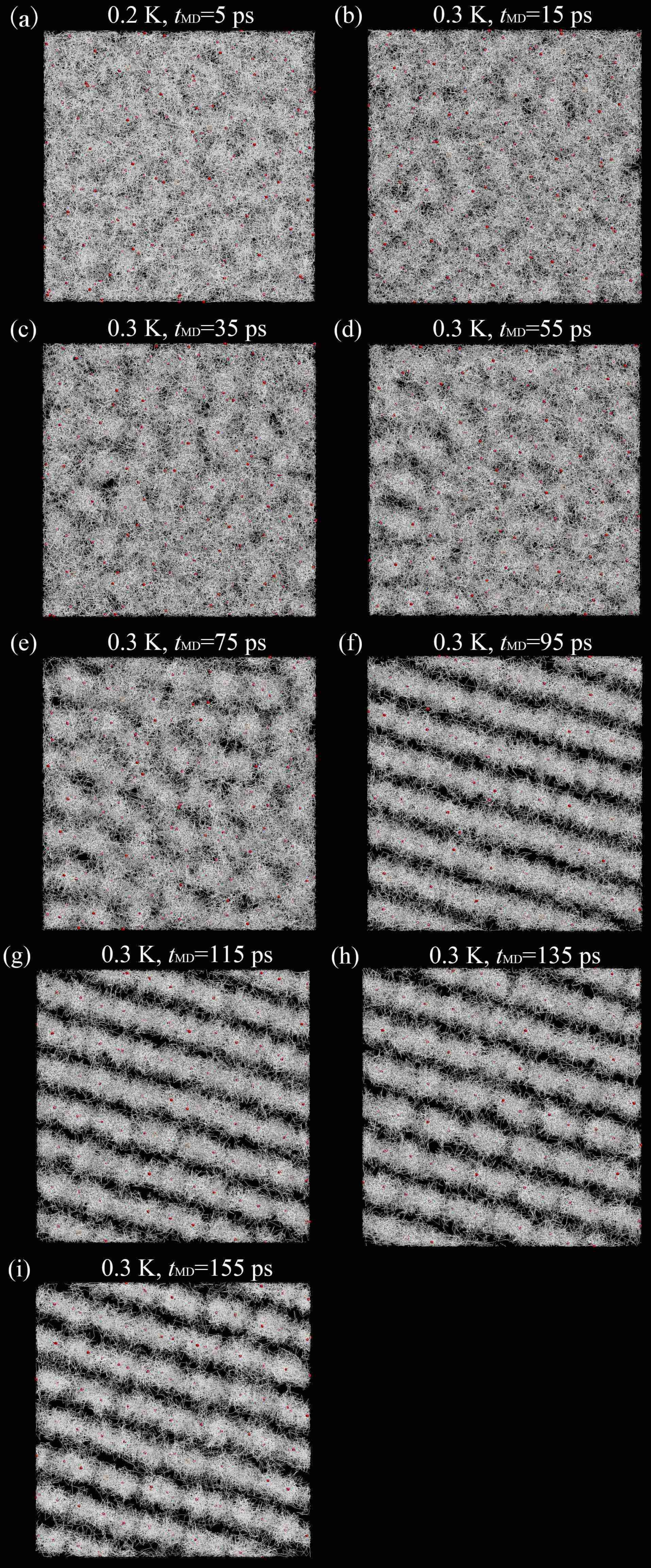}
\caption{\label{fig:FigSC-1} 
The $xy$-projected snapshots of the configuration of the $^4$He atomic necklaces and centroids during the process of isobaric heating from 0.2 K to 0.3 K at 40 bar.
}
\end{figure}
\end{midpage}

\end{document}